\documentclass{aa}  
\usepackage{color}
\usepackage{listings}
\usepackage{caption}
\usepackage{xcolor}
\usepackage{hyperref}
\usepackage[amssymb]{SIunits}
\usepackage{natbib}
\usepackage{supertabular}
\usepackage{cprotect}
\usepackage{placeins}
\usepackage{cancel}
\usepackage{graphicx}
\usepackage{txfonts}

\definecolor{red}{rgb}{1,0,0}
\definecolor{green}{rgb}{0,1,0}
\definecolor{blue}{rgb}{0,0,1}

\begin{document}

\title{Faint end of the $z \sim 3 - 7 $ luminosity function of Lyman-alpha emitters behind lensing clusters observed with MUSE\thanks{Table 4 and the four MUSE cubes used in this work are available in electronic form at the CDS via anonymous ftp to \url{cdsarc.u-strasbg.fr} (\url{130.79.128.5}) or via \url{http://cdsweb.u-strasbg.fr/cgi-bin/qcat?J/A+A/}, or at \url{http://muse-vlt.eu/science/}}}

\titlerunning{Faint end of the $z \sim 3 - 7 $ LAE LF behind lensing clusters observed with MUSE }

   \author{G. de La Vieuville \inst{1} \and
     D. Bina \inst{1} \and
     R. Pello \inst{1} \and
     G. Mahler \inst{2} \and
     J. Richard \inst{2} \and
     A. B. Drake \inst{3} \and
     E. C. Herenz \inst{4} \and
     F. E. Bauer \inst{5,6,7} \and
     B. Clément \inst{2} \and
     D. Lagattuta \inst{2} \and
     N. Laporte \inst{1,8} \and
     J. Martinez \inst{2}\and
     V. Patrício \inst{2,9}\and
     L. Wisotzki \inst{10} \and
     J. Zabl \inst{1}\and
     R. J. Bouwens \inst{11} \and
     T. Contini \inst{1} \and
     T. Garel \inst{2,12} \and
     B. Guiderdoni \inst{2} \and
     R. A. Marino \inst{13} \and
     M. V. Maseda \inst{11} \and
     J. Matthee \inst{13} \and
     J. Schaye \inst{11} \and
     G. Soucail \inst{1}}

   \institute{
            Institut de Recherche en Astrophysique et Planétologie (IRAP), Université de Toulouse, CNRS, UPS, CNES, 14 avenue Edouard Belin, F-31400 Toulouse, France. \email{gdelavieuvil@irap.omp.eu}
            \and
            Univ Lyon, Univ Lyon1, Ens de Lyon, CNRS, Centre de Recherche Astrophysique de Lyon UMR5574, F-69230, Saint-Genis-Laval, France
            \and
            Max Planck Institute für Astronomie, Königstuhl 17, D-69117, Heidelberg, Germany
            \and
            Department of Astronomy, Stockholm University, AlbaNova University Centre, SE-106 91, Stockholm, Sweden
            \and
            Instituto de Astrofísica, Facultad de Física, Pontificia Universidad Católica de Chile, Casilla 306, Santiago 22, Chile
            \and
            Space Science Institute, 4750 Walnut Street, Suite 205, Boulder, Colorado 80301, USA
            \and
            Millenium Institute of Astrophysics, Santiago, Chile
            \and
            Department of Physics and Astronomy, University College London, Gower Street, London WC1E 6BT, UK
            \and
            Dark Cosmology Centre, Niels Bohr Institute, University of Copenhagen, Juliane Maries Vej 30, 2100 Copenhagen, Denmark
            \and
              Leibniz-Institut für Astrophysik Potsdam (AIP), An der Sternwarte 16, D-14482 Potsdam, Germany
            \and
            Leiden Observatory, Leiden University, P.O. Box 9513, 2300 RA, Leiden, The Netherlands
            \and
             Observatoire de Gen\`eve, Université de Gen\`eve, 51 Ch. des Maillettes, 1290 Versoix, Switzerland
             \and
             Department of Physics, ETH Zurich,Wolfgang—Pauli—Strasse 27, 8093 Zurich, Switzerland}

   \date{Received 19 October 2018; accepted 29 May 2019}

 
\abstract
{This paper presents the results obtained with the Multi-Unit Spectroscopic
Explorer (MUSE) at the ESO Very Large Telescope on the faint end of the
Lyman-alpha luminosity function (LF) based on deep observations of four
lensing clusters. The goal of our project is to set strong constraints
on the relative contribution of the Lyman-alpha emitter (LAE) population
to cosmic reionization.
}
{The precise aim of the present study is to further constrain the abundance of LAEs by taking advantage of the magnification provided by lensing clusters to build a blindly selected sample of galaxies which is less biased than current blank field samples in redshift and luminosity. By construction, this sample of LAEs is complementary to those built from deep blank fields, whether observed by MUSE or by other facilities, and makes it possible to determine the shape of the LF at fainter levels, as well as its evolution with redshift.}
{We selected a sample of 156 LAEs with redshifts between $2.9 \le z \le 6.7$ and magnification-corrected luminosities
in the range  $ 39 \lesssim  \log L_{Ly_{\alpha}}$ [erg s$^-1$] $\lesssim 43$. To properly take into account the
individual differences in detection conditions between the LAEs when
computing the LF, including lensing configurations, and spatial and spectral
morphologies, the non-parametric $1/V_{\rm max}$ method was adopted. The price to pay to benefit from magnification is a reduction of the effective volume of the survey, together with a more complex analysis procedure to properly determine the
effective volume $V_{\rm max}$ for each galaxy. In this paper we present a complete procedure for the determination of the LF based on IFU detections in lensing clusters. This procedure, including some new methods for masking, effective volume integration and
(individual) completeness determinations, has been fully automated when possible,
and it can be easily generalized to the analysis of IFU observations in blank fields.}
{ As a result of this analysis, the Lyman-alpha LF has been obtained in four different redshift bins: $2.9 < z < 6,7$, $2.9 < z < 4.0$, $4.0 < z < 5.0,$ and $5.0 < z < 6.7$ with constraints down to $ \log L_{Ly_{\alpha}} = 40.5 $. From our data only, no significant evolution of LF mean slope can be found. When performing a Schechter analysis also including data from the literature to complete the present sample towards the brightest luminosities, a steep faint end slope was measured varying from $\alpha = -1.69^{+0.08}_{-0.08} $ to $\alpha = -1.87^{+0.12}_{-0.12} $ between the lowest and the highest redshift bins.}
{ The contribution of the LAE population to the star formation rate density at $ z \sim 6$ is $\lesssim 50$\% depending on the luminosity limit considered, which is of the same order as the Lyman-break galaxy (LBG) contribution. The evolution of the LAE contribution with redshift depends on the assumed escape fraction of Lyman-alpha photons, and appears to slightly increase with increasing redshift when this fraction is conservatively set to one. Depending on the intersection between the LAE/LBG populations,  the contribution of the observed galaxies to the ionizing flux may suffice to keep the universe ionized at $z \sim 6$.}

   \maketitle
%

\section{Introduction}
\label{sec:introduction}


Reionization is an important change of state of the universe after recombination,
and many resources have been devoted in recent years to understand this process.
The formation of the first structures, stars, and galaxies marked
the end of the dark ages. Following the formation of the first
structures, the density of ionizing photons was high enough to
allow the ionization of the entire neutral hydrogen content of
the intergalactic medium (IGM). It has been established that this state transition was mostly completed by $z \sim 6$ (\citealt{Fan2006, Becker2015}). However
the identification of the sources responsible for this major
transition and their relative contribution to the process is still a
matter of substantial debate.

Although quasars were initially considered as important candidates owing to their ionising
continuum, star-forming galaxies presently appear as the main contributors to the reionization \citep[see e.g.][]{Robertson2013,Robertson2015,Bouwens2015a,Ricci2017}. However a large uncertainty still remains on the actual contribution of quasars, as the faint population of quasars at high redshift remains poorly constrained \citep[see e.g.][]{Willott2010,Fontanot2012,McGreer2013}.
There are two main signatures currently used for the identification
of star-forming galaxies around and beyond the reionization epoch.
The first signature is the Lyman ``drop-out'' in the continuum bluewards with
respect to Lyman-alpha from the combined effect of interstellar
and intergalactic scattering by neutral hydrogen. Different redshift intervals can be defined to select Lyman break galaxies (LBGs) using the appropriate
colour-colour diagrams or photometric redshifts. Extensive literature
is available on this topic since the pioneering work by \citet{Steidel1996}
\citep[see e.g.][and the references therein]{Ouchi2004, Stark2009, McLure2009,Bouwens2015b}. The second method is the detection of the Lyman-alpha line to target
Lyman-alpha emitters (hereafter LAEs). The "classical"
approach is based on wide-field narrow-band (NB) surveys, targeting a precise redshift bin
\citep[e.g.][]{Rhoads2000, Kashikawa2006, Konno2014}. More recent methods made efficient use of 3D/IFU spectroscopy in pencil
beam mode with the Multi-Unit Spectroscopic
Explorer (MUSE) at the Very Large Telecope \citep[VLT;][]{Bacon2015}, which is a technique presently limited to $z \sim$7 in the optical domain.

Based on LBG studies, the UV luminosity function (LF) evolves strongly at $z \ge 4$,
with a depletion of bright galaxies with increasing redshift on one hand, and the slope of the faint end becoming steeper on the other hand
\citep{Bouwens2015b}. This evolution is consistent with the expected evolution of the halo mass function during the galaxy assembly process. Studies of LAEs have found a deficit of strongly emitting ("bright")
Lyman-alpha galaxies at $ z \ge 6.5$, whereas no significant evolution
is observed below $ z \sim 6$  \citep{Kashikawa2006, Pentericci2014, Tilvi2014};
this trend is attributed to either an increase in the fraction of neutral hydrogen in the IGM or an evolution of the parent population, or both. The LBGs and LAEs constitute two different observational approaches to selecting star-forming
galaxies, which are partly overlapping. The prevalence of Lyman-alpha emission in well-controlled samples of star-forming galaxies is also a test for the reionization history.
However, a complete and "as unbiased as possible" census of ionizing sources can only be enabled through 3D/IFU spectroscopy without any photometric preselection.


As pointed out by different authors \citep[see e.g.][]{Maizy2010},
lensing clusters are more efficient than blank fields for detailed (spectroscopic) studies at high
redshift and also to explore the faint end of the LF. In this respect, they are complementary to observations in wide blank fields, which are needed to
set reliable constraints on the bright end of both the UV and LAE LF. Several recent results in
the Hubble Frontier Fields (HFF) \citep{Lotz2017} fully confirm the benefit expected from gravitational magnification
\citep[see e.g.][]{Laporte2014, Atek2014, Infante2015, Ishigaki2015, Laporte2016, Livermore2017}.

   This paper presents the results obtained with MUSE (\citealt{Bacon2010}) at the ESO VLT on the faint end of the LAE LF based on deep observations of four
lensing clusters. The data were obtained as part of the MUSE consortium
Guaranteed Time Observations (GTO) programme and first commissioning run.
The final goal of our project in lensing clusters is to set strong constraints
on the relative contribution of the LAE population to cosmic reionization. As shown in
\citet{Richard2015} for SMACSJ2031.8-4036, 
\citet{Bina2016} for A1689, 
\citet{Lagattuta2017} for A370,
\citet{Caminha2017} for AS1063,
\citet{Karman2017} for MACS1149 and
\citet{Mahler2018} for A2744, MUSE is ideally designed for the study of lensed background sources,
in particular for LAEs at 2.9 $\le z \le$ 6.7. The
MUSE instrument provides a blind survey of the background population, irrespective of
the detection or not of the associated continuum. 
This instrument is also a unique facility capable of deriving the 2D properties
of ``normal'' strongly lensed galaxies, as recently shown by
\citet{Patricio2018}. 
In this project, an important point is that MUSE allows us to reliably recover a greater
fraction of the  Lyman-alpha flux for LAE emitters, as compared to usual long-slit
surveys or even NB imaging.

The precise aim of the present study is to further constrain the abundance of
LAEs by taking advantage of the magnification provided by lensing clusters
to build a blindly selected sample of galaxies which is less biased than
current blank field samples in redshift and luminosity. By construction,
this sample of LAEs is complementary to those built in deep blank fields,
whether observed by MUSE or by other facilities, and makes it possible to
determine in a more reliable way the shape of the LF
towards the faintest levels and its evolution with redshift.   
We focus on four well-known lensing clusters from the GTO sample, namely
Abell 1689, Abell 2390, Abell 2667, and Abell 2744. In this study we present the method and we establish the feasibility of the project before extending this approach to all available lensing clusters observed by MUSE in a future work.


In this paper we present the deepest study of the LAE LF
to date, combining deep MUSE observations with the
magnification provided by four lensing clusters.
In Sect. \ref{sec:data}, we present the MUSE data together with the ancillary \textit{Hubble Space Telescope} (HST)
data used for this project as well as the observational strategy adopted.
The method used to extract LAE sources in the MUSE cubes is presented in Sect. \ref{sec:lae_selection}.
The main characteristics and the references for the four lensing models used
in this article are presented in Sect. \ref{sec:lensing_clusters}, knowing that the
present MUSE data were also used to identify new multiply-imaged systems in these clusters,
and therefore to further improve the mass models. 
The selection of the LAE sample used in this study is presented in
Sect. \ref{sec:sample_description}. Sect. \ref{sec:lf_computation} is devoted to the
computation of the LF. In this
Section we present the complete procedure developed for the determination
of the LF based on IFU detections in lensing clusters; some additional
technical points and examples are given in appendices \ref{annex:create_mask_from_2d_image} to \ref{sec:detailed_volume_schematic}.
This procedure includes novel methods for masking, effective volume integration and
(individual) completeness determination, using as far as possible the true
spatial and spectral morphology of LAEs instead of a parametric approach. 
The parametric fit of the LF by a Schechter function, including data from
the literature to complete the present sample, is
presented in Sect. \ref{sec:lf_fit}. The impact of mass model on the faint end and the contribution of the LAE population to the
star formation rate density (SFRD) are discussed in Sect. \ref{sec:discussion}.
Conclusions and perspectives are given in Sect. \ref{sec:conclusion}. 

Throughout this paper we adopt the following cosmology:
$\Omega_{\Lambda}$ = 0.7, $\Omega_{m}$ = 0.3
and $H_{0}=$ 70\ km\ s$^{-1}$\ Mpc$^{-1}$. Magnitudes are given in the AB
system \citep{Oke1983}. 
All redshifts quoted are based on vacuum rest-frame wavelengths.

\section{Data}
\label{sec:data}

\subsection{MUSE Observations}
\label{subsec:observations}

The sample used in this study consists of four different MUSE cubes of different sizes and exposure times, covering the central regions of well-characterized lensing clusters: Abell 1689, Abell 2390, Abell 2667, and Abell 2744 (resp. A1689, A2390, A2667 and A2744 hereafter). These four clusters already had well constrained mass models before the MUSE observations, as they benefited from previous spectroscopic observations. The reference mass models can be found in \citet{Richard2010} (LoCuSS) for A2390 and A2667, in \citet{Limousin2007} for A1689, and in \citet{Richard2014} for the Frontier Fields cluster A2744.

\begin{figure}
  \centering
  \includegraphics[width=\hsize]{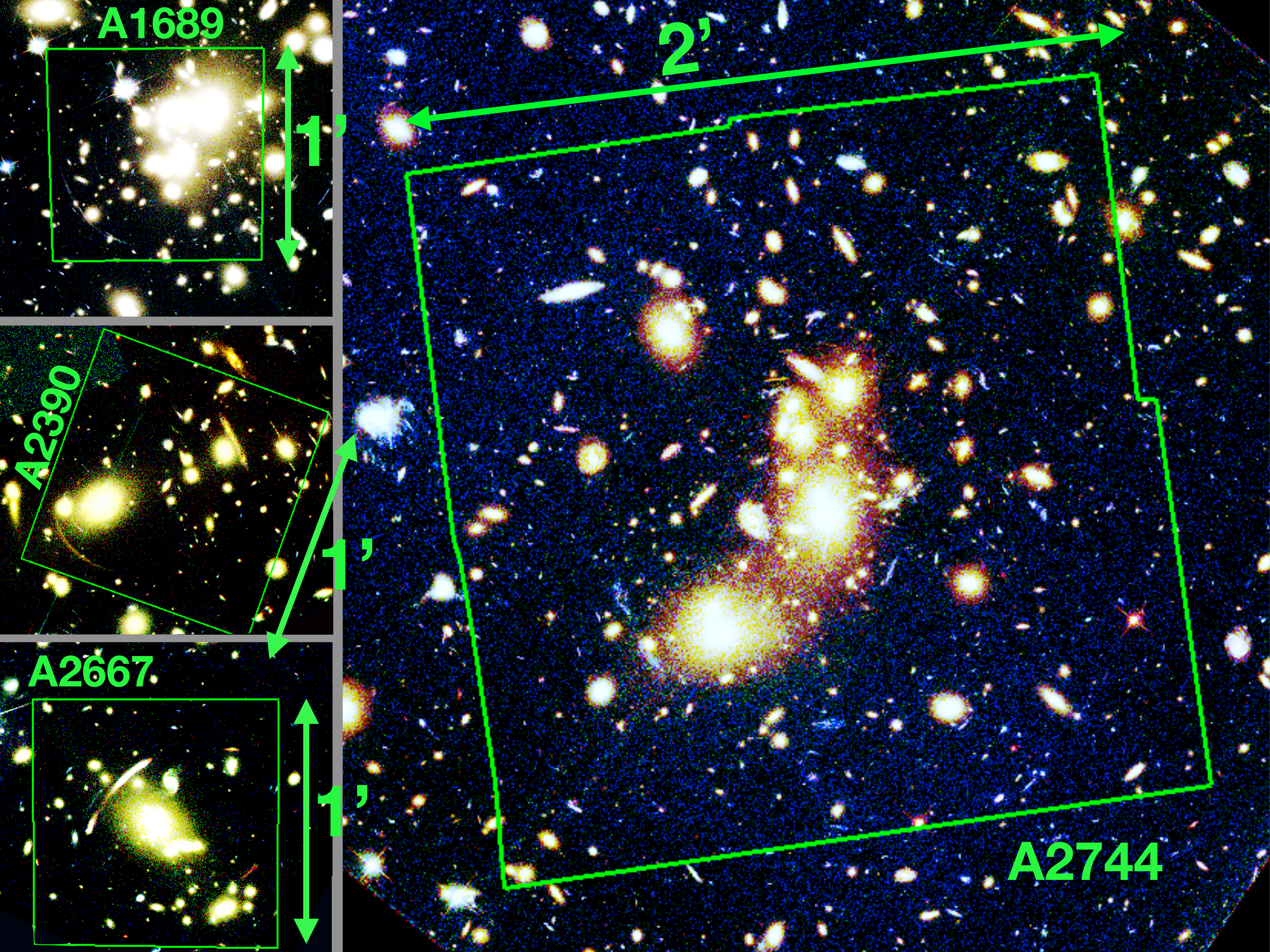}
  \caption{MUSE footprints overlaid on HST deep colour images. North is up and east is to the left. The images are obtained from the F775W, F625W, F475W filters for A1689, from F850LP, F814W, F555W for A2390, from F814W, F606W, F450W for A2667, and from F814W, F606W, F435W for A2744.}
  \label{fig:HST_images}
\end{figure}

The MUSE instrument has a $1\arcmin\times1\arcmin$ field of view (FoV) and a spatial pixel size of $0.2\arcsec$, the covered wavelength range from  $4750 \, \AA$ to $9350 \, \AA$ with a $1.25 \, \AA $ sampling, effectively making the detection of LAEs possible between redshifts of $z=2.9$ and $6.7$. The data were obtained as part of the MUSE GTO programme and first commissioning run (for A1689 only). All the observations were conducted in  the nominal WFM-NOAO-N mode of MUSE. The main characteristics of the four fields are listed in Table \ref{tab:observations_description}. The geometry and limits of the four FoVs are shown on the available HST images, in Fig. \ref{fig:HST_images}.

\paragraph{ \bf A1689:} Observations were already presented in \cite{Bina2016} from the first MUSE commissioning run in 2014.
The total exposure was divided into six individual exposures of $1100\, {\rm s}$. A small linear dither pattern of  $0.2\arcsec$ was applied between each exposure to minimize the impact of the structure of the instrument on the final data. No rotation was applied between individual exposures.

\paragraph{\bf A2390, A2667, and A2744:} The same observational strategy was used for all three cubes: the  individual pointings were divided into exposures of 1800 sec. In addition to a small dither pattern of $1\arcsec$, the position angle was incremented by $90^{\degr}$ between each individual exposure to minimize the striping patterns caused by the slicers of the instrument. A2744 is the only mosaic included in the present sample. The strategy was to completely cover the multiple-image area. For this cluster, the exposures of the four different FoVs are as follows: 3.5, 4, 4, 5 hours of exposure plus an additional 2 hours at the centre of the cluster (see fig. 1 in \citealt{Mahler2018} for the details of the exposure map). For A2390 and A2667, the centre of the FoV was positioned on the central region of the cluster as shown in Table \ref{tab:observations_description} and Fig. \ref{fig:HST_images}.\\

\begin{table*}
\caption{Main characteristics of MUSE observations. The A2744 field was splitted in two (part a and part b) because of the additional pointing covering the centre of the $2\times2$ MUSE mosaic. For A1689 and A2390, the seeing was measured on the white light image obtained from the final datacube. For A2667 and A2744, the seeing was obtained by fitting a MUSE reconstructed F814W image with a seeing convolved HST F814W image (see  \citet{Patricio2018} for A2667 and \citet{Mahler2018} for A2744).}             
\label{tab:observations_description}      
\centering                          
\begin{tabular}{c c c c c c c }        

  \hline\hline                 
         & FoV & Seeing & Integration(h) & RA (J2000)& Dec (J2000) & ESO programme\\   
  \hline
   A1689 & $1\arcmin\times 1\arcmin$ & $0.9\arcsec - 1.1\arcsec$ & 1.8  & $197^\circ 52\arcmin 39\arcsec$ & $-1^\circ 20\arcmin 42\arcsec$ &  60.A-9100(A)\\
   A2390 & $1\arcmin\times 1\arcmin$ & $0.75\arcsec$        & 2    & $328^\circ 23\arcmin 53\arcsec$ & $17^\circ 41\arcmin 48\arcsec$  & 094.A-0115(B)\\
   A2667 & $1\arcmin\times 1\arcmin$ & $0.62\arcsec  $      & 2    & $357^\circ 54\arcmin 50\arcsec$
 & $-26^\circ 05\arcmin 03\arcsec$ & 094.A-0115(A) \\
  A2744 (a) & $2\arcmin\times 2\arcmin$ & $0.58\arcsec$       & 16.5 &  $3^\circ 35\arcmin 14\arcsec$
                                                     &    $-30^\circ 23\arcmin 54\arcsec$ & 094.A-0115(B) \\
  A2744 (b) & $1\arcmin\times 1\arcmin$ & $0.58\arcsec$       & 2 &  $3^\circ 35\arcmin 14\arcsec$
 &    $-30^\circ 23\arcmin 54\arcsec$ & 094.A-0115(B) \\
  \hline
  
\end{tabular}
\end{table*}

\subsubsection{Data reduction}
\label{subsec:data_reduction}
All the MUSE data were reduced using the MUSE ESO reduction pipeline \citep{Weilbacher2012, Weilbacher2014}. This pipeline includes bias subtraction, flat fielding, wavelength and flux calibrations, basic sky subtraction, and astrometry. The individual exposures were then assembled to form a final data cube or a mosaic. An additional sky line subtraction was performed with  the Zurich Atmosphere Purge software (ZAP; \citealt{Soto2016}). This software uses principal component analysis to characterize the residuals of the first sky line subtraction to further remove them from the cubes. Even though the line subtraction is improved by this process, the variance in the wavelength layers affected by the presence of sky lines remains higher, making the source detection more difficult on these layers. For simplicity, hereafter we simply use the term layer to refer to the monochromatic images in  MUSE cubes.

\subsection{Complementary data (HST)}
\label{subsec:ancillary_data}

For all MUSE fields analysed in this paper, complementary deep data from HST are available. They were used to help the source detection process in the cubes but also for modelling the mass distribution of the clusters (see Sect. \ref{sec:lensing_clusters}). A brief list of the ancillary HST data used for this project is presented in Table \ref{tab:available_observations}.
For A1689 the data are presented in \cite{Broadhurst2005}. For A2390 and A2667, a very thorough summary of all the HST observations available are presented in \cite{Richard2008} and more recently  in \cite{Olmstead2014} for A2390. A2744 is part of the HFF programme,  which comprises the deepest observations performed by HST on lensing clusters. All the raw data and individual exposures are available from the Mikulski Archive for Space Telescopes (MAST), and the details of the reduction are addressed in the articles cited above.


\begin{table}
\caption{Ancillary HST observations. From left to right: HST instrument used, filter, exposure time, programme ID (PID), and observation epoch. }             
\label{tab:available_observations}      
\centering                          
\begin{tabular}{l  l l  c c c}        

  \hline\hline                 
  --    & Instrument & Filter           & Exp(ks) & PID & Date \\
  \hline
  A1689 & ACS & F475W   & 9.5  & 9289 & 2002 \\
        & ACS & F625W   & 9.5  & 9289 & 2002 \\
        & ACS & F775W   & 11.8 & 9289 & 2002 \\
        & ACS & F850LP & 16.6 & 9289 & 2002 \\
  A2390 &  WFPC2 & F555W & 8.4  & 5352 & 1994 \\
        &  WFPC2 & F814W & 10.5 & 5352 & 1994 \\
        &  ACS & F850LP & 6.4  & 1054 & 2006 \\
  A2667 & WFPC2 & F450W  & 12 & 8882 & 2001 \\
        & WFPC2 & F606W  & 4  & 8882 & 2001 \\
        & WFPC2 & F814W  & 4  & 8882 & 2001 \\
        & NICMOS & F110W & 18.56 & 10504 & 2006 \\
        & NICMOS & F160W & 13.43  & 10504 & 2006 \\
  A2744  & ACS & F435W    & 45    & 13495 & 2013-14\\
         & ACS & F606W    & 25    & 13495 & 2013-14\\ 
         & ACS & F814W    & 105   & 13495 & 2013-14\\

         & WFC3 & F105W   & 60   & 13495 & 2013-14\\
         & WFC3 & F125W   & 30   & 13495 & 2013-14\\
         & WFC3 & F140W   & 25   & 13495 & 2013-14\\
         & WFC3 & F160W   & 60   & 13495 & 2013-14\\
  
  \hline                                  
\end{tabular}
\end{table}

 \section{Detection of the LAE population}
\label{sec:lae_selection}

\subsection{Source detection}
\label{subsec:source_detection}

The MUSE instrument is very efficient at detecting emission
lines (see for example \citealt{Bacon2017,Herenz2017}).
On the contrary, deep photometry is well suited to detect faint objects with weak continua, with or without emission lines. To build a complete catalogue of the sources in a MUSE cube, we combined a continuum-guided detection strategy based on deep HST images (see Table \ref{tab:available_observations} for the available photometric data) with a blind detection in the MUSE cubes. Many of the sources end up being detected by both approaches and the catalogues are merged at the end of the process to make a single master catalogue. The detailed method used for the  extraction of sources in A1689 and A2744 can be found in \cite{Bina2016} and \cite{Mahler2018} \footnote{The complete catalogue of MUSE sources detected by G. Mahler in A2744 is publicly available at \url{http://muse-vlt.eu/science/a2744/}}, respectively. The general method used for A2744, which contains the vast majority of sources in the present sample, is summarized below.\\

The presence of diffuse intra-cluster light (ICL) makes the detection of faint sources difficult in the cluster core, in particular for multiple images located in this area. A running median filter computed in a window of $1.3\arcsec$ was applied to the HST images to remove most of the ICL. The ICL-subtracted images were then weighted by their inverse variance map and combined to make a single deep image. The final photometric detection was performed by \verb+SExtractor+ \citep{Bertin1996} on the weighted and combined deep images.

For the blind detection on the MUSE cubes, the \verb+Muselet+ software was used (MUSE Line Emission Tracker, written by J. Richard \footnote{Publicly available as part of the python MPDAF package \citep{Piqueras2017} : \url{http://mpdaf.readthedocs.io/en/latest/muselet.html}  }). This tool is based on \verb+SExtractor+ to detect emission-line objects from MUSE cubes. It produces spectrally weighted, continuum-subtracted NB images (NB) for each layer of the cube. The NB images are the weighted average of five wavelength layers, corresponding to a spectral width of $6.25 \AA$. These images form a NB cube, in which only the emission-line objects remain. This \verb+Sextractor+ tool is then applied to each of the NB images. At the end of the process, the individual detection catalogues are merged together and sources with several detected emission lines are assembled as one single source.\\

After building the master catalogue, all spectra were extracted and the redshifts of galaxies were measured. For A1689, A2390, and A2667, 1D spectra were extracted using a fixed $1.5\arcsec$ aperture. For A2744, the extraction area is based on the \verb+SExtractor+ segmentation maps obtained from the deblended photometric detections described above. At this stage, the extracted spectra are only used for the redshift determination. The precise measurement of the total
line fluxes requires a specific procedure, which is described in Sect. \ref{subsec:flux_computation}. Extracted spectra were manually inspected to identify the different emission lines  and accurately measure the redshift.

A system of confidence levels was adopted to reflect the uncertainty in the measured redshifts, following \citet{Mahler2018}, which has some examples that illustrate the different cases. All the LAEs used in the present paper belong to the confidence categories 2 and 3, meaning that they all have fairly robust redshift measurements. For LAEs with a single line and no continuum detected, the wide wavelength coverage of MUSE, the absence of any other line, and the asymmetry of the line were used to validate the identification of the Lyman-alpha emission. For A1689, A2390, and A2667 most of the background galaxies are part of multiple-image systems, and are therefore confirmed high redshift galaxies based on lensing considerations.\\

In total 247 LAEs were identified in the four fields: 17 in A1689, 18 in A2390, 15 in A2667, and 197 in A2744. The important difference between the number of sources found  in the different fields results from a well-understood combination of field size,  magnification regime, and exposure time, as explained in Sect. \ref{sec:sample_description}.

\subsection{Flux measurements}
\label{subsec:flux_computation}

The flux measurement is part of the main procedure developed and presented in Sect. \ref{sec:lf_computation} to compute the LF of LAEs in lensing clusters observed with MUSE. We discuss this in this section to understand the selection of the final sample of galaxies used to build the LF.

For each LAE, the flux measurement in the Lyman-alpha line was done on a continuum subtracted NB image that contains the whole Lyman-alpha emission. For each source, we built a sub-cube centred on the Lyman-alpha emission, plus adjacent blue and red sub-cubes used to estimate the spectral continuum. The central cube is a square of size $10\arcsec$ and the spectral range depends on the spectral width of the line. To determine this width and the precise position of the Lyman-alpha emission, all sources were manually inspected. The blue and red sub-cubes are centred on the same spatial position, with the same spatial extent, and are 20$\AA$ wide in the wavelength direction. A continuum image was estimated from the average of the blue and red sub-cubes and this image was subtracted pixel-to-pixel from the central NB image. For sources with large full width at half maximum (FWHM), the NB used for flux measurement can regroup more than  20 wavelength layers (or equivalently $25\,\AA$).\\

Because \verb+SExtractor+ with \verb+FLUX_AUTO+ is known to provide a good estimate of the total flux of the sources to the ~5\% level (see e.g. the SExtractor Manual, Sect. 10.4, Fig. 8.), it was used to measure the flux and the corresponding uncertainties on the continuum-subtracted images.\ The \verb+FLUX_AUTO+ routine is based on Kron first moment algorithm, and is well suited to account for the extended Lyman-alpha haloes that can be found around many LAEs (see \citealt{Wisotzki2016} for the extended nature of the Lyman-alpha emission). In addition, the automated aperture is useful to account properly for the distorted images that are often found in
lensing fields.
As our sample contains faint, low surface brightness sources, and given that the NB images are not designed to maximize the signal-to-noise ratio (S/N), it is sometimes challenging to extract sources with faint or low-surface brightness Lyman-alpha emission. In order to measure their flux we force the extraction at the position of the source. To do so, the \verb+SExtractor+ detection parameters were progressively loosened until a successful extraction was achieved. An extraction was considered successful when the source was recovered at less than a certain matching radius ($r_{\rm m } \sim 1\arcsec$)  from the original position given by \verb+Muselet+. Such an offset is sometimes observed between the peak of the UV continuum and the Lyman-alpha emission in case of high magnification. A careful inspection was needed to make sure that no errors or mismatches were introduced in the process.

Other automated alternatives to \verb+SExtractor+ exist to measure the line flux (see e.g. \verb+LSDCat+ in \citealt{Herenz2017} or \verb+NoiseChisel+ in \citealt{Akhlaghi2015} or a curve of growth approach as developed in \citet{Drake2017b}). A comparison between these different methods is encouraged in the future but beyond the scope of the present analysis.

\section{Lensing clusters and mass models}
\label{sec:lensing_clusters}

\begin{figure*}
  \centering
  \includegraphics[width=\hsize]{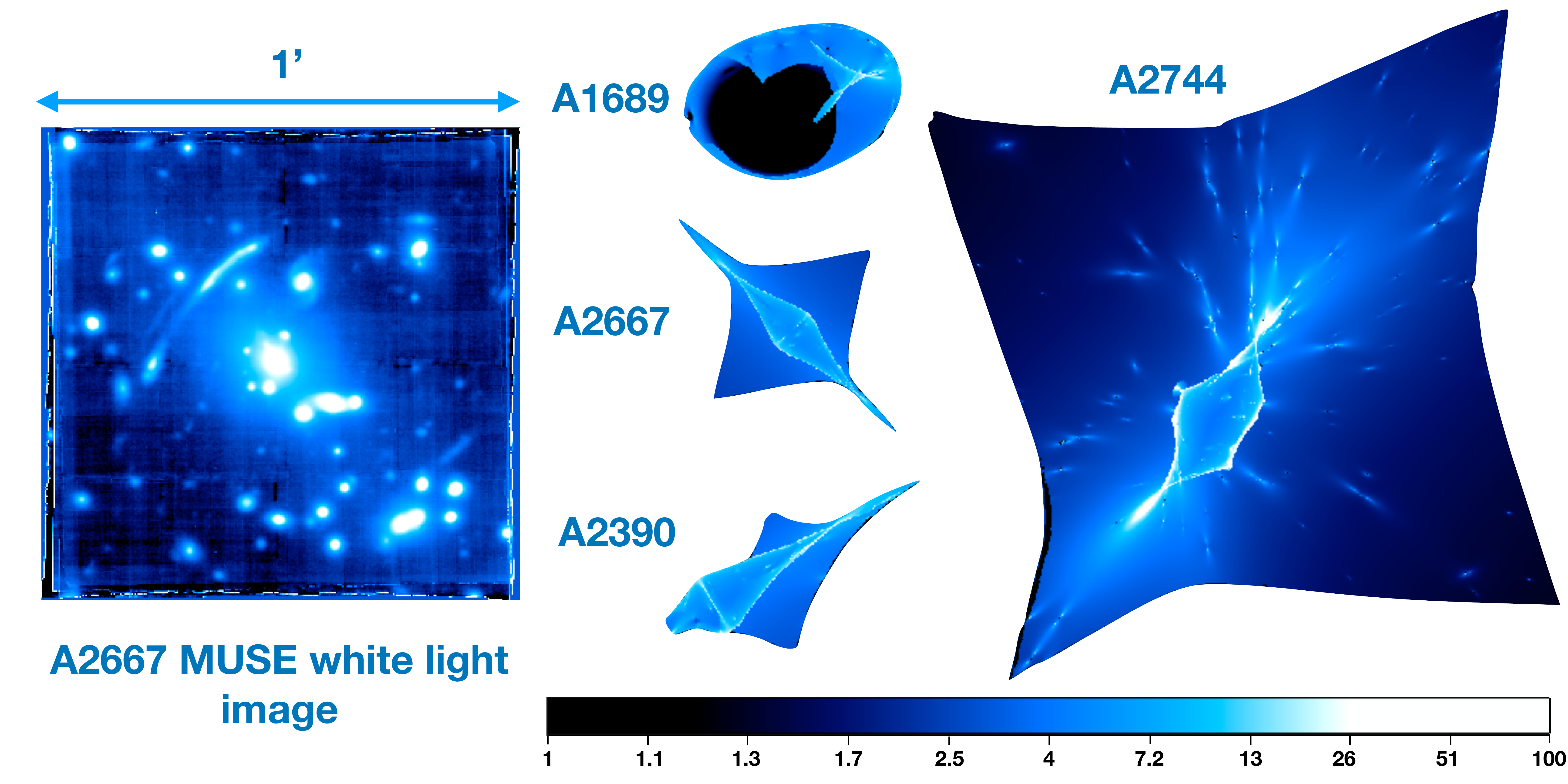}
  \caption{On the left: MUSE white light image of the A2667 field represented with  a logarithmic colour scale. On the right: projection of the four MUSE FoVs in the source plane at $z=3.5$, combined with  the magnification map encoded in the colour. All images on this figure are at the same spatial scale. In the case of multiply imaged area, the source plane magnification values shown correspond to the magnification of the brightest image.}
  \label{fig:mag_map_mosaic}
\end{figure*}

In this work, we used detailed mass models to compute the magnification of each LAE, and the source plane projections of the MUSE FoVs at various redshifts. These projections were needed when performing the volume computation (see Sect. \ref{subsec:volume_computation}). The mass models were constructed with \verb+Lenstool+, using the parametric approach described in \citet{Kneib1996}, \citet{Jullo2007}, and \citet{Jullo2009}. This parametric approach relies on the use of analytical dark-matter (DM) halo profiles to describe the projected 2D mass distribution of the cluster. Two main contributions are considered by \verb+Lenstool+: one for each large-scale structure of the cluster and one for each massive cluster galaxy. The parameters of the individual profiles are optimized through a Monte Carlo Markov Chain (MCMC) minimization. The \verb+Lenstool+ software aims at reducing the cumulative distance in the parameter space between the predicted position of multiple images obtained from the model, and the  observed images. The presence of several robust multiple systems greatly improves the accuracy of the resulting mass model. The use of MUSE is therefore a great advantage as it allowed us to confirm multiple systems through spectroscopic redshifts and also to discover new systems(e.g. \mbox{\citet{Richard2015}}; \mbox{\citet{Bina2016}}; \mbox{\citet{Lagattuta2017}}; \mbox{\citet{Mahler2018}}). Some of the  models used in this study are based on the new constraints provided by MUSE. An example of source plane projection of the MUSE FoVs is provided in Fig. \ref{fig:mag_map_mosaic}. \\

Because of the large number of cluster members, the optimization of each individual galaxy-scale clump cannot be achieved in practice. Instead, a relation combining the constant mass-luminosity scaling relation described in \citet{Faber1976} and the fundamental plane of elliptical galaxies is used by \verb+Lenstool+. This assumption allows us to reduce the parameter space explored during the minimization process, leading to more constrained mass models, whereas individual parameterization of clumps would lead to an extremely degenerate final result and therefore, a poorly constrained mass model. The analytical profiles used were double pseudo-isothermal elliptical potentials (dPIEs) as described in  \citet{Eliasdottir2007}. The ellipticity and position angle of these elliptical profiles were measured for the galaxy-scale clumps with \verb+SExtractor+  taking advantage of the high spatial resolution of the HST images.\\
Because the brightest cluster galaxies (BCGs) lie at the centre of clusters, they are subjected to numerous merging processes and are not expected to follow the same light-mass scaling relation. They are modelled separately in order to not bias the final result. In a similar way, galaxies that are close to the multiple images or critical lines are sometimes   manually optimized because of the significant impact they can have on the local magnification and geometry of the critical lines. \\ 

The present MUSE survey has allowed us to improve the reference models available in previous works. Table \ref{tab:mass_models} summarizes their main characteristics.
For A1689, the model used is an improvement made on the model of \citet{Limousin2007}, previously presented in \citet{Bina2016}.
For A2390, the reference model is presented in \citet{Pello1991}, \citet{Richard2010}, and the recent improvements in Pello et al. (in prep.) For A2667, the original model was  obtained by \citet{Covone2006} and was updated in \citet{Richard2010}. For A2744, the gold model presented in  \citet{Mahler2018} was used, including as novelty the presence of NorthGal and SouthGal, which are two background galaxies included in the mass model because they could have a local influence on the position and magnification of multiple images.

\begin{table*}
  \centering
   \renewcommand{\arraystretch}{1.4} 
\begin{tabular}{l l c c c c c c c c}     
\hline\hline       

  Cluster & Clump & $\Delta \alpha (\arcsec)$ & $\Delta \delta (\arcsec)$ & $e$   & $\theta$ & $r_{\rm core}$(kpc) & $r_{\rm cut}$(kpc) & $\sigma_{\rm 0}\text{(km s}^{-1}$) & Ref\\
  \hline
  
  A1689                & DM1 &$0.6^{+0.2}_{-0.2}$&$-8.9^{+0.4}_{-0.4}$&$0.22^{+0.01}_{-0.01}$&$91.8^{+1.4}_{-0.8}$&$100.5^{+4.6}_{-4.0}$&[1515.7]&$1437.3^{+20.0}_{-11.1}$& (1)\\
  rms = $2.87\arcsec$  & DM2 &$-70.0^{+1.4}_{-1.5}$&$47.9^{+2.3}_{-4.1}$&$0.80^{+0.04}_{-0.05}$&$80.5^{+2.7}_{-2.5}$&$70.0^{+8.0}_{-5.3}$&[500.9]&$643.2^{+0.5}_{-4.5}$&\\
  $n_{\rm const} = 128$ & BCG &$-1.3^{+0.2}_{-0.3}$&$0.1^{+0.4}_{-0.5}$&$0.50^{+0.03}_{-0.05}$&$61.6^{+9.6}_{-4.0}$&$6.3^{+1.2}_{-1.2}$&$132.2^{+42.0}_{-31.5}$& $451.6^{+11.6}_{-12.1}$&\\
  $n_{\rm free} = 33$   & Gal1&[49.1]&[31.5]&0.60$^{+0.07}_{-0.16}$&119.3$^{+6.2}_{-10.0}$&26.6$^{+3.4}_{-4.1}$&179.6$^{+2.5}_{-27.8}$& 272.8$^{+4.5}_{-21.5}$&\\
                      & Gal2 &45.1$^{+0.2}_{-0.9}$ &32.1$^{+0.6}_{-1.1}$&0.79$^{+0.05}_{-0.03}$&42.6$^{+2.3}_{-1.9}$&18.1$^{+0.3}_{-3.4}$&184.8$^{+1.2}_{-11.1}$&432.7$^{+16.6}_{-33.4}$&\\
                &$L^{*}$ Gal & & & & & [0.15]&18.1$^{+0.7}_{-2.2}$&151.9$^{+7.0}_{-0.3}$   \\
  \hline

  A2390               & DM1 & 31.6$^{+1.8}_{-1.3}$&15.4$^{+0.4}_{-1.0}$&0.66$^{+0.03}_{-0.02}$&214.7$^{+0.5}_{-0.3}$&261.5$^{+8.5}_{-5.2}$&[2000.0]&1381.9$^{+23.0}_{-17.6}$& (2) \\
  rms $= 0.33\arcsec$ & DM2 &[-0.9]&[-1.3]&0.35$^{+0.05}_{-0.03}$&33.3$^{+1.2}_{-1.6}$&25.0$^{+1.8}_{-1.1}$&750.4$^{+100.2}_{-65.5}$& 585.1$^{+20.0}_{-9.7}$& (3)\\
  $n_{\rm const} = 45$ & BCG1&[46.8]&[12.8]& 0.11$^{+0.10}_{-0.01}$&114.8$^{+26.8}_{-31.5}$&[0.05] &23.1$^{+3.0}_{-1.6}$& 151.9$^{+5.9}_{-7.5}$& (4)\\
  $n_{\rm free} = 18$  & $L^{*}$ Gal & & & &  & [0.15]   & [45.0]  & 185.7$^{+5.3}_{-3.3}$ & \\    
  \hline

  A2667               &DM1 &0.2$^{+0.5}_{-0.4}$&1.3$^{+0.5}_{-0.4}$&0.46$^{+0.02}_{-0.02}$&-44.4$^{+0.2}_{-0.3}$&79.33$^{+1.1}_{-1.1}$&[1298.7]&1095.0$^{+5.0}_{-3.7}$& (5) \\
  rms $ = 0.47\arcsec$& $L^{*}$ Gal & & & &                   & [0.15] & [45.0]  &91.3$^{+4.5}_{-4.5}$& (3)\\
  $n_{\rm const} = 47$  &  & & &  &  & &  &  & \\
  $n_{\rm free} = 9$&   & & &  &  & &  &  & \\
  \hline

  A2744               & DM1 &-2.1$^{+0.3}_{-0.3}$&1.4$^{+0.0}_{-0.4}$&0.83$^{+0.01}_{-0.02}$&90.5$^{+1.0}_{-1.1}$&85.4$^{+5.4}_{-4.5}$&[1000.0]&607.1$^{+7.6}_{-0.2}$& (6) \\
  rms $= 0.67\arcsec$ & DM2 &-17.1$^{+0.2}_{-0.3}$&-15.7$^{+0.4}_{-0.3}$&0.51$^{+0.02}_{-0.02}$&45.2$^{+1.3}_{-0.8}$&48.3$^{+5.1}_{-2.2}$&[1000.0]&742.8$^{+20.1}_{-14.2}$&\\
  $n_{\rm const} = 134$& BCG1 &[0.0]&[0.0]&[0.21]&[-76.0]&[0.3]&[28.5]& 355.2$^{+11.3}_{-10.2}$\\

  $n_{\rm const} = 30$ & BCG2 &[-17.9]&[-20.0]&[0.38]& [14.8]&[0.3]&[29.5]&321.7$^{+15.3}_{-7.3}$\\
          & NGal &[-3.6]&[24.7]&[0.72]&[-33.0] & [0.1]& [13.2]&175.6$^{+8.7}_{-13.8}$\\

                     & SGal &[-12.7]&[-0.8]&[0.30]&[-46.6]&[0.1]&6.8$^{+93.3}_{-3.2}$& 10.6$^{+43.2}_{-3.6}$\\
                & $L^{*}$ Gal & & & &                  &[0.15]& 13.7$^{+1.0}_{-0.6}$&155.5$^{+4.2}_{-5.9}$\\
\hline                  
\end{tabular}
\cprotect\caption{Summary of the main mass components for the lensing models used for this work. The values of RMS indicated are computed from the position of multiply imaged galaxies in the image plane, $n_{\rm const}$ and $n_{\rm free} $ correspond to the number of constraints passed to \verb+Lenstool+ and the number of free parameters to be optimized, respectively. The coordinates $ \Delta \alpha $ and  $ \Delta \delta$ are in arcsec with respect to the following reference points: {\bf A1689:} $\alpha =    197^\circ 52\arcmin 23\arcsec$, $\delta =    -1^\circ 20\arcmin 28\arcsec $, {\bf A2390:  }
    $\alpha =       328^\circ 24\arcmin 12\arcsec$, $\delta = 17^\circ 41\arcmin 45\arcsec $, {\bf A2667: }
    $\alpha =     357^\circ 54\arcmin 51\arcsec$, $\delta = -26^\circ 05\arcmin 03\arcsec$   {\bf A2744: }
    $\alpha =  3^\circ 35\arcmin 11\arcsec$, $\delta =    -30^\circ 24\arcmin 01\arcsec$. The ellipticity $e$, is defined as $(a^2 - b^2)/(a^2 + b^2)$, where $a$ and $b $ are the semi-major and the semi-minor axes of the ellipse. The position angle, $ \theta$, provides the orientation of the semi-major axis of the ellipse measured counterclockwise with respect to the horizontal axis. Finally, $r_{\rm core}$, $r_{\rm cut}$, and $\sigma_{\rm 0}$ are the core radii, cut radii, and central velocity dispersion, respectively. References are as follows: (1) \citet{Limousin2007}, (2) \citet{Pello1991}, (3) \citet{Richard2010}, (4) Pello et al. (in prep.), (5) \citet{Covone2006}, and (6) the gold model from \citet{Mahler2018} }           
\label{tab:mass_models}
\end{table*}


\section{Selection of the final LAE sample}
\label{sec:sample_description}

To obtain the final LAE sample used to build the LF, only one source per multiple-image system  was retained. The ideal strategy would be to keep the image with the highest S/N, which often coincides with  the image with highest magnification. However, it is more secure for the needs of the LF determination to keep the sources with the most reliable flux measurement and magnification determination. In practice, it means that we often chose the less distorted and most isolated image. The flux and extraction of all sources among multiple systems were manually reviewed to select the best one to be included in the final sample. All the sources for which the flux measurement failed or that were too close to the edge of the FoV were removed from the final sample. One extremely diffuse and low surface brightness source (Id : A2744, 5681) was also removed as it was impossible to properly determine its profile for the completeness estimation in Sect. \ref{subsubsec:estimating_source_profile}. 

The final sample consists of 156 lensed LAEs: 16 in A1689, 5 in A2390, 7 in A2667, and 128 in A2744. Out of these 156 sources, four are removed at a later stage of the analysis for completeness reasons (see Sect. \ref{sec:recovering_mock}) leaving 152 to compute the LFs. The large difference between the clusters on the number of sources detected is expected for two reasons: 
 \begin{itemize}
 \item[-] The A2744 cube is a $2\times 2 $ MUSE FoV mosaic and is deeper than the three other fields: on average four hours exposure time for each quadrant, whereas all the others have two hours or less of integration time (see Table \ref{tab:observations_description}).
 \item[-] The larger FoV allows us to reach further away from the critical lines of the cluster, therefore increasing the probed volume as we get close to the edges of the mosaic.
 \end{itemize}

\begin{figure}
  \centering
  \includegraphics[width=\hsize]{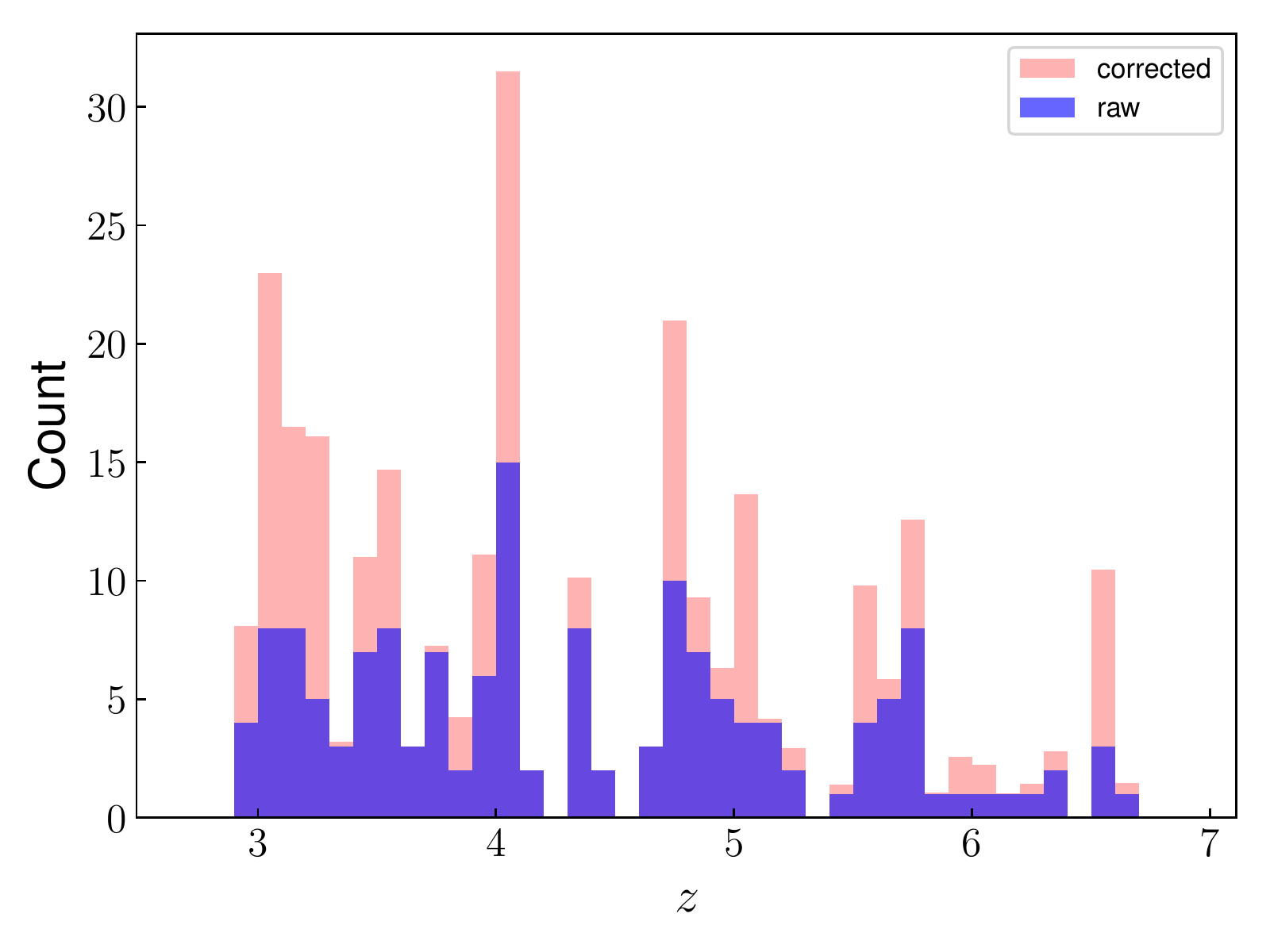}
  \includegraphics[width=\hsize]{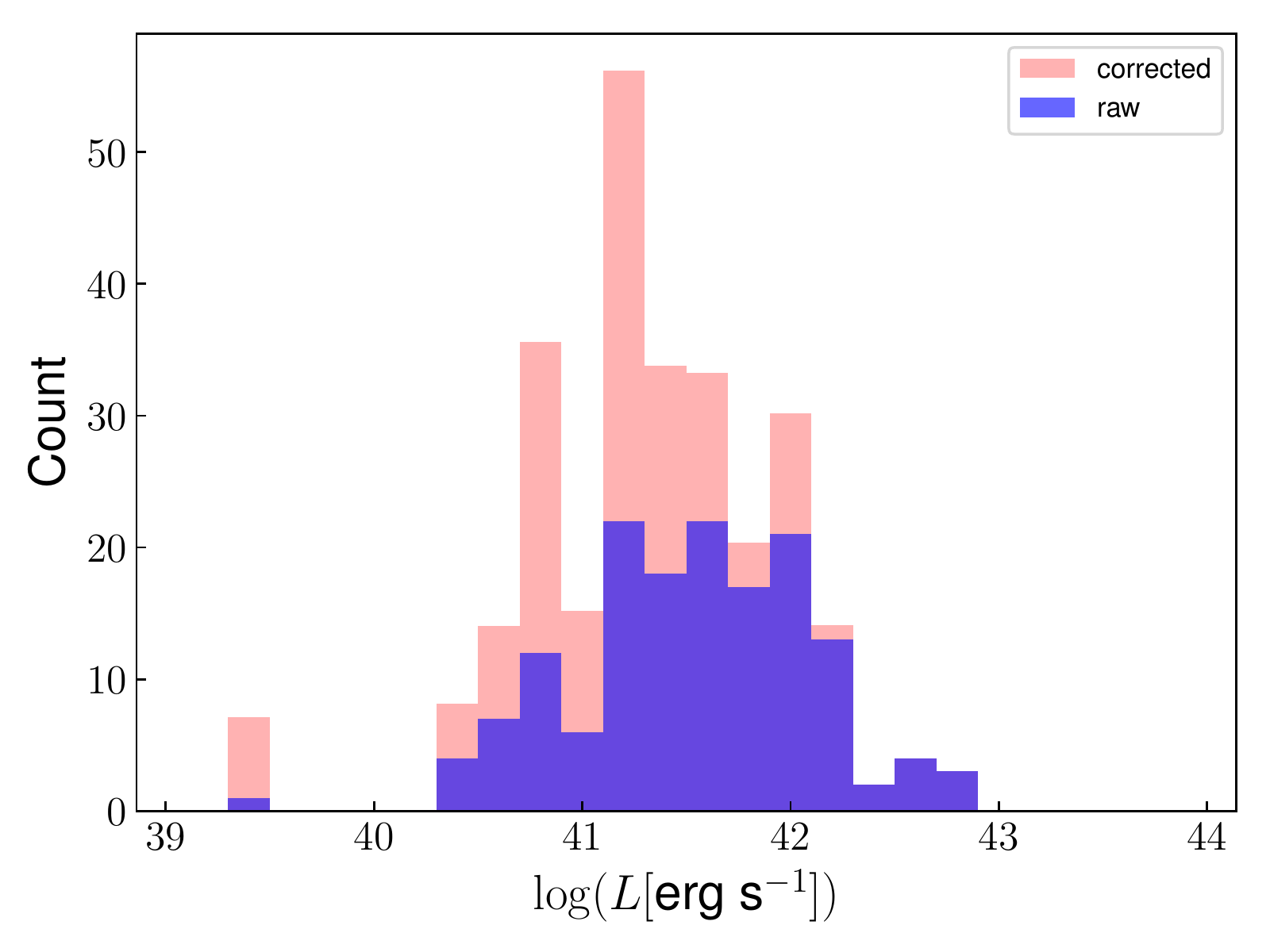}
  \caption{Redshift and magnification corrected luminosity distribution of the 152 LAEs used for the LF computation (in blue). The corrected histograms in light red correspond to the histogram of the population weighted by the inverse of the completeness of each source (see Sect. \ref{subsec:completeness_estimation}). The empty bins seen on the redshift histograms are not correlated with the presence of sky emission lines.} 
  \label{fig:sample_distributions}
\end{figure}

 \noindent This makes the effective volume of universe explored in the A2744 cube much larger (see end of Sect. \ref{subsubsec:volume_integration}) than in the three other fields combined. It is therefore not surprising to find most of the sources in this field. This volume dilution effect is most visible when looking at the projection of the MUSE FoVs in the source plane (see Fig. \ref{fig:mag_map_mosaic}). Even though this difference is expected, it seems that we are also affected by an over-density of background sources at $z = 4 $ as shown in Fig. \ref{fig:sample_distributions}. This over-density is currently being investigated as a potential primordial group of galaxies (Mahler et al. in prep.). The complete source catalogue is provided in Table \ref{tab:sample_tab} and the Lyman-alpha luminosity distribution corrected for magnification can be found on the lower panel of Fig. \ref{fig:sample_distributions}. The corrected luminosity $L_{Ly_{\alpha}}$ was computed from the detection flux $F_{Ly_{\alpha}}$ with
 \begin{equation}
   \label{eq:flux_lum_conversion}
  L_{Ly_{\alpha}} = \frac{F_{Ly_{\alpha}}}{\mu} 4 \pi D_{L}^2 
,\end{equation}

\noindent where $\mu$ and $D_{L}$ are the magnification and luminosity distance of the source, respectively. In this section and in the rest of this work, a flux weighted magnification is used to better account for extended sources and for sources detected close to the critical lines of the clusters where the magnification gradient is very strong. This magnification is computed by sending a segmentation of each LAE in the source plane with \verb+Lenstool+, measuring a magnification for each of its pixels and making a flux weighted average of it. A full probability density of magnification $P(\mu)$ is also computed for each LAE and used in combination with its uncertainties on $F_{Ly_{\alpha}}$ to obtain a realistic luminosity distribution when computing the LFs (see  Sect. \ref{subsec:lf_points}). Objects with the highest magnification are affected by the strongest uncertainties and tend to have very asymmetric $P(\mu)$ with a long tail towards high magnifications. Because of this effect, LAEs with  $\log L < 40 $  should  be considered with great caution.

\defcitealias{Drake2017b}{D17}

\begin{figure}
  \centering
  \includegraphics[width=\hsize]{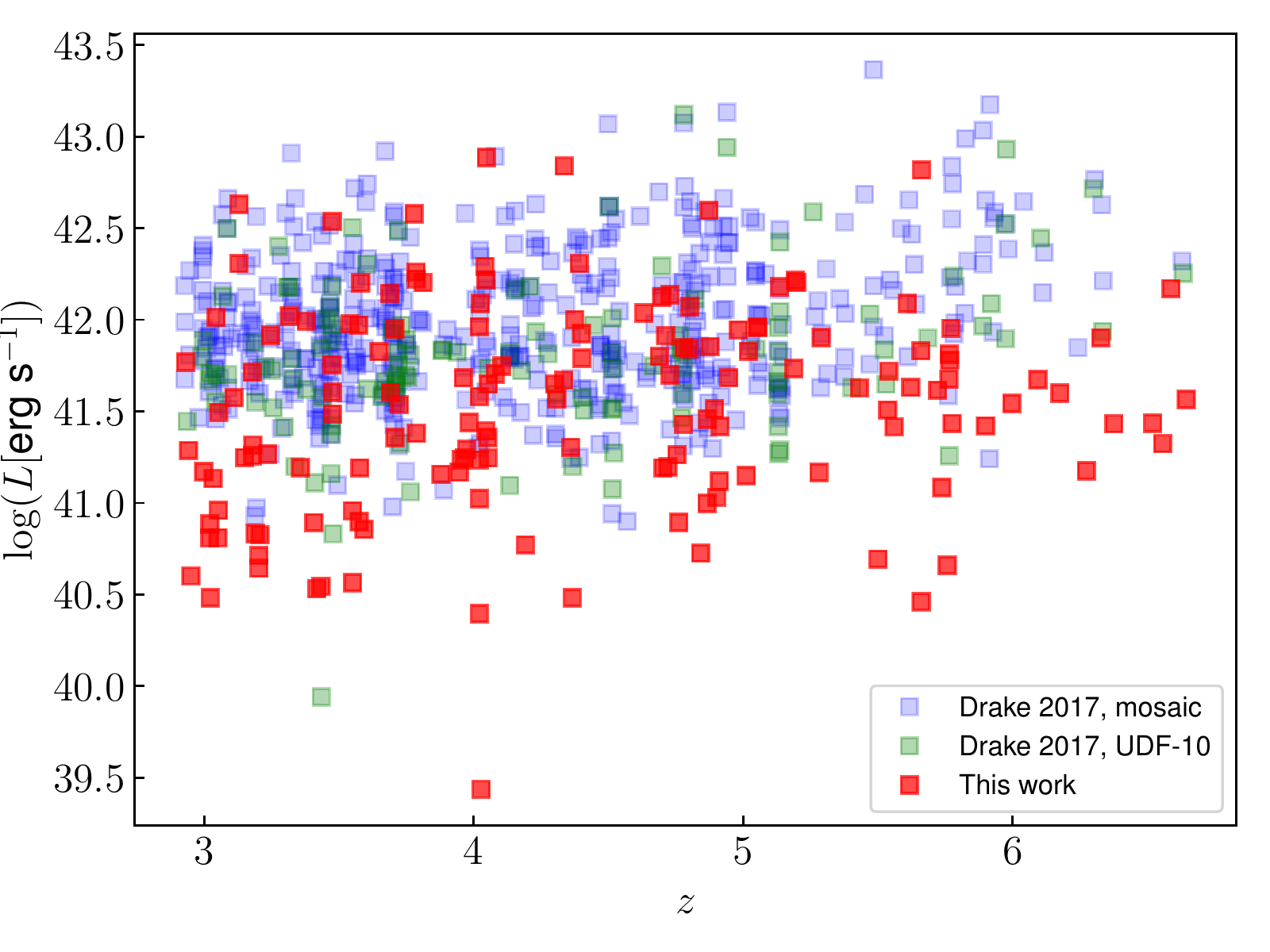}
  \includegraphics[width=\hsize]{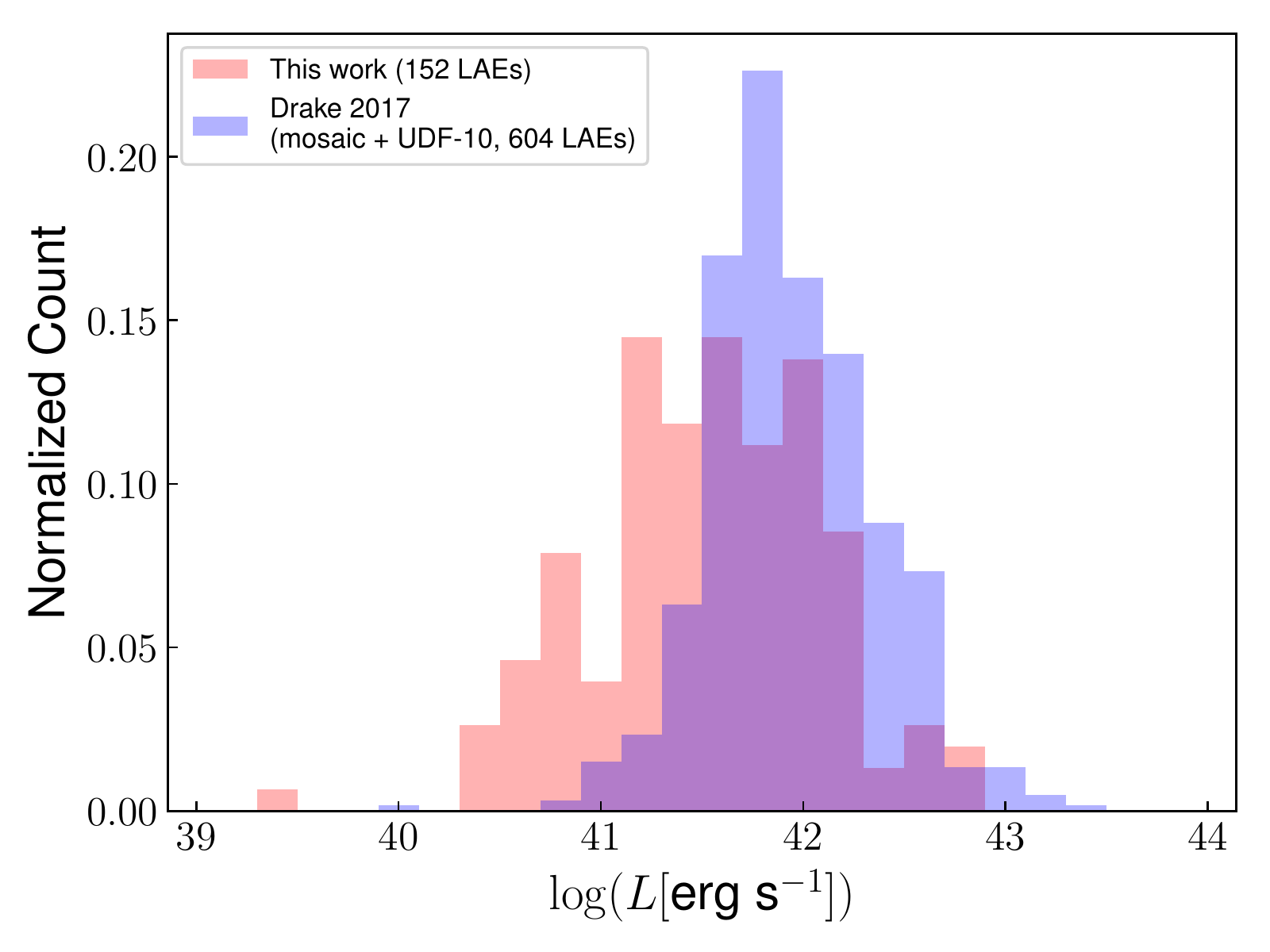}
  \caption{Comparison of the 152 LAEs sample used in this work with \citetalias{Drake2017b}. Upper panel: luminosity vs. redshift; error bars have been omitted for clarity. Lower panel: luminosity distribution of the two samples, normalized using the total number of sources. The use of lensing clusters allows for a broader selection, both in redshift and luminosity towards the faint end.}
  \label{fig:comparison_alyssa}
\end{figure}

Figure \ref{fig:comparison_alyssa} compares our final sample with the sample used in the MUSE HUDF LAE LF study (\citealt{Drake2017b}, hereafter \citetalias{Drake2017b}). The MUSE HUDF \citep{Bacon2017}, with a total of 137 hours of integration, is the deepest MUSE observation to date. It consists of a  $3 \times 3$ MUSE FoV \textit{mosaic}, each of the quadrants being a 10 hours exposure, with an additional pointing (\textit{udf-10}) of 30 hours, overlaid on the mosaic.
The population selected in \citetalias{Drake2017b} is composed of 481 LAEs found in the \textit{mosaic} and 123 in the \textit{udf-10}, for a total of 604 LAEs. On the upper panel of the figure, the plot presents the luminosity of the different samples versus the redshift. Using lensing clusters, the redshift selection tends to be less affected by luminosity bias, especially for higher redshift. On the lower panel, the normalized distribution of the two populations is presented. The strength of the study presented in \citetalias{Drake2017b} resides in the large number of sources selected. However, a sharp drop is observed in the distribution at $\log L \sim 41.5$. Using the lensing clusters, with $\sim 25$ hours of exposure time and a much smaller lens-corrected volume of survey, a broader luminosity selection was achieved. As discussed in the following sections, despite a smaller number of LAEs compared to \citetalias{Drake2017b}, the sample presented in this paper is more sensitive to the faint end of the LF by construction.

\setcounter{table}{4}

   
\section{Computation of the luminosity function}
\label{sec:lf_computation}

Because of the combined use of lensing clusters and spectroscopic data cubes, it is extremely challenging to adopt a parametric approach to determine a selection function. 
By construction, the sample of LAEs used in this paper includes sources coming from very different detection conditions, from intrinsically bright emitters with moderate magnification to highly magnified galaxies that could not have been detected far from the critical lines. To properly take into account these differences when computing the LF, we adopted a non-parametric approach allowing us to treat the sources individually: i.e. the $1/V_{\rm max}$ method \citep{Schmidt1968, Felten1976}. We present in this section the four steps developed to compute the LFs:

\begin{itemize}
  \item[i)] The flux computation, performed for all the \textit{detected} sources. This step was already described in  Sect. \ref{subsec:flux_computation} as the selection of the final sample relies partly on the results of the flux measurements.
  \item[ii)] The volume computation for each of the sources included in the final sample, presented in Sect. \ref{subsec:volume_computation}.
  \item[iii)] The completeness estimation using the real source profiles (both spatial and spectral), presented in Sect. \ref{subsec:completeness_estimation}.
  \item[iv)] The computation of the points of the differential LF, using the results of the volume computation and the completeness estimations, presented in Sect. \ref{subsec:lf_points}.
  \end{itemize}

\subsection{Volume computation in spectroscopic cubes in lensing clusters}
\label{subsec:volume_computation}

The  $ V_{\rm max}$ value is defined as the volume of the survey where an individual source could have been detected. The inverse value, $1 / V_{\rm max}$, is used to determine the contribution of one source to a numerical density of galaxies.
Because this survey consists of several FoV, the $V_{\rm max}$ value for a given source must be determined from all the fields that are part of the survey, including the fields in which the source is not actually present.
The volumes were computed in the source plane to avoid multiple counting of parts of the survey that are multiply imaged. For that, we used \verb+Lenstool+ to get the projection of the MUSE fields in the source plane and then used these projections to compute the volume (see Fig. \ref{fig:mag_map_mosaic} for an example of source plane projection). In this analysis, the volume computation was performed independently from the completeness estimation, focussing on the spectral noise variations of the cubes only.\\



\begin{figure}
  \begin{center}
    \includegraphics[width=\hsize]{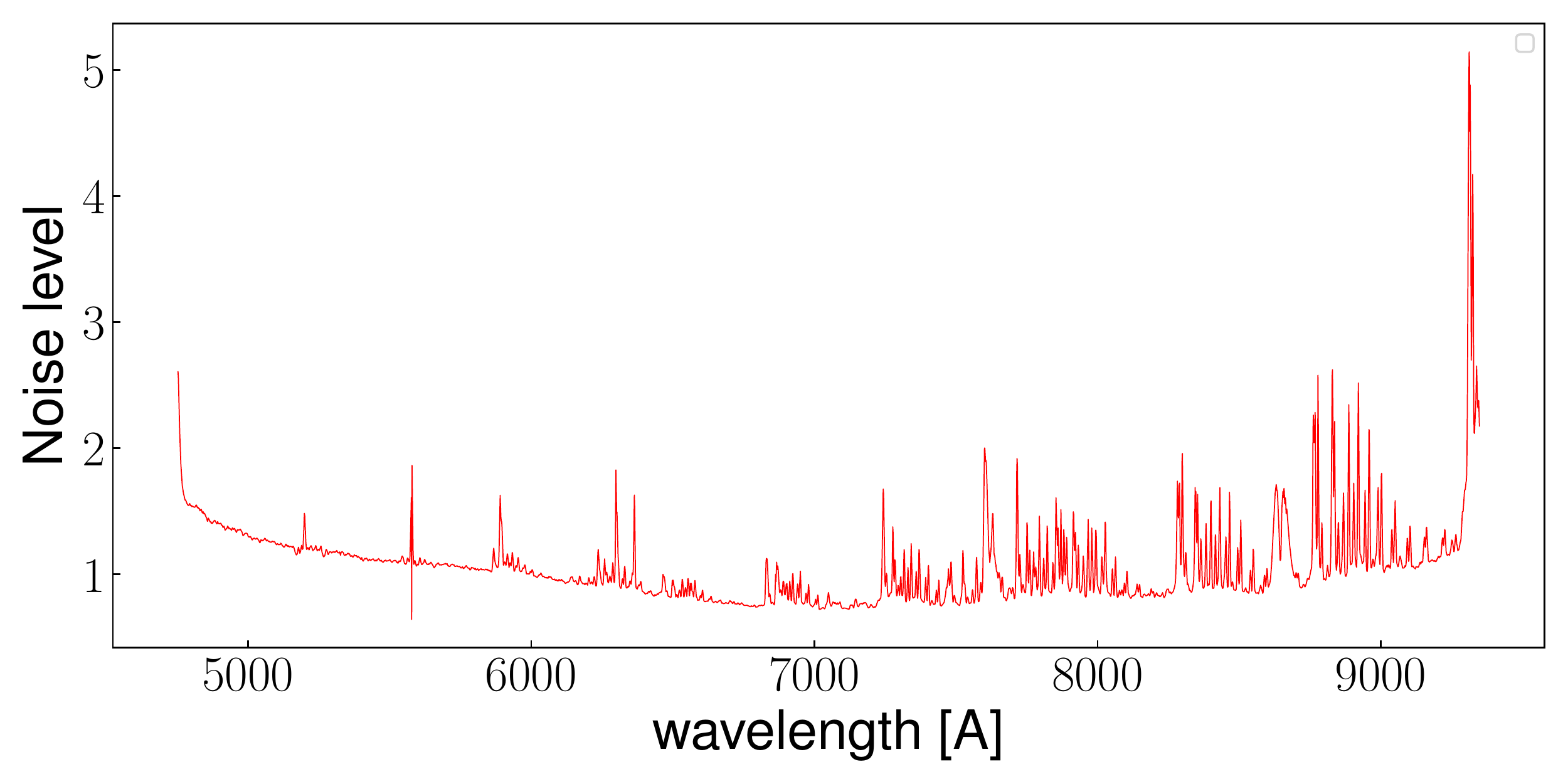}
  \end{center}
  \caption{\label{fig:evolution_of_RMS_level} Evolution of the noise level with wavelength inside the A1689 MUSE cube. We define the noise level of a given wavelength layer of a cube as the spatial median of the RMS layer over a normalization factor. The noise spikes that are more prominent in the red part of the cube are caused by sky lines.}
\end{figure}

The detectability of each LAEs needs to be evaluated on the entire survey to compute $V_{\rm max}$. This task is not straightforward, as the detectability depends on many different factors: 

\begin{itemize}
\item [-] The flux of the source: The brighter the source, the higher the chances to be detected. For a given spatial profile, brighter sources have higher $V_{\rm max}$ values.
\item [-] The surface brightness and line profile of the source: For a given flux, a compact source would have a higher surface brightness value than an extended one, and therefore would be easier to detect. This aspect is especially important as most LAEs have an extended halo (see \citealt{Wisotzki2016}).

\item[-] The local noise level: At first approximation, it depends on the exposure time. This point is especially important for mosaics in which noise levels are not the same on different parts of the mosaic as the noisier parts contribute less to the $V_{\rm max}$ values.

\item [-] The redshift of the source: The Lyman-alpha line profile of a source may be affected by the presence of strong sky lines in the close neighbourhood. The cubes themselves have strong variations of noise level caused by the presence of those sky emission lines (see e.g. Fig. \ref{fig:evolution_of_RMS_level}).
\item [-] The magnification induced by the cluster.: Where the magnification is too small, the faintest sources could not have been detected.
\item [-] The seeing variation from one cube to another.
\end{itemize}

This shows that to properly compute $V_{\rm max}$, each source has to be individually considered. The easiest method to evaluate the detectability of sources is to simply mask the brightest objects of the survey, assuming that no objects could be detected behind them. This can be achieved from a white light image, using a mask generated from a \verb+SExtractor+ segmentation map. The volume computation can then be done on the unmasked pixels and only where the magnification is high enough to allow the detection of the source. However, as shown in Appendix \ref{sec:volume_comparison}, this technique has some limitations to account for the 3D morphologies of real LAEs. For this reason, a method to determine precisely the detectability map (referred to as detection mask or simply masks hereafter) of individual sources has been developed. As the detection process in this work is based on 2D collapsed images, we adopted the same scheme to build the 2D detection masks, and from these,  built the 3D masks in the source plane adapted to each LAE of the sample. Using these individual source plane 3D masks, and as previously mentioned, the volume integration was performed on the unmasked pixels only where the magnification is high enough. In the paragraphs below, we quickly summarize the method adopted to produce masks for 2D images and explain the reasons that lead to the complex method detailed in Sects. \ref{subsubsec:masking_3d_cubes} and \ref{subsubsec:volume_integration}.\\

The basic idea of our method for producing masks for 2D images is to mimic the \verb+SExtractor+ source detection process. For each pixel in the detection image, we determine whether the source could have been detected, had it been centred on this pixel. For this pseudo-detection, we fetch the values of the brightest pixels of the source (hereafter $Bp$) and compare them pixel-to-pixel to the background root mean square maps (RMS maps) produced by \verb+SExtractor+ from the detection image. The pixels where this pseudo-detection is successful are left unmasked, and where it failed, the pixels are masked. Technical details of the method for 2D images can be found in appendix \ref{annex:create_mask_from_2d_image}.\\
The detection masks produced in this way are binary masks and show where the source could have been detected. We use the term ``covering fraction'' to refer to the fraction of a single FoV covered by a mask. A covering fraction of 1 means that the source could not be detected anywhere on the image, whereas a covering fraction of 0 means that the source could be detected on the entire image.

This method of producing the detection masks from 2D images is precise and simple to implement when the survey consists of 2D photometric images. However, when dealing with 3D spectroscopic cubes, its application becomes much more complicated owing to the strong variations of noise level with wavelength in the cubes. Because of these variations, the detectability of a single source through the cubes cannot be represented by a single mask, duplicated on the spectral axis to form a 3D mask. An example of the spectral variations of noise level in a MUSE cube is provided in Fig. \ref{fig:evolution_of_RMS_level}. These spectral variations are very similar for the four cubes. ``Noise level'' is used to refer to the average level of noise on a single layer. It is determined from the RMS cubes, which are created by \verb+SExtractor+ from the detection cube (i.e. the \verb+Muselet+ cube of NB images). For a layer $i$ of the RMS cube, the noise level corresponds to the spatial median of the RMS layer over a normalization factor as follows:

\begin{equation}
  \label{eq:noise_level_definition}
\text{Noise level}(RMS_{\rm i}) = \frac{<RMS_{\rm i}>_{x,y}}{<RMS_{\rm median}>_{x,y}} 
.\end{equation}

\noindent In this equation $<..>_{x,y}$ is the spatial median operator. The 2D median RMS map, $RMS_{\rm median}$, is obtained from a median along the wavelength axis for each spatial pixel of the RMS cube. The normalization is the spatial median value of the median RMS map. The main factor responsible for the high frequency spectral variations of noise level is the presence of sky lines affecting the variance of the cubes.

\begin{figure*}
  \begin{center}
    \includegraphics[width=\hsize]{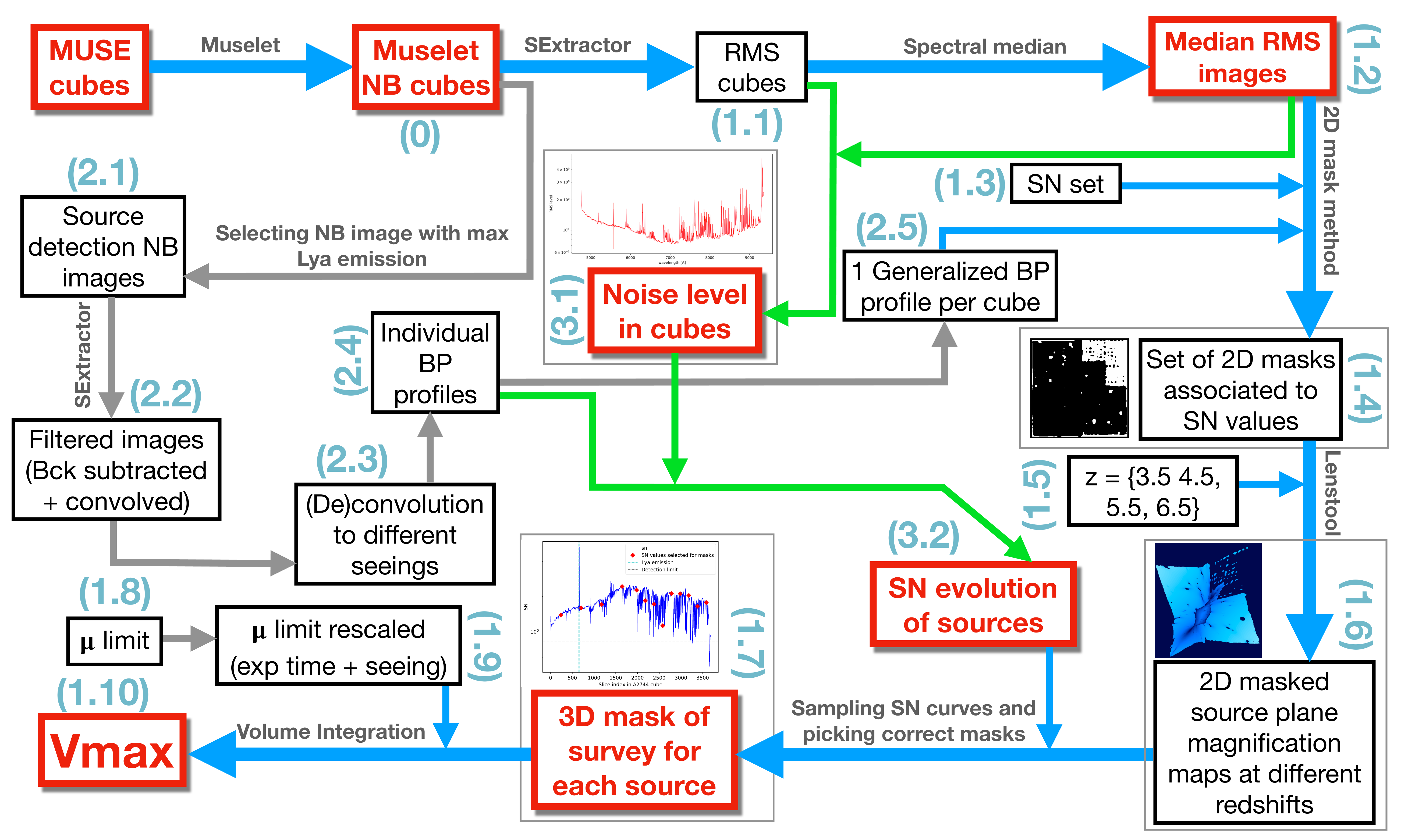}
  \end{center}
  \cprotect\caption{\label{fig:flow_chart_volume} Flow chart of the method used to produce the 3D masks and to compute $V_{\rm max}$. The key points are shown in red and the main path followed by the method is indicated in blue. All the steps related to the determination of the bright pixels are shown in grey. The steps related to the computation of the S/N of each source are indicated in green. The numbered labels in light blue refer to the bullet points in Appendix \ref{sec:detailed_volume_schematic} that briefly sum up all the different steps of this figure.}
\end{figure*}

To properly account for the noise variations, the detectability of each source has to be evaluated throughout the spectral direction of the cubes by creating a series of detection masks from individual layers. These masks are then projected into the source plane for the volume computation. This step is the severely limiting factor, as it would take an excessive amount of computation time. For a sample of 160 galaxies in four cubes, sampling different noise levels in cubes at only ten different wavelengths, we would need to do 6\,400  \verb+Lenstool+ projections. This represents more than 20 days of computation on a 60 CPU computer, and it is still not representative of the actual variations of noise level versus wavelength.
To circumvent this difficulty, we developed a new approach that allows for a fine sampling of the noise level variations while drastically limiting the number of source plane reconstructions. A flow chart of the method described in the next sections is provided in Fig. \ref{fig:flow_chart_volume}.

\subsubsection{Masking 3D cubes}
\label{subsubsec:masking_3d_cubes}

The general idea of the method is to use a S/N proxy of individual sources instead of comparing their flux to the actual noise. In other words, the explicit computation of the detection mask for every source, wavelength layer, and cube is replaced by a set of pre-computed masks for every cube, covering a wide range of S/N values, in such a way that a given source can be assigned the mask corresponding to its S/N in a given layer. Two independent steps were performed before assembling the final 3D masks:
First,  the evolution of S/N values is computed through the spectral dimension of the cubes for each LAE. Second, for each cube, a series of 2D detection masks were created for an independent set of S/N values. This is referred to as the S/N curves hereafter.
These two steps are detailed below. The final 3D detection masks were then assembled by successively picking the 2D mask that corresponds to the S/N value of the source at a given wavelength in a given cube. This process was done for all sources individually.\\

For the first step, the S/N value of a given source was defined as follows, from the bright pixels profile of the source and a RMS map, by comparing  the maximum flux of the brightest pixels profile ($\max(Bp)$) to the noise level of that RMS map.

For each layer of the RMS cube, we computed the S/N value the source would have had at that spectral position in the cube. We point out that this is not a proper S/N value (hence the use of the term ``proxy'') as the normalization used to define the noise levels in Eq. \ref{eq:noise_level_definition} depends on the cube. For a layer $i$ of the RMS cube, the corresponding  $S/N_{\rm i} $ value is given by

\begin{equation}
  \label{eq:sn_definition}
  S/N_{ \rm i} = \frac{\max(Bp)}{\text{Noise level}(RMS_{\rm i})}
.\end{equation}

\begin{figure}
  \centering
  \includegraphics[width=\hsize]{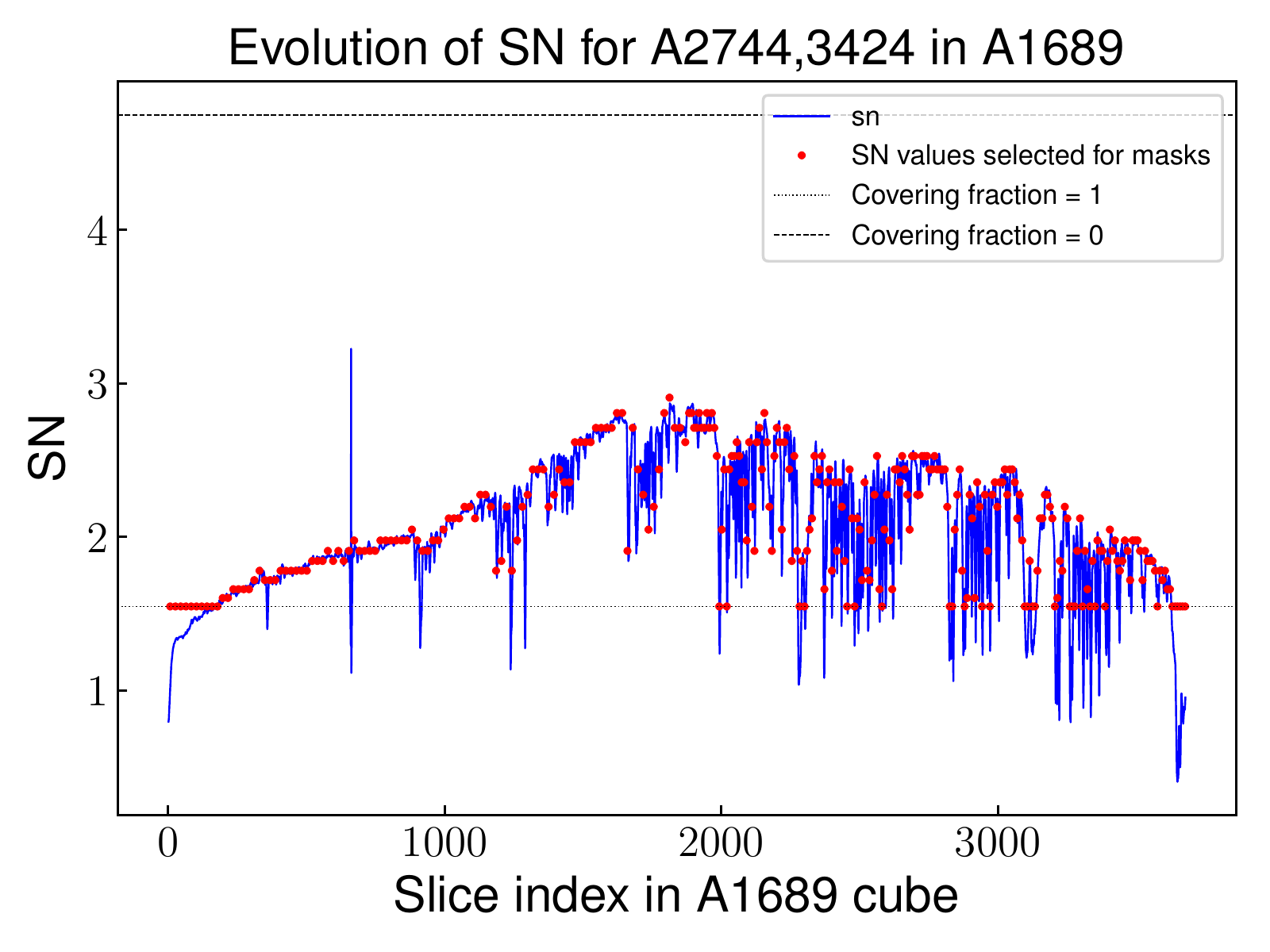}
  \caption{\label{fig:sn_evolution} Example of the 3D masking process. The blue solid line represents the variations of the S/N across the wavelength dimension for the source A2744-3424  in the A1689 cube. The red points over-plotted represent the 2D resampling made on the S/N curve with $\sim 300$ points. To each of these red points, a mask with the closest S/N value is associated. The short and long  dashed black lines represent the  S/N level  for which a covering fraction of 1 (detected nowhere) and 0 (detected everywhere) are achieved, respectively. For all the points between these two lines, the associated masks have a covering fraction ranging from 1 to 0, meaning that the source is always detectable on some regions of the field.}
\end{figure}

An example of  a S/N curve defined this way is provided in Fig. \ref{fig:sn_evolution}. For a given source, this computation was done on every layer of every cube part of the survey. When computing the S/N of a given source in a cube different from the parent cube, the seeing difference (see Table \ref{tab:observations_description}) is accounted for by introducing convolution or deconvolution procedure to set the detection image of the LAE to the resolution of the cube considered. As a result for each LAE, three additional images are produced. The four images (original detection image plus the three simulated ones) are then used to measure the  value of the brightest pixels in all four seeing conditions. For the deconvolution a python implementation of a Wiener filter part of the Scikit-image package \citep{skimage} was used. The deconvolution algorithm itself is presented in \citet{Orieux2010} and for all these computations, the PSF of the seeing is assumed to be Gaussian.\\

For the second step, 2D masks are created from a set of S/N values that encompass all the possible values for our sample. To produce a single 2D mask, the two following inputs are needed: the list of bright pixels of the source $Bp$ and the RMS maps produced from the detection image (in our case, the NB images produced by \verb+Muselet+). To limit the number of masks produced, two simplifications were introduced, the main one being that all RMS maps of a same cube present roughly the same pattern down to a certain normalization factor. This is equivalent to saying that all individual layers of the RMS cube can be approximately modelled and
reproduced by a properly rescaled version of the same median RMS map. The second simplification is the use of four generalized bright-pixel profiles (hereafter $Bp_{\rm g}$). To be consistent with the seeing variations, one profile is computed for each cluster, taking the median of all the individual LAE profiles computed from the detection images simulated in each seeing condition (see Fig. \ref{fig:bright_pixels} for an example of generalized bright pixel profile, also including the effect of seeing). These profiles are normalized in such a way that $\max(Bp_{\rm g}) = 1$. For each value of the S/N set defined, a mask is created for each cluster from its median RMS map and the corresponding $Bp_{\rm g}$, meaning that the 2D detection masks are no longer associated with a specific source, but with a specific S/N value.

Using the definition of S/N adopted in Eq. \ref{eq:sn_definition},  the four $Bp_{\rm g}$ are rescaled to fit any $S/N_{\rm j}$ value of the S/N set and to obtain profiles that are directly comparable to the median RMS maps:
\begin{equation}
  \label{eq:sn_computation}
  S/N_{\rm j} = \frac{\max( c_j \times Bp_{\rm g})}{\text{Noise level}(RMS_{\rm median})}
\end{equation} 
\noindent where $c_{\rm j}$ is the scaling factor. According to Eq. \ref{eq:noise_level_definition}, the noise level of the median RMS maps is just 1, and as mentioned above $\max(Bp_{g}) = 1$. We can see that the scaling factor is simply $ c_{\rm j} = S/N_{\rm j}$. Therefore the four sets of bright-pixels profiles
$S/N_{\rm j} \times Bp_{\rm g}$ and the corresponding median RMS maps are used as input to produce the set of 2D detection masks.\\

After the completion of these two steps, the final 3D detection masks were assembled for every source individually. For this purpose, a subset of wavelength values (or equivalently, a subset of layer index) drawn from the wavelength axis of a MUSE cube was used to resample the S/N curves of individual sources. For each source and each entry of this wavelength subset, the procedure fetches the value in the S/N set that is the closest to the measured value, and returns the associated 2D detection mask, effectively assembling a 3D mask. An example of this 2D sampling is provided in Fig. \ref{fig:sn_evolution}. To each of the red points resampling the S/N curve, a pre-computed 2D detection mask is associated, and the higher the density of the wavelength sampling, the higher the precision on the final reconstructed 3D mask. The important point is that to increase the sampling density, we do not need to create more masks and therefore it is not necessary to increase the number of source plane reconstructions.

\subsubsection{Volume integration}
\label{subsubsec:volume_integration}

In the previous section we presented the construction of 3D masks in the image plane for all sources with a limited number of 2D masks. For the actual volume computation, the same was achieved in the source plane by computing the source plane projection of all the 2D masks, and combining these masks with the magnification maps. Thanks to the method developed in the previous subsection, the number of source plane reconstructions only depends on the length of the S/N set initially defined  and the number of MUSE cubes used in the survey. It depends neither on the number of sources in the sample nor the accuracy of the sampling of the S/N variations. For the projections, we used \verb+PyLenstool+ \footnote{Python module written by G. Mahler, publicly available at \url{http://pylenstool.readthedocs.io/en/latest/index.html}}, which allows for an automated use of \verb+Lenstool+. Reconstruction of the source plane was performed for different redshift values to sample the variation of both the shape of the projected area and the magnification. In practice, the variations are very small with redshift and we reduce the redshift sampling to $z = 3.5, 4.5, 5.5,$ and  $6.5$.

In a very similar way to what is described at the end of the previous section, 3D masks were assembled and combined with magnification maps, in the source plane. In addition to the closest S/N value, the procedure also looks for the closest redshift bin in such a way that, for a given point $(\lambda_{\rm k},S/N_{\rm k})$ of the resampled S/N curve, the redshift of the projection is the closest to  $z_{\rm k} = \frac{\lambda_{\rm k}}{\lambda_{\rm Ly_{\alpha}}} - 1 $.

The last important aspect to take into account when computing $V_{\rm max}$ is to limit the survey to the regions where the magnification is such that the source could have been detected. The condition is given by

\begin{equation}
  \label{eq:mu_lim_definition}
  \frac{\mu_{\rm lim}}{\mu}\frac{F_{\rm d}}{\delta F_{\rm d}} = 1
,\end{equation}

\noindent where $\mu$ is the flux weighted magnification of the source, $F_{\rm d}$ the detection flux, and $\delta F_{\rm d}$ the uncertainty on the detection which reflects the local noise properties. This condition simply states that $\mu_{\rm lim}$ is the magnification that would allow for a S/N of 1 under which the detection of the source would be impossible. It is complex to find  a S/N criterion to use that would be coherent with the way \verb+Muselet+ works on the detection images, since the images used for the flux computation are different and of variable spectral width compared to the \verb+Muslet+ NBs. Therefore, this criterion for the computation of $\mu_{\rm lim}$ is intentionally conservative to avoid overestimating the steepness of the faint end slope.

To be consistent with the difference in seeing values and in exposure time from cube to cube, $\mu_{\rm lim}$ is computed for each LAE and for each MUSE cube (i.e. four values for a given LAE). A source only detected because of very high magnification in a shallow and bad seeing cube (e.g. A1689) would need a much smaller magnification to be detected in a deeper and better seeing cube (e.g. A2744). For the exposure time difference, the ratio of the median RMS value of the entire cube is used, and for the seeing the ratio of the squared seeing value is used. In other words, the limiting magnification in A2744 for a source detected in A1689 is given by
\begin{equation}
  \label{eq:mu_lim_rescaled}
    \mu_{\rm lim, A2744} =
    \frac{<RMS_{\rm A274}>_{x,y,\lambda}}{<RMS_{\rm A1689}>_{x,y,\lambda}}
    \frac{s_{A2744}^2}{s_{A1689}^2} \times \mu_{\rm lim, A1689}
  ,\end{equation}

\noindent where $<..>_{x,y,\lambda}$ is the median operator over the three axis of the RMS cubes and $s$ is the seeing. The exact same formula can be applied to compute the limit magnification of any source in any cube. This simple approximation is sufficient for now as only the volume of the rare LAEs with very high magnification are dominated by the effects of the limiting magnification.

The volume integration is performed from one layer of the source plane projected (and masked) cubes to the next, counting only pixels with $\mu > \mu_{\rm lim}$. For this integration, the following cosmological volume formula was used: 
\begin{equation}
  \label{eq:def_cosmological_volumes}
  V = \omega  \frac{c}{H_0} \int_{z_{\rm min}}^{z_{\rm max}}
  \frac{D_L^2(z')}{(1 + z')^2 E(z')} dz'
,\end{equation}
\noindent where $\omega$ is the angular size of a pixel, $D_{L}$ is the luminosity distance, and $E(z)$ is given by 
\begin{equation}
E(z) = \sqrt{\Omega_{\rm m}(1 + z)^3 + (1-\Omega_{\rm m} - \Omega_{\rm \Lambda})(1 + z)^2 + \Omega_{\rm \Lambda}}
.\end{equation}

In practice, and for a given source, when using more than 300 points to resample the S/N curve along the spectral dimension, a stable value is reached for the volume (i.e. less than 5\% of variation with respect to a sampling of 1\,000 points).
A comparison is provided in appendix \ref{sec:volume_comparison} between the results obtained with this method and the equivalent findings when a simple mask based on \verb+SExtractor+ segmentation maps is adopted instead.
The maximum co-volume explored between $2.9 < z < 6.7 $, accounting for magnification, is about $16\,000 Mpc^3$, distributed as follows among the four clusters:  $\sim 900$ Mpc$^3$ for A1689,  $\sim 800$ Mpc$^3$ for A2390,  $\sim 600$ Mpc$^3$ for A2667, and  $\sim 13\,000$ Mpc$^3$ for A2744.

\subsection{Completeness determination using real source profiles}
\label{subsec:completeness_estimation}

Completeness corrections account for the sources missed during the selection process. Applying the correction is crucial for the study of the LF. The procedure used in this article separates, on one hand, the contribution to incompleteness due to S/N effects across the detection area, and the contribution due to masking across the spectral dimension on the other hand (see $V_{\rm max }$ in Sect. \ref{subsec:volume_computation}).

The 3D masking method presented in the previous section aims to map precisely the volume where a source could be detected. However, an additional completeness correction was needed to account for the fact that a source does not have
a 100\% chance of being detected on its own wavelength layer.
In the continuity of the non-parametric approach developed for the volume computation, the completeness was determined for individual sources. To better account for the properties of sources, namely their spatial and spectral profiles, simulations were performed using their real profiles instead of parameterized realizations. Because the detection of sources was done in the image plane, the simulations were also performed in the image plane on the actual masked detection layer of a given source (i.e the layer of the NB image cube containing the peak of the Lyman-alpha emission of the source). The mask used on the detection layer was picked using the same method  as described in \ref{subsubsec:masking_3d_cubes}, leaving only the cleanest part of the layer available for the simulations. 

\subsubsection{Estimating the source profile}
\label{subsubsec:estimating_source_profile}

To get an estimate of the real source profile, we used the \verb+Muselet+ NB image that captures the peak of the Lyman-alpha emission (called the max-NB image hereafter).
Using a similar method to that presented in Sect.  \ref{subsec:flux_computation}, the extraction of sources on the max-NB images were forced by progressively loosening the detection criterion. The vast majority of our sources were successfully detected on the first try using the original parameters used by \verb+Muselet+
for the initial detection of the sample: \verb+DETECT_THRES = 1.3 + and \verb+MIN_AREA = 6+.

\begin{figure}
  \centering

  \includegraphics[width=\hsize]{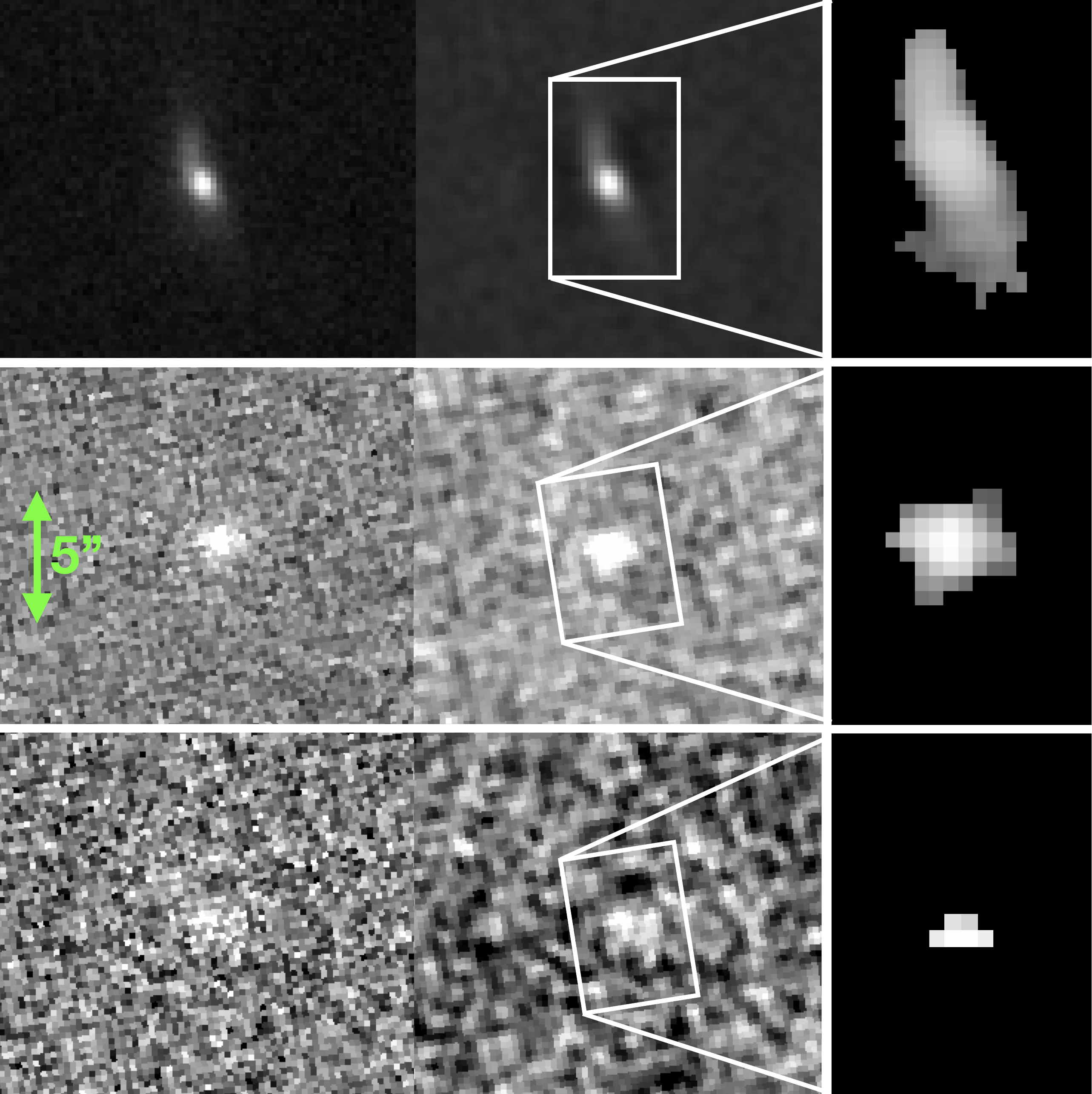}
  \cprotect\caption{Example of source profile recovery for three representative LAEs. Left column: detection image of the source in the \verb+Muselet+ NB cube (i.e. the max-NB image). Middle column: filtered image (convolved and background-subtracted) produced by \verb+SExtractor+ from the image in the left column. Right column: recovered profile of the source obtained by applying the segmentation map on the filtered image. The spatial scale is not the same as for the two leftmost columns. All the sources presented in this figure have a flag value of 1.}
  \label{fig:extraction_mosaic}
\end{figure}

To recover the estimated profile of a source, the pixels belonging to the source  were extracted on the filtered image according to the segmentation map. The filtered image is the convolved and background-subtracted image that \verb+SExtractor+ uses for the detection. The use of filtered images allowed us to retrieve a background-subtracted and smooth profile for each LAE. Fig. \ref{fig:extraction_mosaic} presents examples of source profile recovery for three representative LAEs.

A flag was assigned to each extracted profile to reflect the quality of the extraction, based on a predefined set of parameters (detection threshold, minimum number of pixels, and matching radius)  used for the successful extraction of the source. A source with flag 1 is extremely trustworthy, and was recovered with the original set of parameters used for initial automated detection of the sample. A source with flag 2 is still a robust extraction and a source with flag 3 is doubtful and is not used for the LF computation. Of the LAEs, 95\% were properly recovered with a flag value of 1. The summary of flag values is shown in Table \ref{tab:extraction_flag}. The three examples presented in Fig. \ref{fig:extraction_mosaic} have a flag value of 1 and were recovered using \verb+DETECT_THRESH = 1.3+, \verb+MIN_AREA=6+ and a matching radius of $0.8\arcsec$. Objects with flag $> 1 $ are less than  5\%  of the total sample. For the few sources with an extraction flag above 1, several possible explanations are found, listed by order of importance as follows: 
\begin{itemize}
\item[-] The image used to recover the profiles ($30\arcsec$) is smaller than the entire max-NB image. As the \verb+SExtractor+ background estimation depends on the size of the input image, this may slightly affect the detection of some objects. This is most likely the predominant reason for a flag value of two.
\item[-] There is a small difference in the coordinates between the recovered position and listed position. This may be due to a change in morphology with wavelength or bandwidth. By increasing the matching radius to recover the profile, we obtained a successful extraction but we also increased the value of the extraction flag.
\item[-] The NB used does not actually correspond to the NB that leads the source to be detected. By picking the NB image that catches the maximum of the Lyman-alpha emission we do not necessarily pick the layer with  the cleanest detection. For example the peak could fall in a very noisy layer of the cube, whereas the neighbouring layers would provide a much cleaner detection.

\item[-] The source is extremely faint and was actually detected with relaxed detection parameters or manually detected.
\end{itemize}

We checked that we did not include LAEs that were expected to be at a certain position as part of multiple-image system. This is to say, we did not select the noisiest images in multiple-image systems.

\begin{table}
  \caption{Summary of the extraction flag values for sources in the different lensing fields (see text for details).}             
\label{tab:extraction_flag}      
\centering                          
\begin{tabular}{l|  c c c c| c}        

  \hline\hline                 
  Flag    & A1689  & A2390 & A2667 & A2744 & All Sample\\
  \hline
  1       & 16     & 5     & 7     & 121   & 149      \\
  2       & 0      & 0     & 0     & 6     & 6        \\
  3       & 0      & 0     & 0     & 1     & 1        \\
  \hline
  Total  & 16      & 5     & 7     & 128   & 156      \\
  \hline                                  
\end{tabular}
\end{table}

\subsubsection{Recovering mock sources}
\label{sec:recovering_mock}

Once a realistic profile for all LAEs was obtained, source recovery simulations were conducted. For this step, the detection process was exactly the same as that initially used for the sample detection. However, since we limited the simulations to the max-NB (see Sect. \ref{subsubsec:estimating_source_profile}) images and not the entire cubes, we did not need to use the full \verb+Muselet+ software.  To gain computation time, we only used \verb+SExtractor+ on the max-NB images, using the same configuration files that \verb+Muselet+ uses, to reproduce the initial detection parameters. In this section, the set of parameters were also \verb+DETECT_THRESH = 1.3+ and \verb+MIN_AREA = 6+.\\

To create the mock images, we used the masked max-NB images.
Each source profile was randomly injected many times on the corresponding detection max-NB image, avoiding overlapping. After running the detection process on the mocks, the recovered sources were matched to the injected sources based on their position. The completeness values were derived by comparing the number of successful matches to the number of injected sources. The process was repeated 40 times to derive the associated uncertainties.

\begin{figure}
  \centering
  \includegraphics[width=\hsize]{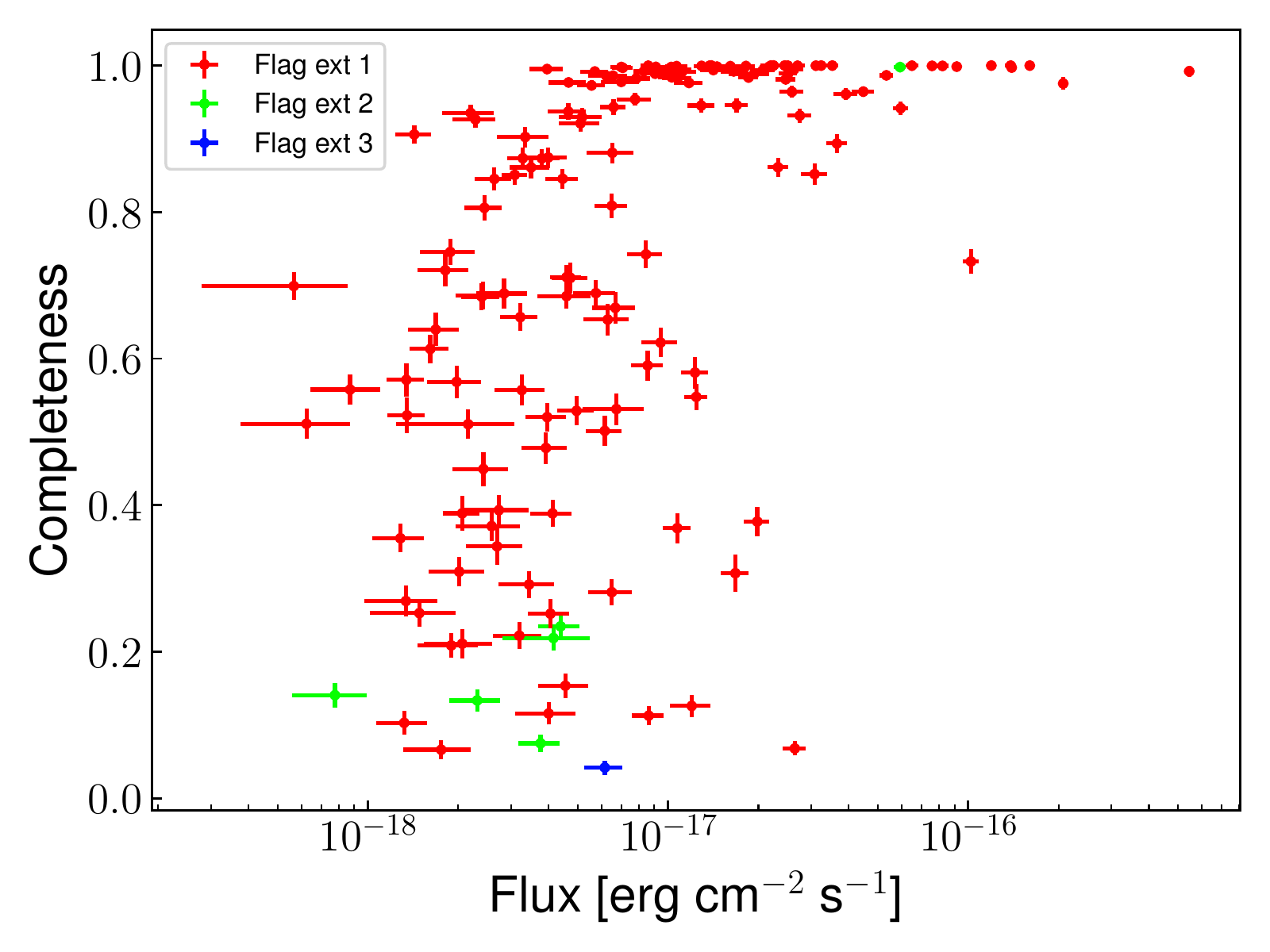}
  \caption{Completeness value for LAEs vs. their detection flux. Colours indicate the detection flags. We note that only the incompleteness  owing to S/N on the unmasked regions of the detection layer is plotted  in this graph (see Sect. \ref{subsec:completeness_estimation}).}
  \label{fig:completeness_vs_flux}
\end{figure}

The results of the completeness obtained for each source of the sample are shown in Fig. \ref{fig:completeness_vs_flux}. The average completeness value over the entire sample is ~0.74 and the median value is 0.90. The values are this high because we used masked NB images, effectively making  source recovery simulations on the cleanest part of the detection layer only. As seen on this figure, there is no well-defined trend between completeness and detection flux. At a given flux, a compact source detected on a clean layer  of the cube has a higher completeness than a diffuse source with the same flux detected on a layer affected by a sky line. Four LAEs with a flag value of 3 or with a completeness value less than 10\% are not used for the computation of the LFs in Sect. \ref{subsec:lf_points}.

\begin{figure}
  \centering

  \includegraphics[width=\hsize]{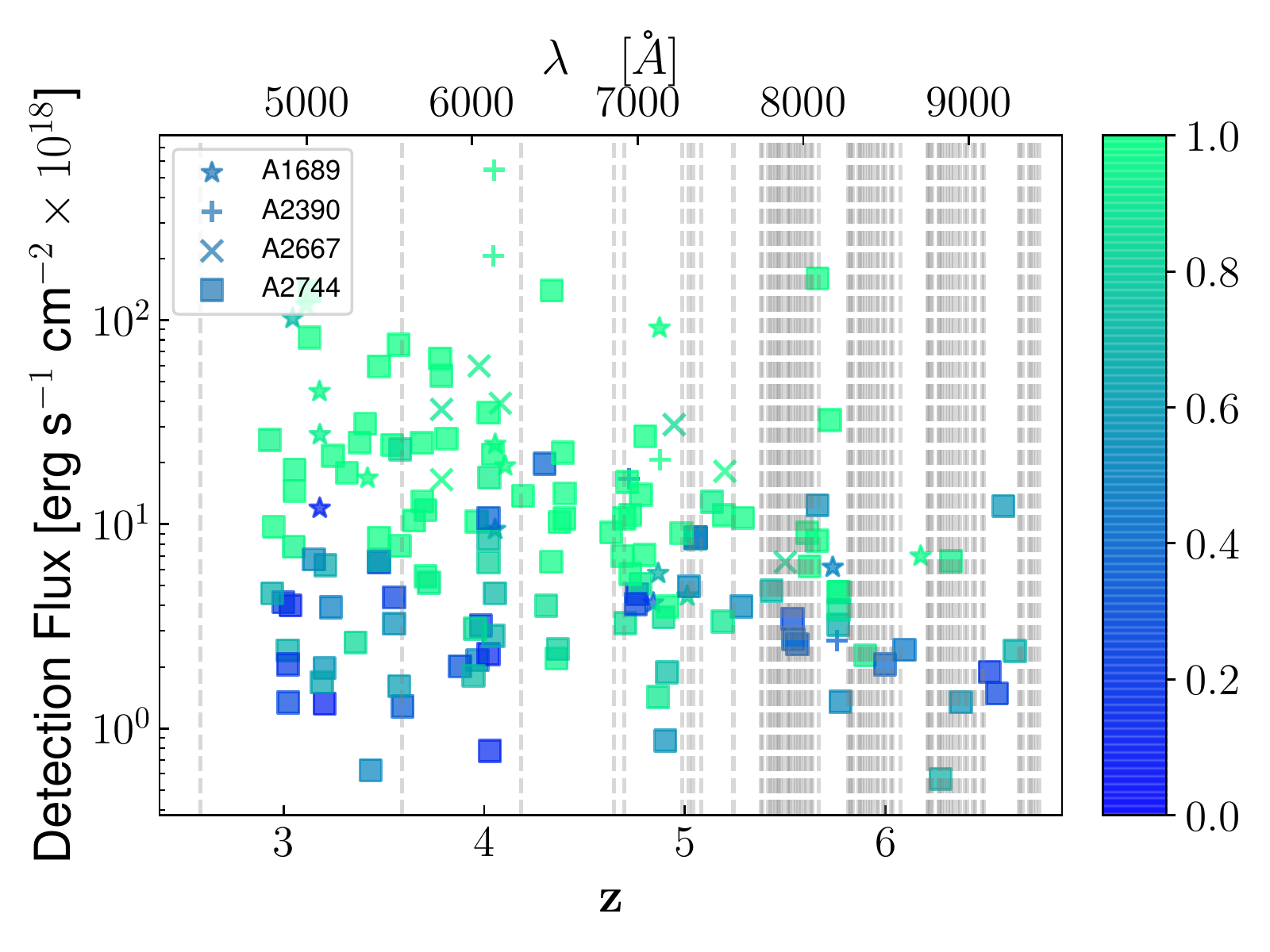}
  \caption{Completeness (colour bar) of the sample as a function of redshift and detection flux. Each symbol indicates a different cluster. The  light grey vertical lines are indicated by the main sky lines. There is no obvious correlation in our selection of LAEs between the completeness and the position of the sky lines.}
  \label{fig:comp_flux_z}
\end{figure}

A more popular approach to estimate the completeness would be to perform heavy  Monte\ Carlo (MC) simulations for each of the cubes in the survey to get a parameterized completeness (see \citealt{Drake2017b} for an example). The classical approach consists in injecting sources with parameterized spatial and spectral morphologies and retrieving the completeness as a function of redshift and flux. This method is extremely time consuming, in particular for IFUs where the extraction process is lengthy and tedious.
The main advantage of computing the completeness based on the real source profile is that it allows us to accurately account for the different shapes and surface brightnesses of individual sources. And because the simulations are done on the detection image of the source in the cubes, we are also more sensitive to the noise increase caused by sky lines. As seen in Fig.\ref{fig:comp_flux_z}, except from the obvious flux--completeness correlation, it is difficult to identify correlations between completeness and redshift or sky lines. This tends to show that the profile of the sources is a dominant factor when it comes to  estimating the completeness properly. The same conclusion was reached in  \citetalias{Drake2017b} and  in \citet{Herenz2019}. A non-parametric approach of completeness is therefore better suited in the case of lensing clusters, where a proper parametric approach is almost impossible to implement because of the large number of parameters to take into account (e.g. spatial and spectral morphologies including distortion effects, lensing configuration, and cluster galaxies).

\subsection{Determination of the luminosity function}
\label{subsec:lf_points}

To study the possible evolution of the LF with redshift, the 152 LAE population has been subdivided into several redshift bins: $z1 : 2.9 < z < 4.0 $, $z2 : 4.0 < z < 5.0, $ and $z3 : 5.0 < z < 6.7 $. In addition to these three LFs, the global LF for the entire sample $z_{\rm all} : 2.9 < z < 6.7$ was also determined.
For a given redshift and luminosity bin, the following expression to build the points of the differential LFs was used:

\begin{equation}
  \label{eq:def_Phi}
  \Phi(L_{\rm i}) = \frac{1}{\Delta \log L_{\rm i}} \sum_{j}
  \frac{1}{C_{\rm j} V_{\rm max,  j }}
,\end{equation}

\noindent where $\Delta \log L_{\rm i}$ corresponds to the width of the luminosity bin in logarithmic space, $j$ is the index corresponding to the sources falling in the  bin indexed by $i$, and $C_{\rm j}$ stands for the completeness correction of the source j.

To account for the uncertainties affecting each LAE properly, MC iterations are performed to build 10\,000 catalogues from the original catalogue. For each LAE in the parent catalogue, a random magnification is drawn from its $P(\mu)$, and a random flux and completeness values are also drawn assuming a Gaussian distribution of width fixed by their respective uncertainties. A single value of the LF was obtained at each iteration following Eq. \ref{eq:def_Phi}. The distribution of LF values obtained at the end of the process was used to derive the average in linear space and to compute asymmetric error bars. The MC iterations are well suited to account for LAEs with poorly constrained luminosities. This happens for sources close, or even on, the critical lines of the clusters. Drawing random values from their probability density and uncertainties for magnification and flux results in a luminosity  distribution (see Eq. \ref{eq:flux_lum_conversion}), which allows these sources to have a diluted contribution across several luminosity bins.\\

For the estimation of the cosmic variance, we used the cosmic variance calculator presented in \citet{Trenti2008}. Lacking other options, a single compact geometry made of the union of the effective areas of the four FoVs is assumed and used as input for the calculator. The blank field equivalent of our survey is an angular area of about $1.2\arcmin\times 1.2\arcmin$. Given that a MUSE FoV is a square of size $1\arcmin$, the observed  area of the present survey is roughly $7\arcmin\times 7\arcmin$ square.  Our survey is therefore roughly equivalent to a bit more than only one MUSE FoV in a blank field. The computation is done for all the bins as the value depends on the average volume explored in each bin as well as on the intrinsic number of sources.
The uncertainty due to  cosmic variance on the intrinsic counts of galaxies in a luminosity bin typically range from 15\% to 20\% for the global LF and from 15\% to 30\% for the LFs computed in redshift bins. For $\log (L) \lesssim 41 $, the total error budget is dominated by the MC dispersion, which is mainly caused by  objects with poorly constrained luminosity jumping from one bin to another during the MC process. The larger the bins the lesser this effect because a given source is less likely to jump outside of a larger bin.
For $ 41 \lesssim \log (L) \lesssim 42 $ the Poissonian uncertainty is slightly larger than the cosmic variance but does not completely dominate the error budget. Finally for $ 42 \lesssim \log (L)$, the Poissonian uncertainty is the dominant source of error due to the small volume and therefore the small number of bright sources in the survey.

The data points of the derived LFs and the corresponding error bars are listed in Table \ref{tab:lf_points}. These LF points provide solid constraints on the shape of the faint end of the LAE distribution. In the following sections, we elaborate on these results and discuss the  evolution of the faint end slope as well as the implications for cosmic reionization.

 \begin{table*}
 \caption{Luminosity bins and LF points used in Fig. \ref{fig:fit_lf}. The value $ <N> $ is the average number of sources in the luminosity bin and $N_{\rm corr}$ is the average number corrected for completeness. The value $< V_{\rm max} >$ is the average $V_{\rm max}$ for the sources in the bin. The average values are taken across the multiple MC iterations used to compute the statistical errors on the LF points. The uncertainties  on $\log(\Phi)$ are 68\% error bars, combining Poissonian error, MC iterations, and an estimation of the cosmic variance.}
 \label{tab:lf_points}
 \centering 
 \renewcommand{\arraystretch}{1.4} 
 \begin{tabular}{c c c c c} \hline \hline 
 
  $\log(L)$ &  $\log(\Phi)$ & $< N >$ &  $ < N_{\rm corr} >$ &  $ <V_{\rm max} >$ \\
  erg s$^{-1}$ &  $ (\Delta(\log(L)) = 1)^{-1} \text{Mpc}^{-3}$ & & & Mpc$^3$ \\\hline 
 \multicolumn{5}{c}{\textbf{2.9 < z < 6.7 }} \\ \hline 
 $ 39.00 < \textbf{39.63} < 40.25$ & $ -1.28_{-0.44}^{+0.21} $ & 2.05 & 8.97 & 124.68\\ 
 $ 40.25 < \textbf{40.38} < 40.50$ & $ -1.57_{-0.40}^{+0.41} $ & 3.52 & 7.04 & 4971.62\\ 
 $ 40.50 < \textbf{40.63} < 40.75$ & $ -1.64_{-0.43}^{+0.33} $ & 9.43 & 24.83 & 10977.19\\ 
 $ 40.75 < \textbf{40.88} < 41.00$ & $ -1.45_{-0.07}^{+0.09} $ & 12.77 & 33.27 & 12063.96\\ 
 $ 41.00 < \textbf{41.13} < 41.25$ & $ -1.74_{-0.20}^{+0.10} $ & 18.68 & 48.11 & 12816.23\\ 
 $ 41.25 < \textbf{41.38} < 41.50$ & $ -1.79_{-0.15}^{+0.11} $ & 23.28 & 48.07 & 12991.31\\ 
 $ 41.50 < \textbf{41.63} < 41.75$ & $ -1.89_{-0.13}^{+0.10} $ & 26.81 & 39.75 & 13926.47\\ 
 $ 41.75 < \textbf{41.88} < 42.00$ & $ -1.97_{-0.16}^{+0.10} $ & 26.15 & 35.60 & 14658.58\\ 
 $ 42.00 < \textbf{42.13} < 42.25$ & $ -2.22_{-0.16}^{+0.17} $ & 18.08 & 21.32 & 15017.49\\ 
 $ 42.25 < \textbf{42.38} < 42.50$ & $ -2.96_{-0.38}^{+0.18} $ & 4.22 & 4.28 & 15696.11\\ 
 $ 42.50 < \textbf{42.63} < 42.75$ & $ -3.01_{-0.34}^{+0.19} $ & 3.94 & 3.95 & 16060.71\\ 
 $ 42.75 < \textbf{42.88} < 43.00$ & $ -3.13_{-0.41}^{+0.21} $ & 3.00 & 3.01 & 16141.73\\ 
 \hline 
 \multicolumn{5}{c}{\textbf{2.9 < z < 4.0}} \\ \hline 
 $ 40.00 < \textbf{40.25} < 40.50$ & $ -2.48_{-0.72}^{+0.35} $ & 1.90 & 4.73 & 4430.41\\ 
 $ 40.50 < \textbf{40.75} < 41.00$ & $ -1.64_{-0.15}^{+0.11} $ & 14.99 & 38.65 & 4145.63\\ 
 $ 41.00 < \textbf{41.25} < 41.50$ & $ -1.66_{-0.15}^{+0.11} $ & 18.37 & 45.65 & 4468.50\\ 
 $ 41.50 < \textbf{41.75} < 42.00$ & $ -2.12_{-0.17}^{+0.14} $ & 14.53 & 18.14 & 5178.73\\ 
 $ 42.00 < \textbf{42.25} < 42.50$ & $ -2.47_{-0.25}^{+0.15} $ & 8.17 & 8.69 & 5216.12\\ 
 $ 42.50 < \textbf{42.75} < 43.00$ & $ -2.96_{-0.46}^{+0.22} $ & 2.95 & 2.95 & 5437.33\\ 
 \hline 
 \multicolumn{5}{c}{\textbf{4.0 < z < 5.0 }} \\ \hline 
 $ 39.00 < \textbf{39.25} < 39.50$ & $ -0.49_{- \infty}^{+0.33} $ & 0.76 & 5.47 & 44.11\\ 
 $ 39.50 < \textbf{40.00} < 40.50$ & $ -1.33_{-0.71}^{+0.54} $ & 1.79 & 3.71 & 939.22\\ 
 $ 40.50 < \textbf{40.75} < 41.00$ & $ -1.52_{-0.09}^{+0.09} $ & 4.83 & 14.76 & 2818.30\\ 
 $ 41.00 < \textbf{41.25} < 41.50$ & $ -1.76_{-0.24}^{+0.13} $ & 13.72 & 28.05 & 3706.94\\ 
 $ 41.50 < \textbf{41.75} < 42.00$ & $ -1.96_{-0.17}^{+0.12} $ & 19.40 & 21.96 & 4113.33\\ 
 $ 42.00 < \textbf{42.25} < 42.50$ & $ -2.39_{-0.27}^{+0.17} $ & 8.49 & 8.58 & 4254.24\\ 
 $ 42.50 < \textbf{42.75} < 43.00$ & $ -2.87_{-0.47}^{+0.22} $ & 3.00 & 3.02 & 4430.02\\ 
 \hline 
 \multicolumn{5}{c}{\textbf{5.0 < z < 6.7}} \\ \hline 
 $ 40.00 < \textbf{40.25} < 40.50$ & $ -1.21_{- \infty}^{+0.39} $ & 0.66 & 1.25 & 50.28\\ 
 $ 40.50 < \textbf{40.75} < 41.00$ & $ -1.78_{-0.65}^{+0.64} $ & 2.43 & 4.84 & 2985.57\\ 
 $ 41.00 < \textbf{41.25} < 41.50$ & $ -1.99_{-0.23}^{+0.15} $ & 9.88 & 22.43 & 4763.46\\ 
 $ 41.50 < \textbf{41.75} < 42.00$ & $ -1.81_{-0.19}^{+0.13} $ & 19.06 & 35.27 & 5087.77\\ 
 $ 42.00 < \textbf{42.25} < 42.50$ & $ -2.46_{-0.28}^{+0.30} $ & 5.61 & 8.29 & 5469.76\\ 
 $ 42.50 < \textbf{42.75} < 43.00$ & $ -3.49_{- \infty}^{+0.31} $ & 1.00 & 1.00 & 6187.25\\ 
 \hline 
 \hline 
 \end{tabular} 
 \end{table*}


\section{Parametric fit of the luminosity function}
\label{sec:lf_fit}

\begin{figure}
  \includegraphics[width=\hsize]{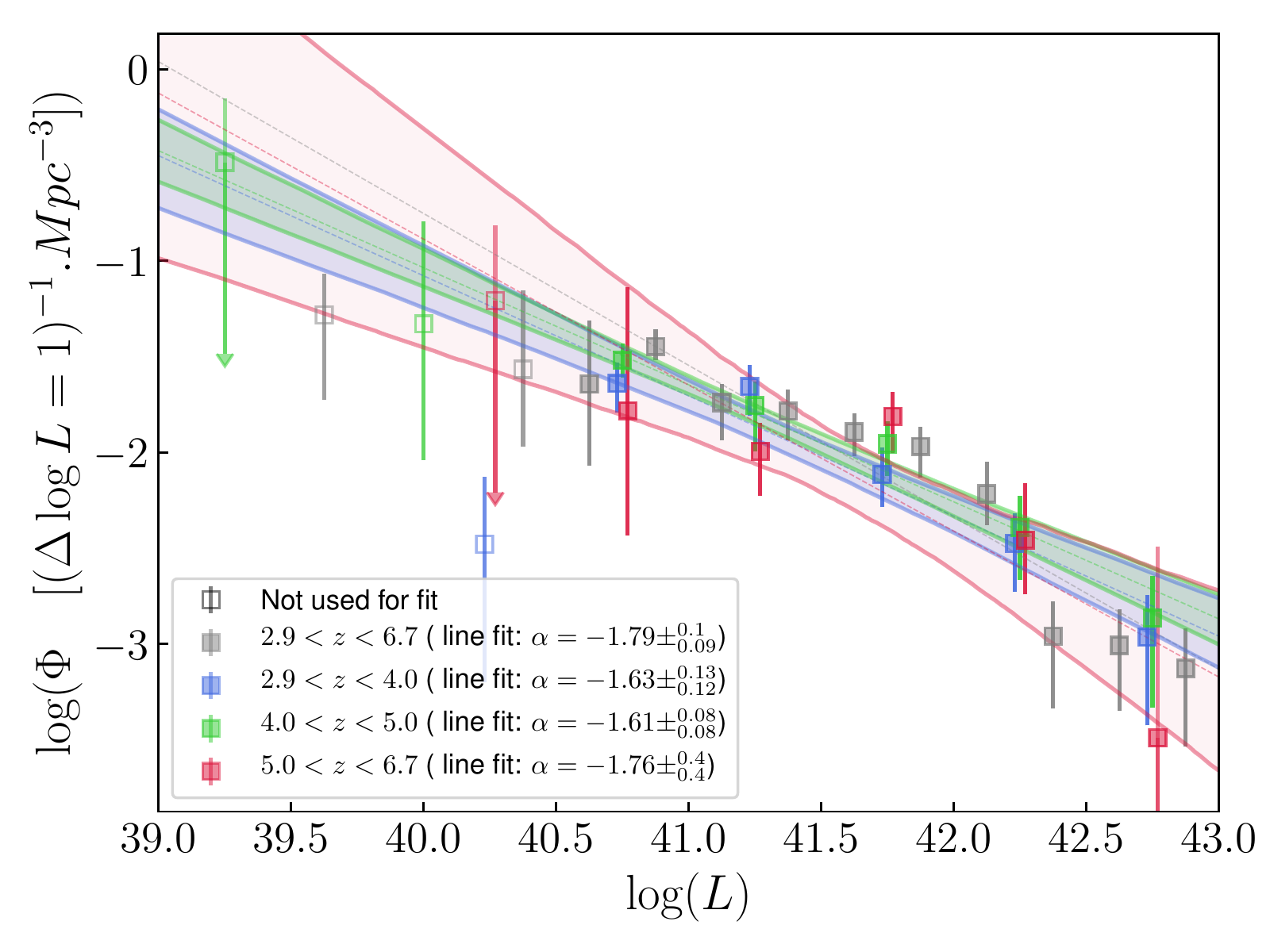}
  \caption{Luminosity function points computed for the four redshift bins. Each LF was fitted with a straight dotted line and the shaded areas are the 68\% confidence regions derived from these fits. For the clarity of the plot, the confidence area derived for $z_{\rm all}$ is not shown and a slight luminosity offset is applied to the LF points for $z_1$ and $z_3$.}
  \label{fig:lf_points}
\end{figure}

The differential LFs are presented in Fig. \ref{fig:lf_points} for the four redshift bins. Some points in the LF, shown as empty squares, are considered as unreliable and presented for comparison purpose only. Therefore, they are not used in the subsequent parametric fits. An LF value is considered unreliable when it is dominated by the contribution of a single source, with either a small $V_{\rm max}$ or a low completeness value, due to luminosity and/or redshift sampling. These unreliable points are referred to as ``incomplete'' hereafter. The rest of the points are fitted with a straight line as a visual guide, the corresponding 68\% confidence regions are represented as shaded areas. For $z_3$, the exercise is limited owing to the large uncertainties and the lack of constraints on the bright end. The measured mean slope for the four LFs are as follows: $\alpha=-1.79_{-0.09}^{+0.1}$ for $z_{\rm all}$, $\alpha=-1.63_{-0.12}^{+0.13}$ for $z_1$, $\alpha=-1.61_{-0.08}^{+0.08}$ for $z_2$ and $\alpha = -1.76_{-0.4}^{+0.4}$ for $z_3$. These values are consistent with no evolution of the mean slope with redshift.

In addition, and because the integrated value of each LF is of great interest regarding the constraints they can provide on the sources of reionization, the different LFs were fitted with the standard Schechter function \citep{Schechter1976} using the formalism described in \citet{Dawson2007}. The Schechter function is defined as
\begin{equation}
  \Phi(L)dL = \frac{\Phi_*}{L_*} \left(\frac{L}{L_*}\right)^{\alpha}
  \exp\left( - \frac{L}{L_*} \right) dL 
,\end{equation}

\noindent where $\Phi_*$ is a normalization parameter, $L_*$ a characteristic luminosity  that defines the position of the transition from the power law to the exponential law at high luminosity, and $\alpha$ is the slope of the power law at low luminosity. In logarithmic scale the Schechter function is written as

\begin{equation}
  \Phi_{\rm log}(L) d(\log L) = \left(\frac{L}{\log e} \right)
  \left( \frac{\Phi_*}{L_*}\right)
  \left( \frac{L}{L_*}\right)^{\alpha}
  \exp\left( -\frac{L}{L_*}\right) d(\log L)
.\end{equation}

This function represents the numerical density per logarithmic luminosity interval.
The fits were done using the Python package \textit{Lmfit} \citep{Newville2014}, which is specifically dedicated to nonlinear optimization and provides robust estimations for confidence intervals. We define an objective function, accounting for the strong asymmetry in the error bars, whose results are then minimized in the least-squares sense, using the default Levenberg-Marquardt method provided by the package. The results of this first minimization are then passed to a MCMC process\footnote{\textit{Lmift} uses the emcee algorithm implementation of the emcee Python package (see \citealt{Foreman-Mackey2013})}  that uses the same objective function. The uncertainty on the three parameters of the Schechter function ($ \alpha , L_*, \Phi_*$) are recovered from the resulting individual posterior distributions.
The minimization in the least-square sense is an easy way to fit our data but is not guaranteed to give the most probable parameterization for the LFs. A  more robust method would be the maximum-likelihood method. However, because of the non-parametric approach used in this work to build the points of the LF, taking into account the specific complexity of the lensing regime, the implementation of a maximum-likelihood approach such as those developed in \citetalias{Drake2017b} or in \citet{Herenz2019} could not be envisaged.

Because of the use of lensing clusters, the volume of Universe explored is smaller than in  blank field surveys. The direct consequence is that we are not very efficient in probing the transition area around $L_*$ and the high luminosity regime of the LF. Instead, the lensing regime is more efficient in selecting faint and low luminosity galaxies and is therefore much more sensitive to the slope parameter. To properly determine the three best parameters, additional data are needed to constrain the bright end of the LFs. To this aim, previous LFs from the literature are used and combined together into a single average LF with the same luminosity bin size as the LFs derived in this work. This last point is important to ensure that the fits are not dominated by the literature data points that are more numerous with smaller bin sizes and uncertainties. In this way we determine the three Schechter parameters while properly sampling the covariance between them.

The choice of the precise data sets used for the Schechter fits is expected to have a significant impact on the results, including possible systematic effects. To estimate the extent of this effect and its contribution to uncertainties, different series of data sets were used to fit the LF, among those available in a given redshift interval (see Fig. \ref{fig:fit_lf}). The best-fit parameters recovered are found to be always consistent within their own error bars.

\begin{figure*}
  \centering
  \includegraphics[width=0.49 \hsize]{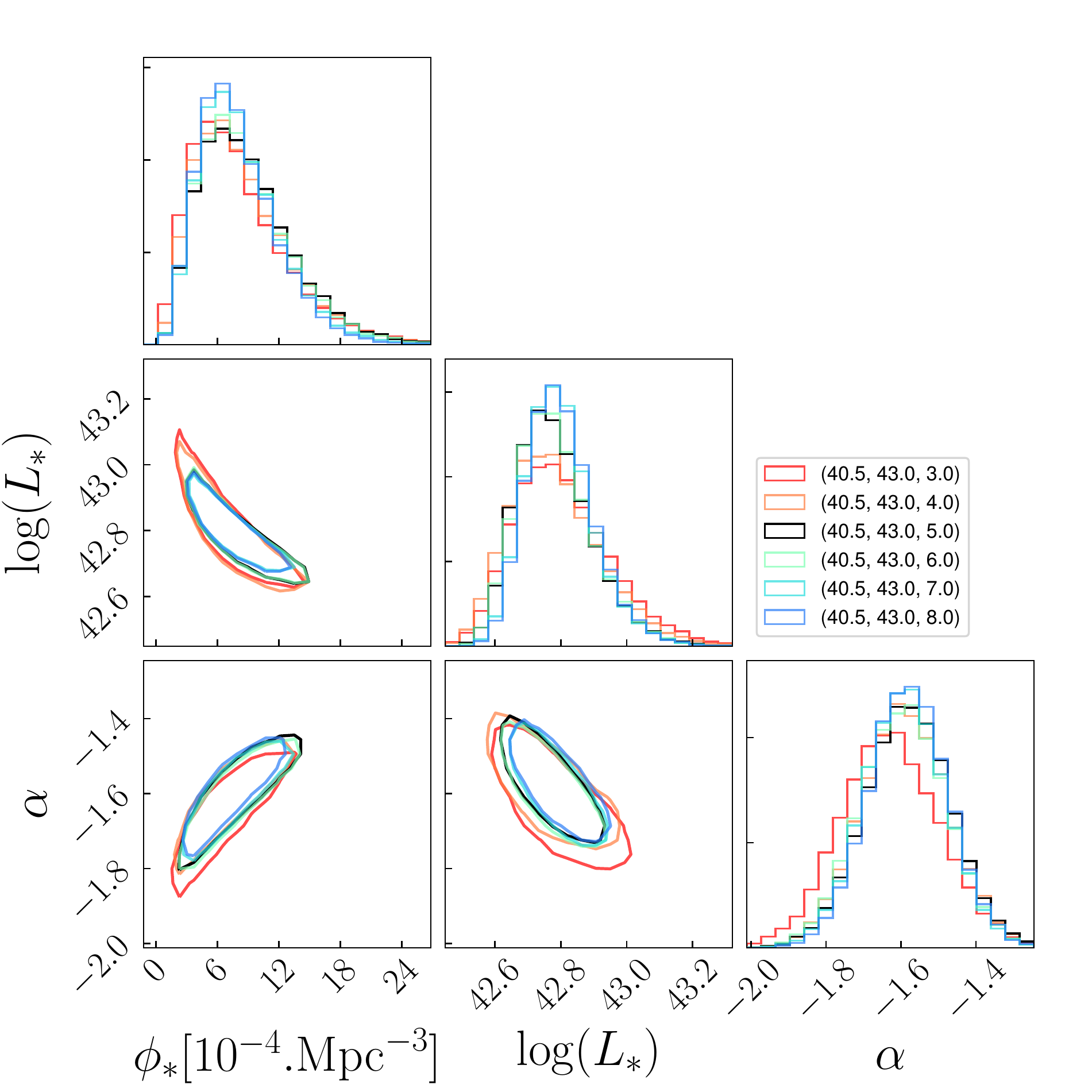}
  \includegraphics[width=0.49 \hsize]{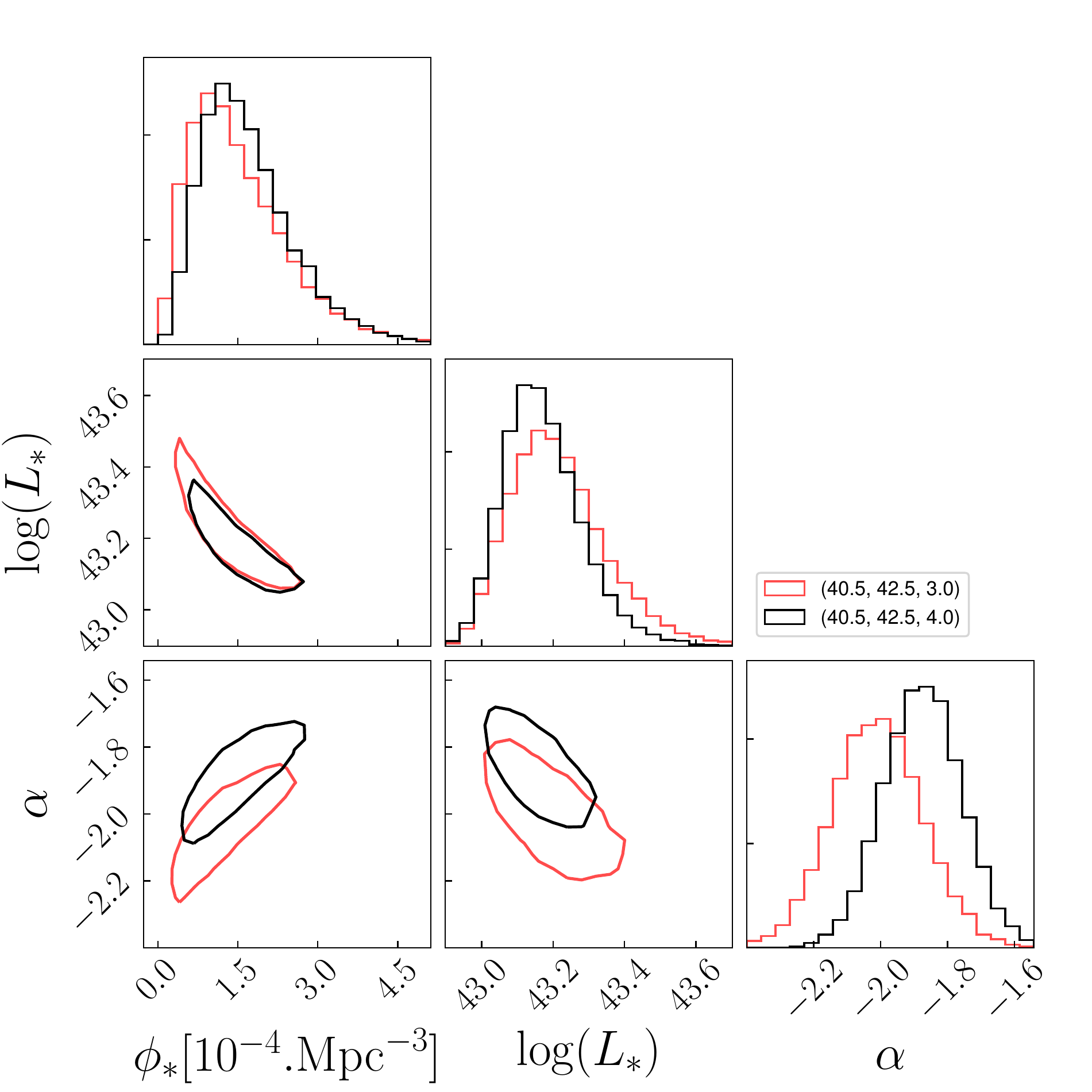}
  \caption{Areas of 68\% confidence derived on the Schechter parameters when testing different binnings. Left panel shows the results for $2.9 < z < 4.0$ and the right panel those for $5.0 < z < 6.7 $. The legends on the plots indicate, from left to right, $\log(L)_{\rm min}$, $\log(L)_{\rm max}$ and the number of bins considered for the fit between these two limits. The black lines show the results obtained from the optimal bins adopted in this work.}
  \label{fig:bin_test_results}
\end{figure*}

\begin{figure*}
  \includegraphics[trim=0cm 0.3cm 0.3cm 0.3cm,clip,width=0.5\hsize] {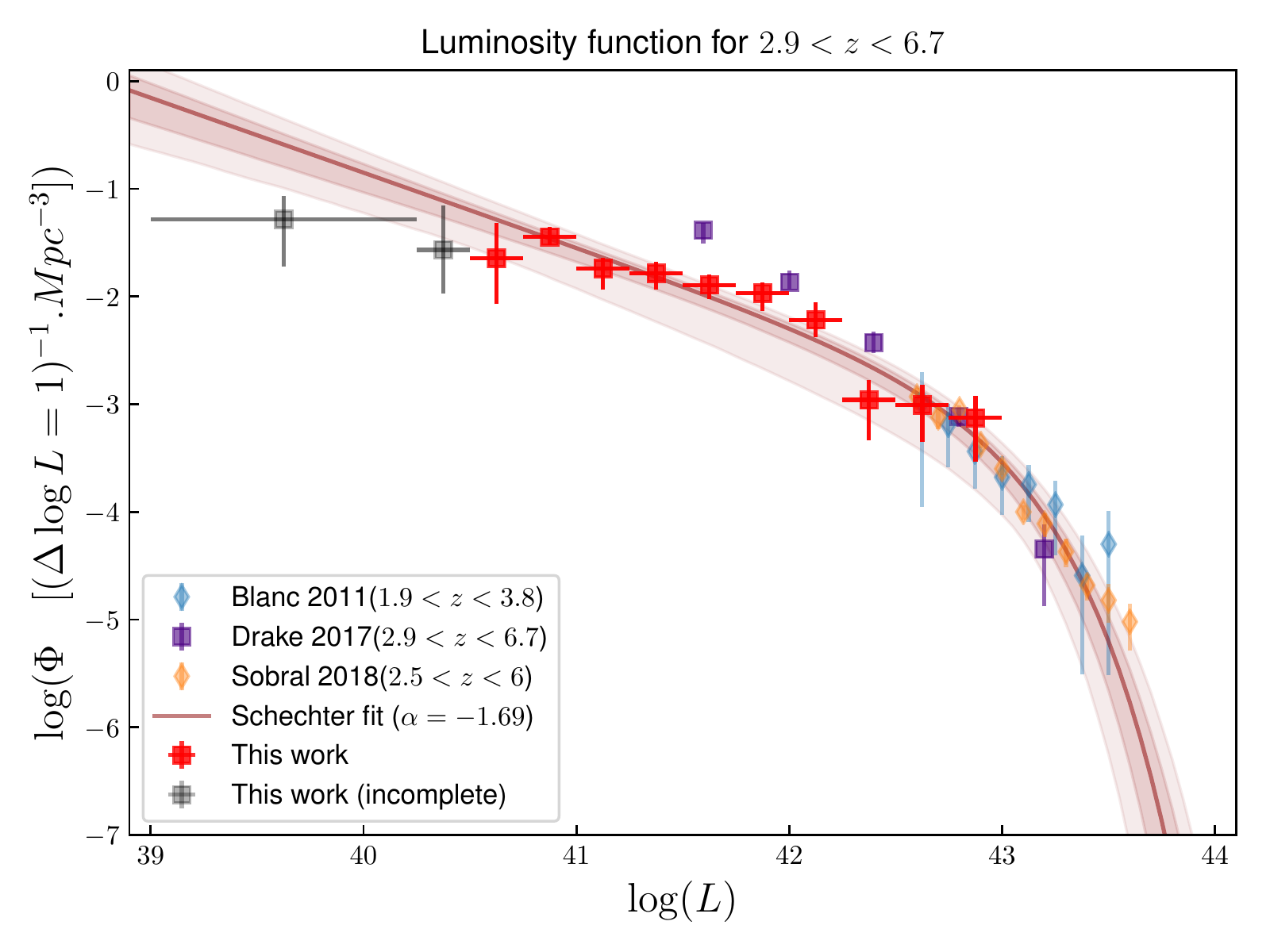}
  \includegraphics[trim=0cm 0.3cm 0.3cm 0.3cm,clip,width=0.5\hsize]
  {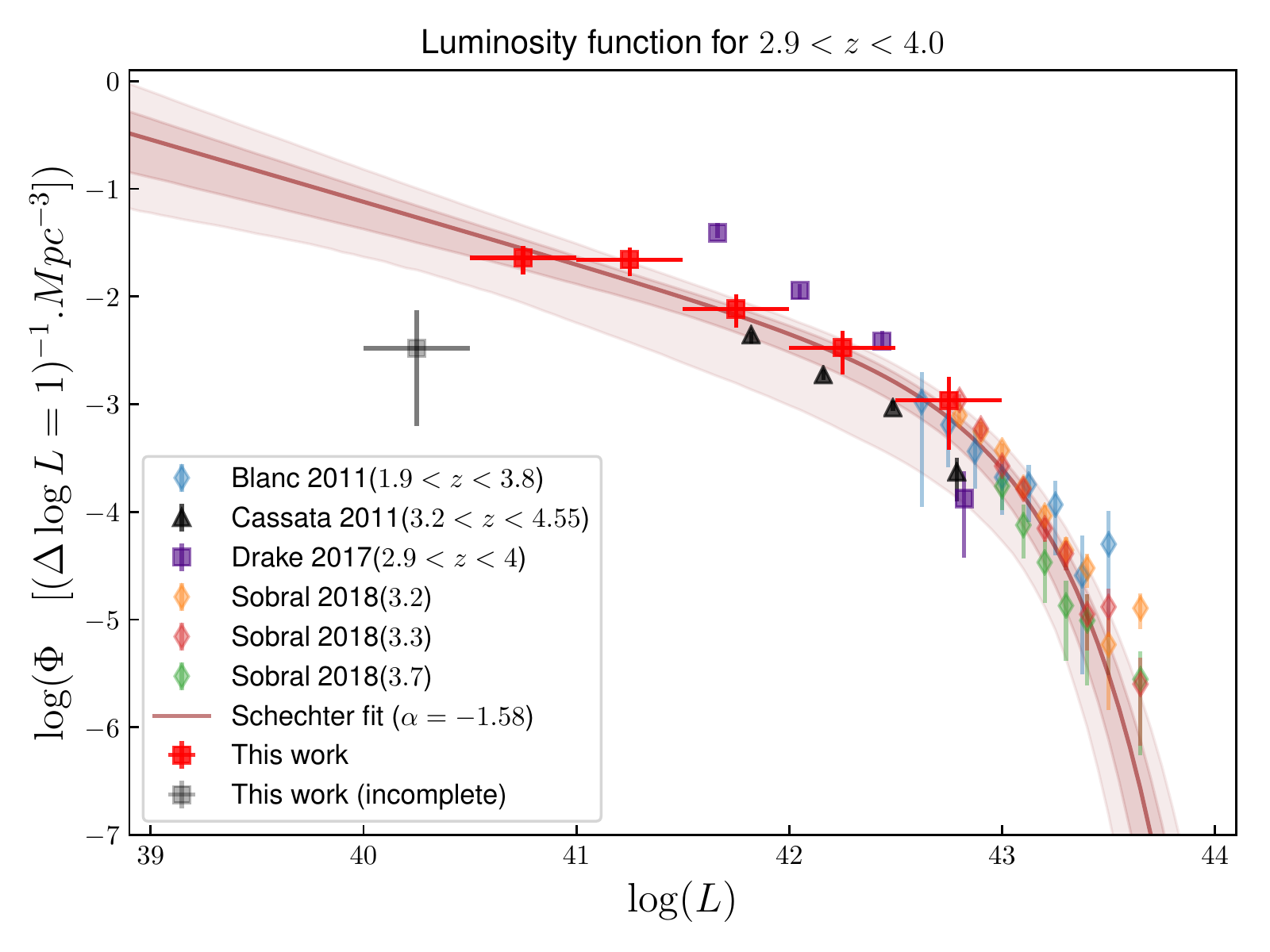}
  \includegraphics[trim=0cm 0.3cm 0.3cm 0.3cm,clip,width=0.5\hsize]
  {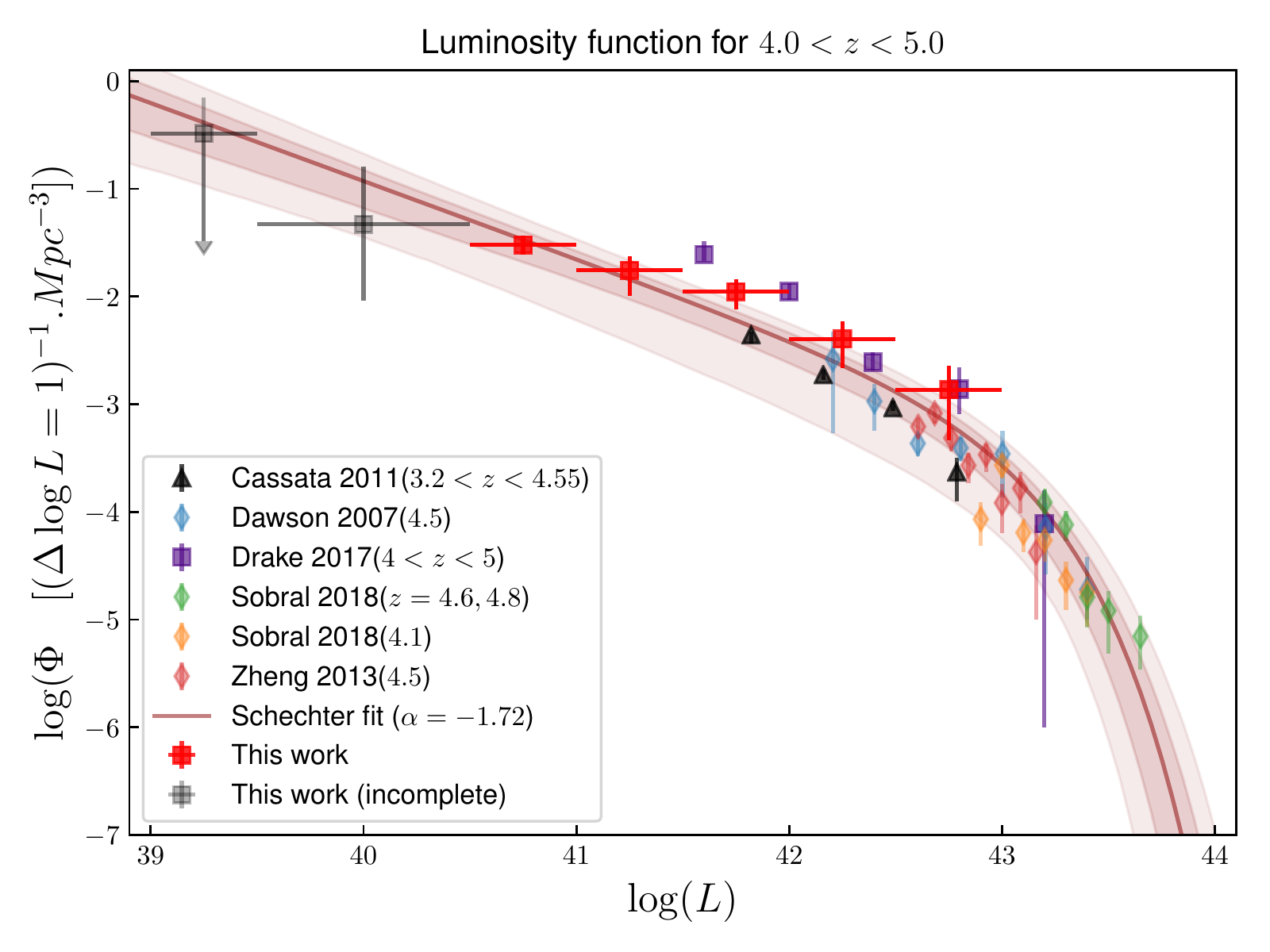}
  \includegraphics[trim=0cm 0.3cm 0.3cm 0.3cm,clip,width=0.5\hsize]
  {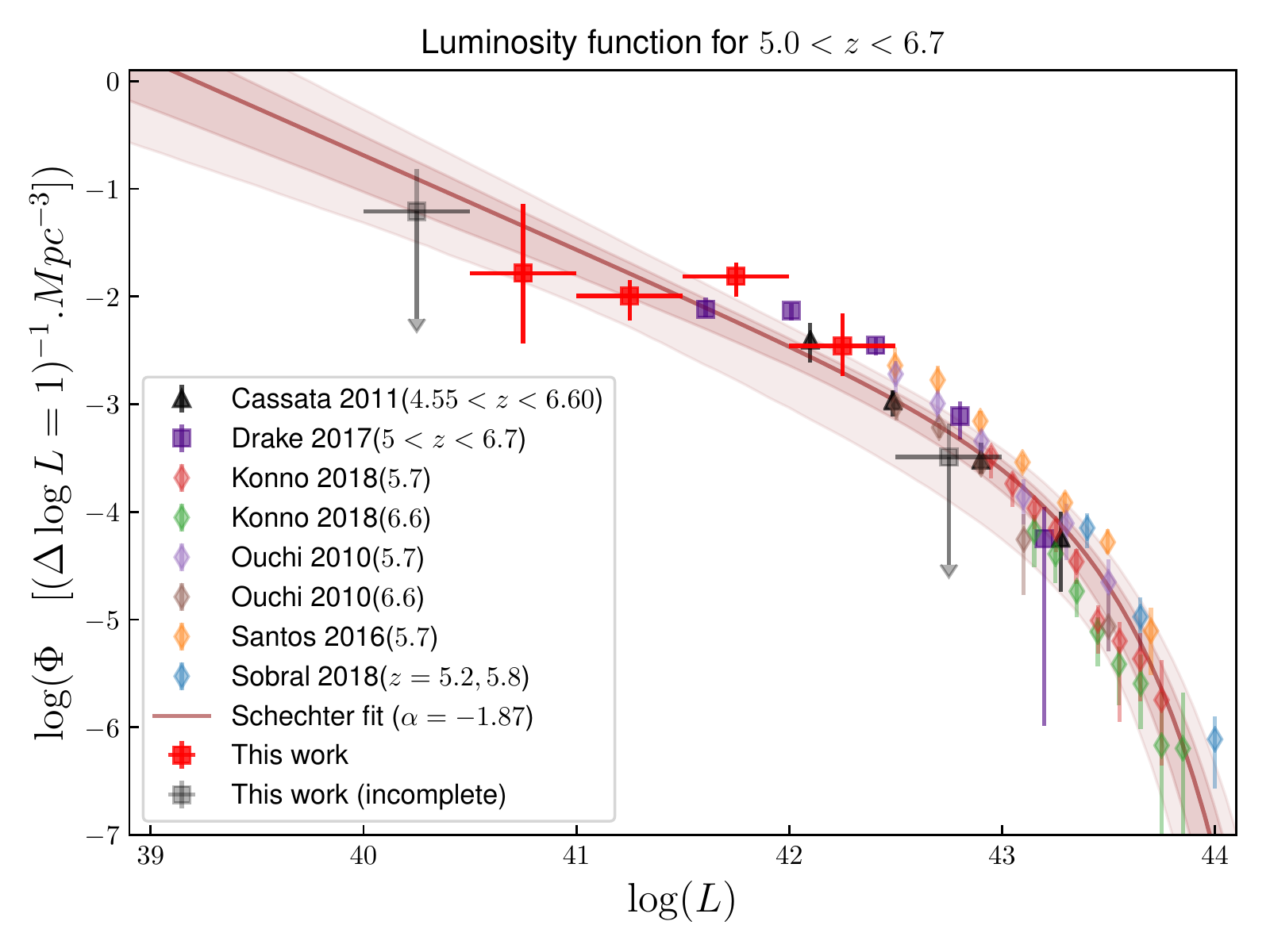}
  \caption{Luminosity functions and their respective fits for the 4 different redshift bins considered in this study. The red and grey squares represent the points derived in this work, where the grey squares are considered incomplete and are not used in the different fits. The literature points used to constrain the bright end of the LFs are shown as lightly coloured diamonds. The black points represent the results obtained by \citet{Cassata2011}, which were not used for the fits. The purple squares represent the points derived using the $V_{\rm max}$ method in \citetalias{Drake2017b} and are only shown for comparison. The best Schechter fits are shown as a solid line and the 68\% and 95\% confidence areas as dark red coloured regions, respectively.}
  \label{fig:fit_lf}
\end{figure*}

In addition, the error bars do not account for the error introduced by the binning of the data. To further test the robustness of the slope measurement and to recover more realistic error bars, different bins were tested for the construction of the LF. The exact same fit process was applied to the resulting LFs. The confidence regions derived from these tests are shown in Fig. \ref{fig:bin_test_results} for $z_1$ and $z_3$. The bins used hereafter to build the LFs are identified in this figure as black lines. We estimate that these bins are amongst the most reliable possibilities, and in the following they are referred to as the "optimal" bins. They were determined in such a way that each bin is properly sampled in both redshift and luminosity, and has a reasonable level of completeness. Figure  \ref{fig:bin_test_results} shows that $\alpha$ is very stable for $z_1$ and that all the posterior distributions are very similar. Because we are able to probe very low luminosity regimes far below $L_*$, the effect of binning on the measured slope is negligible for $z_{\rm all}$ because of the increased statistics. As redshift increases as a consequence of lower statistics and higher uncertainties, the effects of binning on the measured slope increases. For $z_2$ the LF is affected by a small overdensity of LAEs at $z \sim 4$  resulting in a higher dispersion on the faint end slope value when testing different binnings. It was ensured that the optimal binning allowed this fit to be consistent with the fit made for $z_{\rm all}$: in both cases the points at $41.5 \lesssim \log L \lesssim 42 $, affected by the same sources at $z \sim 4$, are treated as a small overdensity with respect to the Schechter distribution. Finally, for  $z_3$, the lack of statistics seriously limits the possibilities of binnings to test. The only viable options are the two presented on the right panel of Fig. \ref{fig:bin_test_results}: in both cases the quality of the fit is poor compared to the other redshift bins, but the measured slopes are consistent within their own error bars.\\

\begin{figure}
  \centering
  \includegraphics[width=\hsize]{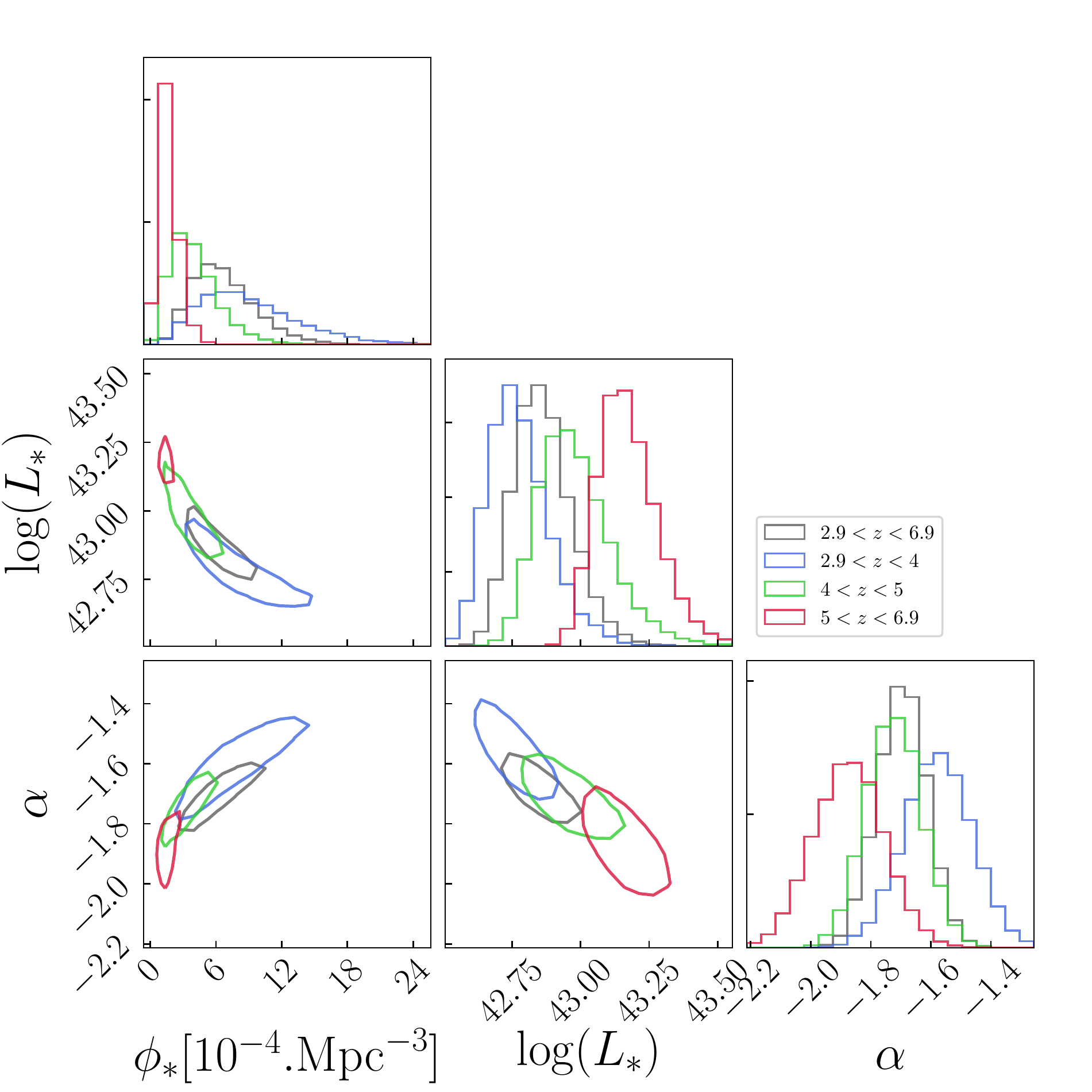}
  \caption{Evolution of the Schechter parameters with redshift. The contours plotted correspond to the limits of the 68\%  confidence areas determined from the results of the fits.}
  \label{fig:evolution_conf_interval}
\end{figure}

The LF points from the literature used to constrain the bright end are taken from \citet{Blanc2011} and \citet{Sobral2018} for $z_{\rm all}$ and $z_1$, 
\citet{Dawson2007}, \citet{Zheng2013}, and \citet{Sobral2018} for $z_2$, and finally
\citet{Ouchi2010}, \citet{Santos2016}, \citet{Konno2018}, and \citet{Sobral2018} for $z_3$. The goal is to extend our own data towards the highest luminosities using available high-quality data with enough overlap to check the consistency with the present data set. The best fits and the literature data sets used for the fits are also shown in Fig. \ref{fig:fit_lf} as full lines and lightly coloured diamonds, respectively. The dark red coloured regions indicate the 68\% and 95\% confidence areas for the Schechter fit. The best Schechter parameters are listed in Table \ref{tab:fit_results}. In addition, this Table contains the results obtained when the exact same method of LF computation is applied to the sources of A2744 as an independent data set. This is done to assess the robustness of the method and to see whether or not the addition of low volume and high magnification cubes add significant constraints on the faint end slopes. All four fits made using the complete sample are summed up in Fig. \ref{fig:evolution_conf_interval}, which shows the evolution of the confidence regions  for $\alpha$, $\Phi_*$, and $L_*$ with redshift.

\begin{table*} 
\caption{Results of the fit of the Schechter function in the different redshift intervals. The last two columns list the Lyman-alpha flux density and the SFRD as a function of redshift, obtained from the integration of the LFs derived in Sect. \ref{sec:lf_fit}. The errors on the parameters of the Schechter function correspond to  68\% confidence interval. The values $\rho_{Ly_{\rm \alpha}}$  are computed using a lower integration limit $\log(L) = 40.5,$ which is considered to be the completeness limit of this work. For each redshift bin, the Schechter parameters are measured from the the LFs computed from the entire sample and from the LAEs of A2744 only.}
\label{tab:fit_results}
\centering     
\renewcommand{\arraystretch}{1.5} 
\begin{tabular}{l l | c c c c c c c} 
\hline \hline 
 & & N$_{\rm obj}$ & N$_{\rm corrected}$ & $\Phi_*$ & $\log L_* $ & $\alpha$ & $\log \rho_{Ly\alpha}$ & $\log SFRD $\\ 
  & &              &                  &  $10^{-4} $Mpc$^{-3}$   & erg s$^{-1}$  &     & erg s$^{-1}$ Mpc$^{-3}$    & M$_{\odot}$ yr$^{-1}$ Mpc$^{-3}$ \\  
\hline \hline 
$2.9 < z < 6.7$&All clusters&152&278&$6.38_{-2.46}^{+3.26}$&$42.85_{-0.10}^{+0.11}$&$-1.69_{-0.08}^{+0.08}$&$40.08_{-0.04}^{+0.04}$&$-1.94_{-0.04}^{+0.04}$ \\ 
&A2744 only&125&235&$3.40_{-1.59}^{+2.33}$&$42.97_{-0.12}^{+0.15}$&$-1.85_{-0.08}^{+0.08}$&$40.14_{-0.04}^{+0.04}$&$-1.88_{-0.04}^{+0.04}$ \\ 
\hline 
$2.9 < z < 4.0  $&All clusters&61&119&$8.29_{-3.66}^{+5.25}$&$42.77_{-0.10}^{+0.12}$&$-1.58_{-0.11}^{+0.11}$&$39.99_{-0.07}^{+0.07}$&$-2.03_{-0.07}^{+0.07}$ \\ 
&A2744 only&40&102&$7.51_{-3.43}^{+4.97}$&$42.78_{-0.10}^{+0.13}$&$-1.58_{-0.12}^{+0.12}$&$39.97_{-0.07}^{+0.07}$&$-2.05_{-0.07}^{+0.07}$ \\ 
\hline 
$4.0  < z < 5.0  $&All clusters&52&86&$3.67_{-1.72}^{+2.51}$&$42.96_{-0.11}^{+0.14}$&$-1.72_{-0.09}^{+0.09}$&$39.99_{-0.06}^{+0.06}$&$-2.03_{-0.06}^{+0.06}$ \\ 
&A2744 only&40&68&$1.52_{-0.87}^{+1.45}$&$43.12_{-0.15}^{+0.20}$&$-1.96_{-0.09}^{+0.08}$&$40.13_{-0.05}^{+0.05}$&$-1.89_{-0.05}^{+0.05}$ \\ 
\hline 
$5.0  < z < 6.7  $&All clusters&39&73&$1.53_{-0.68}^{+0.96}$&$43.16_{-0.10}^{+0.12}$&$-1.87_{-0.12}^{+0.12}$&$40.03_{-0.09}^{+0.11}$&$-1.99_{-0.09}^{+0.11}$ \\ 
&A2744 only&33&64&$1.40_{-0.64}^{+0.91}$&$43.18_{-0.10}^{+0.12}$&$-1.90_{-0.12}^{+0.12}$&$40.05_{-0.11}^{+0.12}$&$-1.97_{-0.11}^{+0.12}$ \\ 
\hline 
\hline 
\end{tabular}  
\end{table*}

Table \ref{tab:fit_results} shows that the results are very similar for $z1$ and $z3$ when considering A2744 only or the full sample. For $z_{\rm all}$ and $z_2$ the recovered slopes exhibit a small difference at the $\lesssim 2 \sigma$ level. This difference is caused by one single source with $40.5 \lesssim \log L  \lesssim 41 $, which has a high contribution to the density count. When adding more cubes and sources, the contribution of this LAE is averaged down because of the larger volume and the contribution of other LAEs. This argues in favour of a systematic  underestimation of the cosmic variance in this work. Using the results of cosmological simulations to estimate a proper cosmic variance is out of the scope of this paper. For the higher redshift bin, even though the same slope is measured when using only the LAEs of A2744, the analysis can only be pushed down to $\log L = 41$ (instead of $\log L = 40.5$ for the other redshift bins or when using the full sample). This shows the benefit of increasing the number of lensing fields to avoid a sudden drop in completeness at high redshift. The effect of increasing the number of
lensing fields will be addressed in a future article in preparation. In the following, only the results obtained with the full sample are discussed

The values measured for $L_*$ are in good agreement with the literature (e.g. $\log(L_*) = 43.04 \pm 0.14$ in \cite{Dawson2007} for $z \simeq 4.5$,  $\log(L_*) = 43.25^{+0.09}_{-0.06}$  in \cite{Santos2016} for $z \simeq 5.7$ and a fixed value of $\alpha=-2.0$, and $\log(L_*) = 43.3^{+0.5}_{-0.9}$ in \cite{Hu2010} for $z \simeq 5.7$ and a fixed value of $\alpha = -2.0$) and these values tend to increase with redshift.
This is not a surprise as this parameter is most sensitive to the data points from the literature used to fit the Schechter functions. Given the large degeneracy and therefore large uncertainty affecting the normalization parameter $\phi_*$, a direct comparison and discussion with previous studies is difficult and not so relevant.
  Regarding the $\alpha$ parameter, the Schechter analysis reveals a steepening of the faint end slope with increasing redshift, which in itself means an increase in the observed number of low luminosity LAEs with respect to the bright population with redshift. However, this is a \hbox{$\sim 1 \sigma $} trend that can only be seen in the light of the Schechter analysis, with a solid anchorage of the bright end, and cannot be seen using only the points derived in this work (see e.g. Fig. \ref{fig:lf_points}).

Taking advantage of the unprecedented level of constraints on the low luminosity regime, the present analysis has confirmed a steep faint end slope varying from $\alpha = -1.58^{+0.11}_{-0.11}$  at $ 2.9 < z < 4$ to $\alpha = -1.87^{+0.12}_{-0.12}$ at $5 < z < 6.7$. The result for the lower redshift bin is not consistent with  $\alpha=-2.03^{+1.42}_{-0.07}$ measured using the maximum-likelihood technique in \citetalias{Drake2017b}. At higher redshift, the slopes measured in \citetalias{Drake2017b} are upper limits, which are consistent with all the values in Table \ref{tab:fit_results}. The points in purple in Fig. \ref{fig:fit_lf} are the points derived with the $V_{\rm max}$ from this same study. It can be seen that there is a systematic difference, increasing at lower luminosity for $z_{\rm all}$, $z_1$ and $z_2$. This difference, taken at face value, could be evidence for a systematic underestimation of the cosmic variance both in this work and in \citetalias{Drake2017b}. This aspect clearly requires further investigation in the future. Faint end slope values of $\alpha=-2.03^{+0.4}_{-0.3}$ for $ z = 5.7$  and
  $\alpha  = -2.6^{+0.6}_{-0.4}$ for $z \sim 5.7 $ ($\alpha = -2.5^{+0.5}_{-0.5}$ for $z \sim 6.6$) were found in  \citet{Santos2016} and \citet{Konno2018}, respectively. These values are reasonably consistent with our measurement made for $z_3$. In this case again, the comparison with the literature is quite limited as the faint end slope is often  fixed \mbox{\citep[see e.g.][]{Dawson2007,Ouchi2010}} or the luminosity range probed is not adequate leading to poor constraints on $\alpha$.

From Fig. \ref{fig:fit_lf}, we see that the Schechter function provides a relatively good fit for $z_{\rm all}$, $z_1$, and $z_2$. The over-density in number count at $z \sim 4$ for $ 41.5 \lesssim \log L \lesssim 42 $ is indeed seen as an over-density with respect to the Schechter distribution. For $z_3$ however, the fit is not as good with one point well above the $1\sigma$ confidence area. The final goal of this work is not the measurement of the Schechter slope in itself, but to provide a solid constraint on the shape of the faint end of the LF. Furthermore it is not certain that such a low luminosity population is expected to follow a Schechter distribution. Some studies have already explored the possibility of a turnover in the LF of UV selected galaxies \citep[e.g.][]{Bouwens2017,Atek2018}, and the same possibility is not to be excluded for the LAE population. For the specific needs of this work, it remains convenient to  adopt a parametric form as it makes the computation of proper integrations with correct error transfer easier (see Sect. \ref{sec:discussion}) and facilitates the comparison with previous and future works. When talking about integrated LFs, any reasonable deviations from the Schechter form is of little consequence as long as the fit is representative of the data. In other words, as long as no large extrapolation towards low luminosity is made, our Schechter fits provide a good estimation of the integrated values.


\section{Discussion and contribution of LAEs to reionization}
\label{sec:discussion}

In this section, before going to the integration of the LFs and the constraints and implications for reionization, we discuss the  uncertainties introduced by the use of lensing.
As part of the HFF programme, several good quality mass models were produced and made publicly available by different teams, using different methodologies. The uncertainties introduced by the use of lensing fields when measuring the faint end of the UV LF are discussed in detail in \citet{Bouwens2017} and \citet{Atek2018} through simulations. A more general discussion on the reason why mass models of the same lensing cluster may differ from one another can be found in \citet{Priewe2017}. And finally, a thorough comparison of the mass reconstruction produced by different teams with different methods from simulated lensing clusters and HST images is done in \citet{Meneghetti2017}. The uncertainties are of two types:
\begin{itemize}
\item The large uncertainties for high magnification values. This aspect is well treated in this work through the use of $P(\mu),$ which allows any source to have a diluted and very asymmetric contribution to the LF over a large luminosity range. This aspect was already addressed in Sect. \ref{sec:sample_description}.
\item The possible systematic variation from one mass model to another. This aspect is more complex as it has an impact on both the individual magnification of sources and on the total volume of the survey. 
\end{itemize}

\begin{figure}
  \centering
  \includegraphics[width=\hsize]{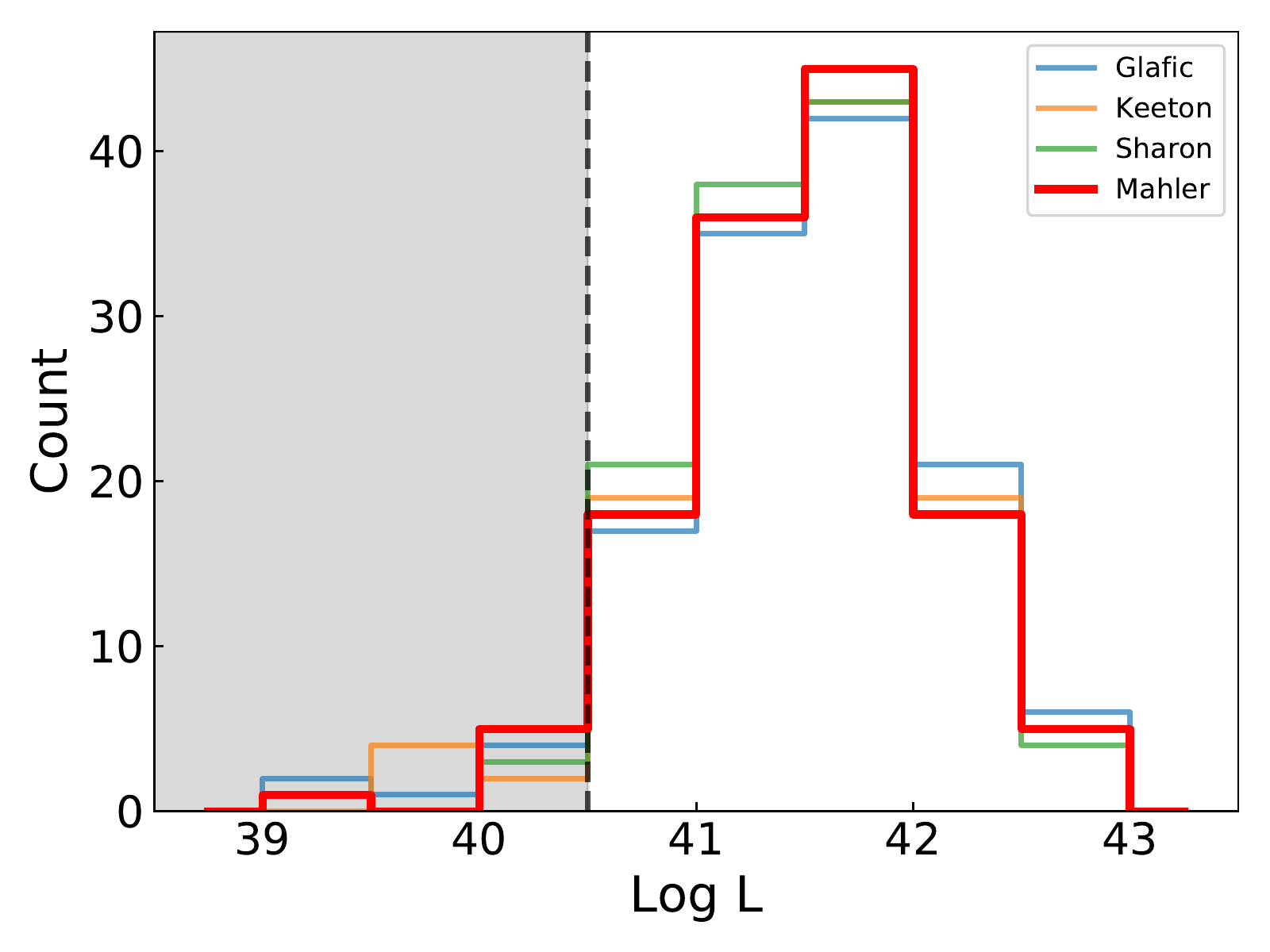}
  \cprotect\caption{Comparative Lyman-alpha luminosity histograms obtained using the magnification resulting from different mass models. The grey area represents the completeness limit of this work.}
  \label{fig:mass_model_comp}
\end{figure}

Figure \ref{fig:mass_model_comp} illustrates the problem of variation of individual magnification from one mass model to another, using the V4 models produced by the GLAFIC team \citep{Kawamata2016, Kawamata2018}, Sharon \& Johnson \citep{Johnson2014}, and Keeton that are publicly available on the HFF website \footnote{\url{https://archive.stsci.edu/prepds/frontier/lensmodels/}}. Since we are restricted to the HFF, this comparison can only be done for the LAEs of A2744. The figure shows the Lyman-alpha luminosity histograms when using alternatively the individual magnification provided by these three additional models. The bin size is $\Delta \log L = 0.5, $ which is the bin size used in this work for the LFs at $z_1$,$z_2$ and $z_3$. For $\log L > 40.5$ the highest dispersion is of the order of 15\%. This shows that even though there is a dispersion when looking at the magnification predicted by the four models, the underlying luminosity population remains roughly the same. Regarding the needs of the LF, this is the most important point.

Figure 10 of \citet{Atek2018} shows an example of the variations of volume probed with rest-frame UV magnitude using different mass models for the lensing cluster MACS1149. This evolution is very similar for the models derived by the Sharon and Keeton teams and, in the worst case scenario, implies a factor of $\sim$ 2 of difference among the models compared in this figure. These important variations are largely caused by the lack of constraints on the mass distribution outside of the multiple image area: a small difference in the outer slope of the mass density  affects the overall mass of the cluster and therefore the total volume probed. However, unlike other lensing fields from the HFF programme, A2744 has an unprecedented number of good lensing constraints at various redshifts thanks to the deep MUSE observations. These constraints were shared between the teams and are included in all the V4 models used for comparison in this work. These four resulting mass models are robust and coherent, at the state of the art of what can be achieved with the current facilities. It has also been shown by \citet{Meneghetti2017} based on simulated cluster mass distributions, that the methodology employed by the  CATS (the CATS model for A2744 is the model presented in \citet{Mahler2018}) and GLAFIC teams are among the best to recover the intrinsic mass distribution of galaxy clusters. To test the possibility of a systematic error on the survey volume, the surface of the source plane reconstruction of the MUSE FoV is compared at $z=4.5$ using the same four models as in Fig. \ref{fig:mass_model_comp}. The surfaces are $(1.23\arcmin)^2$, $(1.08\arcmin)^2$,$(1.03\arcmin)^2$, and $(0.94\arcmin)^2$ using the mass models of Mahler, GLAFIC, Keeton, and Sharon, respectively. The strongest difference is observed between the models of Mahler and Sharon and corresponds to a relatively small difference of only 25\%.

Given the complex nature of the MUSE data combined with the lensing cluster analysis, precisely assessing the effect of a possible total volume bias is nontrivial and out of the scope of this paper. From this discussion it seems clear that the use of lensing fields introduces an additional uncertainty on the faint end slope. However the luminosity limit under which this effect becomes dominant remains unknown as all the simulations \citep{Bouwens2017,Atek2018} were only done for the UV LF for which the data structure is much simpler.\\

In order to estimate the contribution of the LAE population to the cosmic reionization, its SFRD was computed. From the best parameters derived in the previous section, the integrated luminosity density $\rho_{Ly_{\rm \alpha}}$ was estimated.
The SFRD produced by the LAE population can be estimated using the following prescription for the \citep{Kennicutt1998} assuming the case B for the recombination \citep{Osterbrock2006}:

\begin{equation}
  SFRD_{Ly_{\rm \alpha}} [\text{M}_{\odot} \text{yr}^{-1} \text{Mpc}^{-3}] =
  L_{Ly_{\rm \alpha}}
  [\text{erg s}^{-1}\text{ Mpc}^{-3}] / 1.05 \times 10^{42}
.\end{equation}

This equation assumes an escape fraction of the Lyman-alpha photons ($f_{Ly_{\alpha}}$) of 1 and is therefore a lower limit for the SFRD. Uncertainties on this integration were estimated with MC iterations, by perturbing the best-fit parameters within their rescaled error bars, neglecting the correlations between the parameters. 
The values obtained for the $SFRD_{Ly_{\alpha}}$ and $\rho_{Ly_{\alpha}}$ are presented in Table \ref{tab:fit_results} for a lower limit of integration of \textbf{$\log(L) = 40.5,$} which corresponds to the lowest luminosity points used to fit the LFs (i.e. no extrapolation towards lower luminosities). The equation \textbf{$\log(L) = 44$} is used as upper limit for all integrations. The upper limit has virtually no impact on the final result because the LF drops so steeply at higher luminosity.

\begin{figure*}
  \centering
  \includegraphics[width=\hsize]{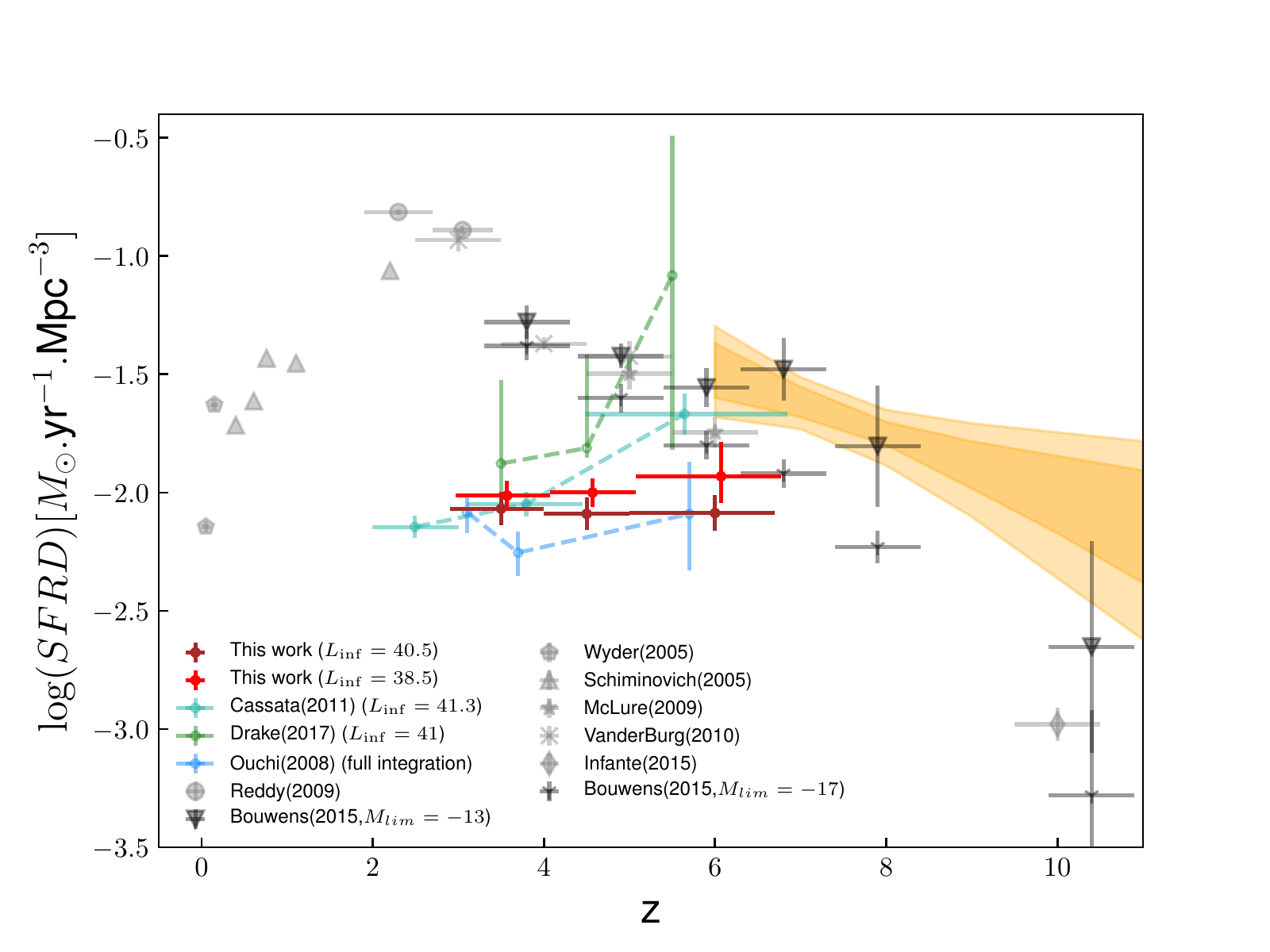}
  \caption{Evolution of the SFRD with redshift with different lower limits of integration. The limit $\log L = 38.5$ corresponds to a 2 dex extrapolation with respect to the completeness limit in this work. Our results (in red / brown) are compared to SFRD in the literature computed for LBGs (in light grey) and from previous studies of the LAE LF (in green / blue). For the clarity of the plot, a small redshift offset was added to the points with $L_{\rm inf} = 38.5$. The darker grey points correspond to the SFRD derived from the LFs in \citet{Bouwens2015b} for a magnitude limit of  integration of $M_{\rm UV} = -17$ corresponding to the observation limit, and $M_{\rm UV} =  -13$. The points reported by \citet{Cassata2011} are corrected for IGM absorption. The yellow area corresponds to  the $1 \sigma $ and  $2 \sigma $ estimations of the total SFRD corresponding to the cosmic emissivity derived in \citet{Bouwens2015a}.}
  \label{fig:evolution_sfrd}
\end{figure*}

We show in Fig. \ref{fig:evolution_sfrd} the results obtained using different lower limits of integration and how they compare to previous studies of both LBG and LAE LFs. The yellow area corresponds to the $1\sigma$ and $2\sigma$ SFRD needed to reionize the universe fully, which is estimated from the cosmic ionizing emissivity derived  in \citet{Bouwens2015a}. The cosmic emissivity was derived using a clumping factor of 3, the conversion to UV luminosity density was done assuming $\log (\xi_{\rm ion} f_{escp})$  = 24.50, where $f_{escp}$ is the escape fraction of UV photons and $\xi_{\rm ion}$ is the Lyman-continuum photon production efficiency. Finally the conversion to SFRD was done with the following relation: $SFRD [\text{M}_{\odot}.\text{yr}^{-1}] = \rho_{\rm UV} / (8.0 \times 10^{27})$  (see \citealt{Kennicutt1998, Madau1998}). Because all the slopes are over $\alpha = -2$ (for $\alpha < - 2$ the integral of the Schechter parameterization diverges), the integrated values increase relatively slowly when decreasing the lower luminosity limit. On the same plot, the SFRD computed from the integration of the LFs derived in \cite{Bouwens2015b} are shown in darker grey for two limiting magnitudes: $M_{\rm UV} = -17$  (which is the observation limit) and $M_{\rm UV} = -13,$ which is thought to be the limit of galaxy formation (e.g. \citealt{Rees1977}, \citealt{MacLow1999} and \citealt{Dijkstra2004}).

\defcitealias{Drake2017b}{D17}
\defcitealias{Cassata2011}{C11}
\defcitealias{Ouchi2008}{O08}
\defcitealias{Herenz2019}{H19}

From this plot, and with $f_{Ly_{\alpha}} = 1 $, we see that the  observed LAE population only is not enough to reionize the universe fully at $z \sim 6$, even with a large extrapolation of 2 dex down to $\log L = 38.5$. However, a straightforward comparison is dangerous: an escape fraction $f_{Ly_{\alpha}} \gtrsim 0.5$ would be roughly enough to match the cosmic ionizing emissivity needed for reionization at $z \sim 6$. Moreover, in this comparison, we implicitly assumed that the LAE population has the same properties ($\log(f_{\rm escp} \xi_{\rm ion}) = 24.5$) as the LBG population in \citet{Bouwens2015b}. A recent study on the typical values of $\xi_{\rm ion}$ and its scatter for typical star-forming galaxies at $z \sim 2$ by \citet{Shivaei2018} has shown that $\xi_{\rm ion}$ is highly uncertain as a consequence of galaxy-to galaxy variations on the stellar population and UV dust
attenuation, while most current estimates at high-$z$ rely on (too) simple prescriptions from stellar population models. The SFRD obtained from LAEs when no evolution in $f_{Ly_{\alpha}}$ is introduced remains roughly constant as a function of redshift when no extrapolation is introduced and slightly increases with redshift when using $L_{\rm inf} = 38.5$. Figure \ref{fig:evolution_sfrd} shows in green/blue, the $SFRD_{Ly_{\alpha}}$ values derived in previous studies of the LAE LF, namely \citet{Ouchi2008}, \citet{Cassata2011} (hereafter, \citetalias{Ouchi2008}, \citetalias{Cassata2011}), and \citetalias{Drake2017b}.
In \citetalias{Cassata2011}, a basic correction for IGM absorption was performed assuming $f_{Ly_{\alpha}}$ varying from 15\% at $z = 3$ to 50\% at $z = 6$ and using a simple radiative transfer prescription from \citet{Fan2006}. This correction can easily explain the clear trend of increase of SFRD with redshift and the discrepancy with our points at higher redshift. At lower redshifts, the IGM correction is lower and the points are in a relatively good agreement. The points in \citetalias{Ouchi2008} are the result of a full integration of the LFs with a slope fixed at $\alpha = -1.5$ and are in reasonable agreement for all redshift domains. The two higher redshift points derived in \citetalias{Drake2017b} are inconsistent with our measurements.  This is not a surprise as the slopes derived in \citetalias{Drake2017b} are systematically steeper and inconsistent with this work.


The use of an IFU (MUSE) in \citetalias{Drake2017b}, in \citet{Herenz2019} (hereafter \citetalias{Herenz2019}), and this survey ensures that we better recover the total flux, even though we may still miss the faintest part of the extended Lyman-alpha haloes (see \citealt{Wisotzki2016}). This is not the case for NB \citepalias[e.g.][]{Ouchi2008} or slit-spectroscopy \citep[e.g.][]{Cassata2011} surveys in which a systematic loss of flux is possible for spatially extended sources or broad emission lines because of the limited aperture of the slits or the limited spectral width of NB filters. It is noted in \citetalias{Herenz2019} that the $3.2 < z < 4.55$ LF estimates in \citetalias{Cassata2011} tend to be lower than most literature estimates (including those in \citetalias{Herenz2019}). One possible explanation would be a systematic loss of flux, which results in a systematic shift of the derived LF towards lower luminosities. Interestingly, when assuming point-like sources to compute the selection function, \citetalias{Herenz2019} manages to recover very well the results of \citetalias{Cassata2011} for this redshift domain. It is also interesting to see that as luminosity decreases, the LF estimates from \citetalias{Cassata2011} become more and more consistent with the points and Schechter parameterization derived in this work. For $z_3$, the \citetalias{Cassata2011} LF is even fully consistent with the Schechter parameterization across the entire luminosity domain (see Fig. \ref{fig:fit_lf}). The following line of thought could explain the concordance of this work with the \citetalias{Cassata2011} estimates at lower luminosity and higher redshift:
 At lower luminosity and higher redshift, a higher fraction of LAEs detected are point-like sources, making the \citetalias{Cassata2011} LFs more consistent with our values; and at higher luminosity and lower redshift, more extended LAEs are detected and a more complex correction is needed to get a realistic LF estimate.

The second advantage of using an IFU is linked to the selection of the LAE population. The \citetalias{Ouchi2008} authors used a NB photometric selection of sources with spectroscopic follow-up to confirm the LAE candidates. This results in an extremely narrow redshift  window which is likely to lead to lower completeness of the sample due to the two-step selection process.
The studies by \citetalias{Drake2017b} and \citetalias{Herenz2019}, adopt the same approach as this work: a blind spectroscopic selection of sources. In addition, as shown in Fig. \ref{fig:comparison_alyssa} and stated in Sect. \ref{sec:lf_fit} when discussing the differences in slope between A2744 alone and the full sample, the use of highly magnified observations allows for a more complete source selection at increasing redshift. The sample used in the present work could arguably have a higher completeness level than other previous studies.

To summarize the above discussion, the observational strategy adopted in this study by combining the use of MUSE and lensing clusters has allowed us to

\begin{itemize}
\item reach fainter luminosities, providing better constraints on the faint end slope of the LF, while still taking advantage of the previous studies to constrain the bright end;
\item recover a greater fraction of flux for all LAEs;
\item cover a large window in redshift and flux;
\item reach a higher level of completeness, especially at high redshift.
\end{itemize}

A steepening of the faint end slope is observed with redshift, which follows what is usually expected. This trend can be explained by a higher proportion of low luminosity LAEs observed at higher redshift owing to higher dust content at lower redshift. On the other hand, the density of neutral hydrogen is expected to increase across the $ 5<z<6.7 $ interval, reducing the escape fraction of Lyman-alpha photons, a trend affecting LAEs in a different way depending on large-scale structure. 
While an increase of SFRD with redshift is observed, the evolution of the observed $SFRD_{Ly_{\alpha}}$ is also affected by $f_{Ly_{\alpha}}$. From the point of view of the literature, the expected evolution of $f_{Ly_{\alpha}}$ is  an  increase with redshift up to $z \sim 6-7$ and then a sudden drop at higher redshift (see e.g. \citealt{Clement2012}, \citealt{Pentericci2014}). For $z < 6$, the increase of $f_{Ly_{\alpha}}$ is generally explained by the reduced amount of dust at higher redshift. And for $z \sim 6 - 7$ and above, we start to probe the reionization era and owing to the increasing amount of neutral hydrogen and the resonant nature of the $Ly_{\rm \alpha}$, the escape fraction is expected to drop at some point. It has been suggested  in  \citet{Trainor2015} and \citet{Matthee2016} that the escape fraction would decrease with an increasing SFRD. This would only increase the significance of the trend observed, as it means the points with the higher SFRD would have a larger correction.

Furthermore the derived LFs and the corresponding SFRD values could be affected by bubbles of ionized hydrogen, especially in the last redshift bin. In our current understanding of the phenomenon, reionization is not a homogeneous process \citep{Becker2015, Bosman2018}. It could be that the expanding areas of ionized hydrogen develop faster in the vicinity of large structures with a high ionising flux, leaving other areas of the universe practically untouched. There is increasing observational evidence of this effect (see e.g. \citealt{Stark2017}). It was shown in \citet{Matthee2015}, using a simple toy model, that an increased amount of neutral hydrgen in the IGM could produce a flattening of the faint end shape of the LF. This same study also concluded that the clustering of LAEs had a large impact on the individual escape fraction, which makes it difficult to estimate a realistic correction, as the escape fraction should be estimated on a source to source basis.

As previously discussed, it is neither certain nor expected that the LAE population alone is enough to reionize the universe at  $z \sim 6$. However, the LBG and the LAE population have roughly the same level of contribution to the total SFRD at face value. Depending on the intersection between the two populations, the observed LAEs and LBGs together could produce enough ionizing flux to maintain the ionized state of the universe at $z \sim 6 $.

This question of the intersection is crucial in the study of the sources of reionization. Several authors have addressed the prevalence of LAE among LBG galaxies, and have shown that the fraction of LAE increases
for low luminosity UV galaxies till $z \sim 6$, whereas the LAE fraction strongly decreases towards $z \sim 7$ (see e.g. \citealt{Stark2010}, \citealt{Pentericci2011}). The important point however is to precisely determine the contribution of the different populations of star-forming galaxies within the same volume, which is a problem that requires the use of 3D/IFU
spectroscopy. As a preliminary result, we estimate that $ \sim 20$\% of the sample presented in this study have no detected counterpart
on the deep images of the HFFs. A similar analysis is being conducted on the deepest observations of MUSE on the Hubble Ultra Deep Field \citep{Maseda2018}.


\section{Conclusions}
\label{sec:conclusion}

The goal of this study was to set constraints on the sources of cosmic reionization by studying the LAE LF. Taking advantage of the great capabilities of the MUSE instrument and using lensing clusters as a tool to reach lower luminosities, we blindly selected behind four lensing clusters a population of 156 spectroscopically identified LAEs that have $2.9 < z < 6.7 $ and magnification corrected luminosities $39 \lesssim \log L \lesssim 43 $.

Given the complexity in combining the spectroscopic data cubes of MUSE with gravitational lensing, and taking into account that each source needs an appropriate treatment to properly account for its magnification and representativity, the computation of the LF needed a careful implementation, including some original developments. For these needs, a specific procedure was developed, including the following new methods: First, we created a precise $V_{\rm max}$ computation for the sources found behind lensing clusters is based on the creation of 3D masks. This method allows us to precisely map the detectability of a given source in  MUSE spectroscopic cubes. These masks are then used to compute the cosmological volume in the source plane. This method could be easily adapted to be used in blank field surveys.
Second, we developed a completeness determination based on simulations using the real profile of the sources. Instead of performing a heavy parametric approach based on MC source injection and recovery simulations, which is not ideally suited for lensed galaxies, this method uses the real profile of sources to estimate their individual completeness. The method is faster, more flexible, and accounts in a better way for the specificities of individual sources, both in the spatial and spectral dimensions.

After applying this procedure to the LAE population, the Lyman-alpha LF has been built for different redshift bins using 152 of the 156 detected  LAEs. Four LAEs were removed because their contribution was not trustworthy. Because of the observational strategy, this study provides the most reliable constraints on the shape of the faint end of the LFs to date and therefore, a more precise measurement of the integrated SFRD associated with the LAE population. The results and conclusions can be summarized as follows:

\begin{itemize}
\item The LAE population found behind the four lensing clusters was split in four redshift bins: $2.9 < z < 6.7$, $2.9 < z < 4.0$, $4.0 < z < 5.9, $ and $5.0 < z < 6.7$. Because of the lensing effect, the volume of universe probed is greatly reduced in comparison to blank field studies. The estimated average volume of universe probed in the four redshift bins are $\sim 15\,000$ Mpc$^3$, $\sim 5\,000$ Mpc$^3$, $\sim 4\,000$ Mpc$^3$, and $\sim 5\,000$ Mpc$^3$, respectively.

\item The LAE LF was computed for the four redshift bins. By construction of the sample, the derived LFs efficiently probe the low luminosity regime and the data from this survey alone provide solid constraints on the shape of the faint end of the observed LAE LFs. No significant evolution in the shape of the LF with redshift is found using these points only. These results have to be taken with caution given the complex nature of the lensing analysis, on the one hand,
and the small effective volume probed by the current sample on the other hand. Our results argue towards a possible systematic underestimation of cosmic variance in the present and other similar works.
  
\item A Schechter fit of the LAE LF was performed by combining the LAE LF computed in this analysis with data from previous studies to constrain the bright end.  As a result of this study, a steep slope was measured for the faint end, varying with redshift between $\alpha = -1.58^{+0.11}_{-0.11}$  at $2.9 < z < 4$ and $\alpha = -1.87^{+0.12}_{-0.12}$ at $ 5 < z < 6.7 $
  
\item The $SFRD_{Ly_{\alpha}}$ values were obtained as a function of redshift by the integration of the corresponding Lyman-alpha LF and compared to the levels needed to ionize the universe as determined in \citet{Bouwens2015a}. No assumptions were made regarding the escape fraction of the Lyman-alpha photons and the $SFRD_{Ly_{\alpha}}$  derived in this work correspond to the observed values. Because of the well-constrained LFs and a better recovery of the total flux, we estimate that the present results are more reliable than previous studies. Even though the LAE population undoubtedly contributes to a significant fraction of the total SFRD, it remains unclear whether this population alone is enough to ionize the universe at $z \sim 6$. The results depend on the actual escape fraction of Lyman-alpha photons.

\item The LAEs and the LBGs have a similar level of contribution at $z \sim 6$ to the total SFRD level of the universe. Depending on the intersection between the two populations, the union of both the LAE and LBG populations may be enough to reionize the universe at $z \sim 6$.
\end{itemize}

Through this work, we have shown that the capabilities of the MUSE instrument make it an ideal tool to determine the LAE LF.
Being an IFU, MUSE allows for a blind survey of LAEs, homogeneous in redshift, with a better recovery of the total flux as compared to
classical slit facilities. The selection function is also better understood as compared to NB imaging.

About $20\%$ of the present LAE sample have no identified photometric counterpart, even on the deepest surveys to date, i.e. HFF. This is an important point to keep in mind as this is a first element of response regarding the intersection between the LAE and LBG populations. Further investigation is needed to better quantify this intersection. Also the extension of the method presented in this paper to other lensing fields should make it possible to improve the determination of the Lyman-alpha LF and to make the constraints on the sources of the reionization more robust.

   
\begin{acknowledgements}
We thank the anonymous referee for their critical review and useful suggestions. This work has been carried out thanks to the support of the OCEVU Labex
(ANR-11-LABX-0060) and the A*MIDEX project (ANR-11-IDEX-0001-02) funded by the
"Investissements d'Avenir" French government programme managed by the ANR.
Partially funded by the ERC starting grant CALENDS (JR, VP, BC, JM),
the Agence Nationale de la recherche bearing the
reference ANR-13-BS05-0010-02 (FOGHAR), and the
``Programme National de Cosmologie and Galaxies'' (PNCG) of CNRS/INSU, France.
GdV, RP, JR, GM, JM, BC, and VP also acknowledge support by the Programa de
Cooperacion Cientifica - ECOS SUD Program C16U02. NL acknowledges funding from the European Research Council (ERC) under the European Union's Horizon 2020 research and innovation programme (grant agreement No 669253), ABD acknowledges support from the ERC advanced grant ``Cosmic Gas',' LW acknowledges support by the Competitive Fund of the Leibniz Association through grant SAW-2015-AIP-2, and TG acknowledges support from the European Research Council under grant agreement ERC-stg-757258 (TRIPLE)..

Based  on  observations  made  with  ESO  Telescopes  at
the  La  Silla  Paranal  Observatory  under  programme  IDs
060.A-9345,  094.A-0115,  095.A-0181,  096.A-0710,  097.A0269,  100.A-0249,  and  294.A-5032.  Also  based  on  observations  obtained  with  the  NASA/ESA  Hubble  Space Telescope, retrieved from the Mikulski Archive for Space Telescopes  (MAST)  at  the  Space  Telescope  Science  Institute
(STScI). STScI  is  operated  by  the  Association  of  Universities for Research in Astronomy, Inc. under NASA contract
NAS 5-26555. This research made use of Astropy,
a community-developed core Python package for Astronomy
\citep{Astropy2013}. All plots in this paper were created using Matplotlib \citep{Matplotlib}.
\end{acknowledgements}

\bibliographystyle{aa}
\nocite{Blanc2011} \nocite{Sobral2018} \nocite{Zheng2013} \nocite{Ouchi2010} \nocite{Konno2018} \nocite{Schiminovich2005} \nocite{McLure2009} \nocite{Vanderburg2010}

\bibliography{bibliography.bib}

\begin{appendix}

\begingroup
\section{Method to create a mask for a 2D image}
\label{annex:create_mask_from_2d_image}
 
In this section we describe the generic method used to create a mask from the detection image of one given source. The goal is to produce a binary mask or detection mask that indicate where the source could have been detected. The details on how this generic method can be adapted to produce masks for spectroscopic cubes can be found in Sect. \ref{subsec:volume_computation}. The method relies on the detection process itself. For each pixel of the detection image, this approach checks whether the object would have been detected had it been centred on that pixel. This is done by comparing the local noise to the signal of the brightest pixels of the source used as input.\\

The method is based on \verb+SExtractor+. To perform the source detection, \verb+SExtractor+ uses a set of parameters, the most important of which are the \verb+DETECT_THRESH+ and \verb+MIN_AREA+.  The first parameter corresponds to a detection threshold and the second to a minimal number of neighbouring pixels. \verb+SExtractor+ works on a convolved and background subtracted image called the filtered image. A source is only detected if at least \verb+MIN_AREA+ neighbouring pixels are \verb+DETECT_THRESH+ times above the background RMS map (shortened to only RMS map) produced from the detection image. This RMS map is the noise map of the background image also computed by \verb+SExtractor+. The comparison between the filtered image and the RMS map is done pixel to pixel meaning that \verb+filtered[x,y]+ is compared to \verb+RMS[x,y]+ \\

The detection mask computation method is based on the same two parameters: \verb+DETECTION_THRESH+ and \verb+MIN_AREA+. From the filtered image, the procedure selects only the \verb+MIN_AREA+ brightest pixels of the source, (we call this list of values \verb+Bp+) and compares these to the RMS map. The bright pixels profiles of our LAE sample are shown on Fig. \ref{fig:bright_pixels} for illustration purpose. This list contains all the information related to the spatial features of the input source needed by the method.
The adopted criterion is close to that applied by \verb+SExtractor+ for the detection even though it is not, strictly speaking, the same:

\begin{figure}
  \centering
  \includegraphics[trim=0cm  0.8cm 1.2cm 0.9cm,clip, width=\hsize] {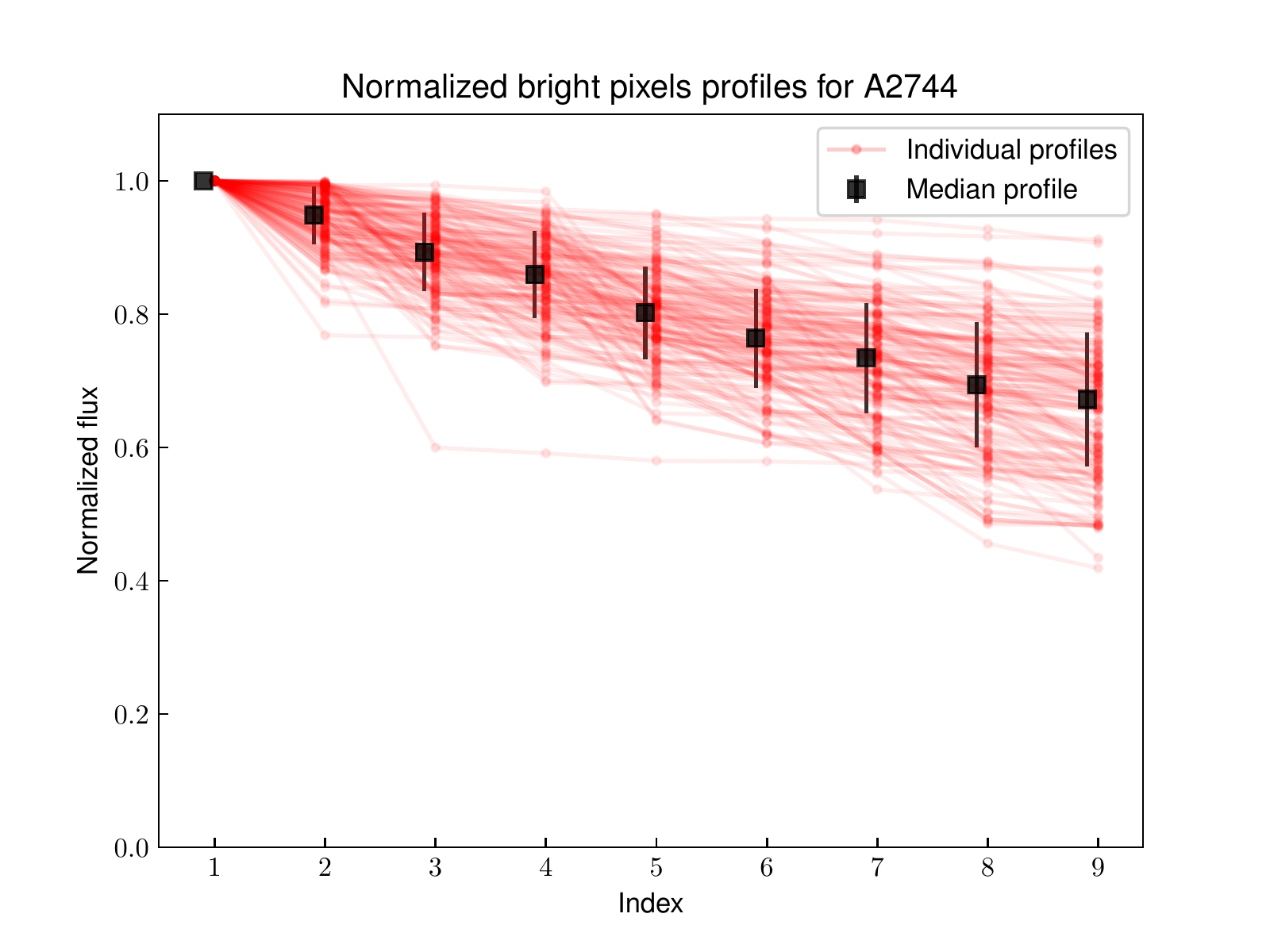}
  \includegraphics[trim=0cm  0.8cm 1.2cm 0.9cm,clip, width=\hsize] {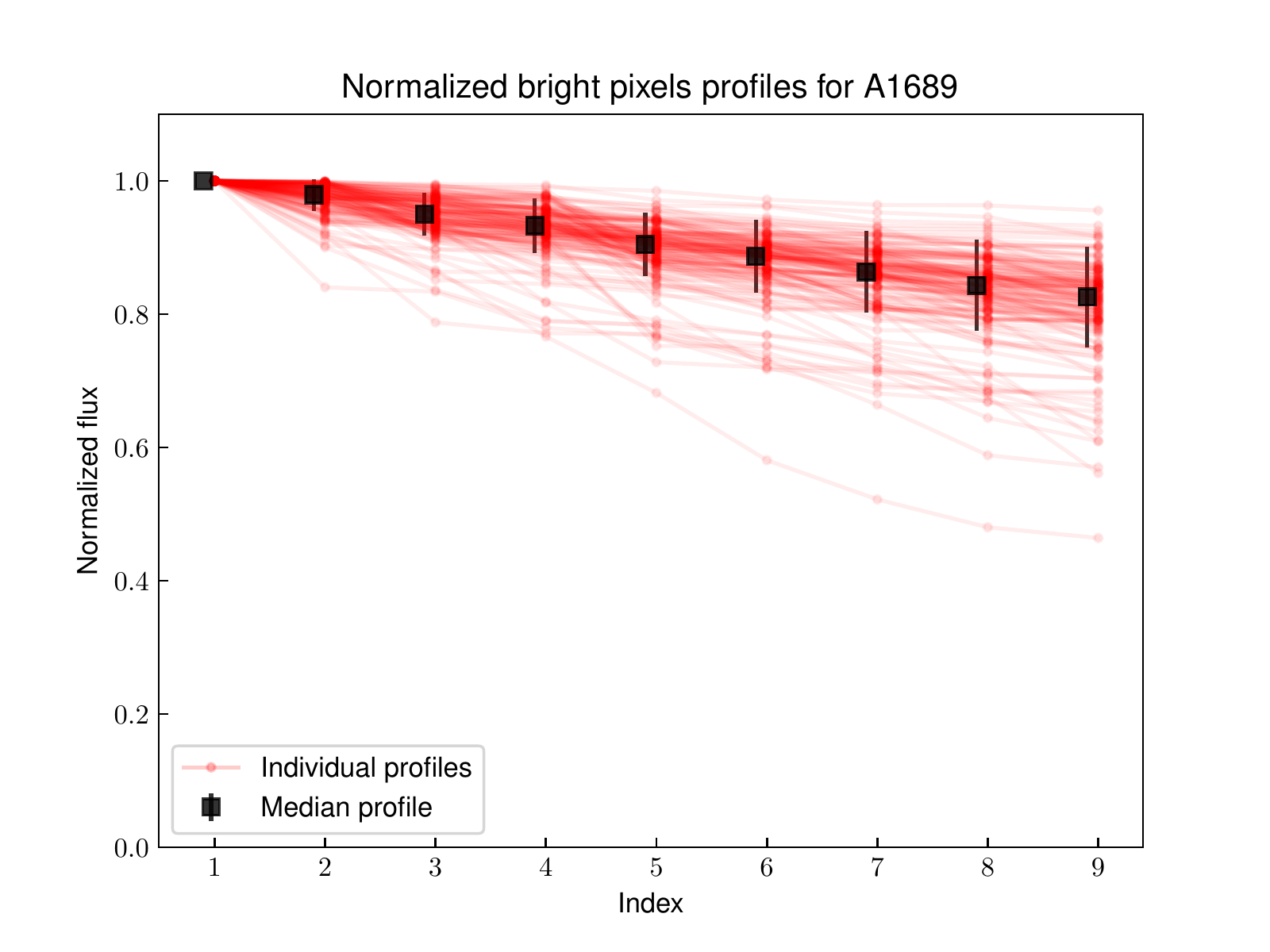}
  \cprotect\caption{Individual bright pixel profiles of all LAEs computed in the seeing condition of A2744 (top) and A1689 (bottom). We note that these are not spatial profiles as two consecutive pixels may not be adjacent on the image. Only the \verb+MIN_AREA+-th first pixels are necessary to compute a mask (\verb+MIN_AREA = 6+ was used in this work).}

  \label{fig:bright_pixels}
\end{figure}

\begin{itemize}
\item [-] For each pixel \verb+[x,y]+ of the RMS map, a list of nine RMS pixels is created; the list contains the central RMS pixel and the eight connected neighbouring RMS pixel values. We call this list \verb+local_noise[x,y]+.
\item [-] From the \verb+Bp+ list that contains the brightest pixel of the input source, \verb+min(Bp)+ is determined and only this value used for the comparison to \verb+local_noise+. For the comparison, the following criterion is used: if any value in \verb+local_noise[x,y]+ fulfils the condition \verb+min(Bp) / DETECT_THRESH < local_noise[x,y]+, then the pixel \verb+[x,y]+ is  masked. In all of the other cases, the central pixel remains unmasked. This criterion is a bit looser than that used by SExtractor as the comparison is only done for \verb+min(Bp)+ and not for all the pixels. However assuming that the noise in a certain small area is not too drastically different, the \verb+SExtractor+ criterion and the criterion we use are still very close. If \verb+min(Bp)+ fulfils the criterion, is it very likely that the other bright pixels, who all have higher signal values, also fulfils the same criterion at some point on the nine pixel area.
\item [-] The operation is performed for each pixel of the RMS map.
\end{itemize}

An example of  application is given in figure \ref{fig:pixel_criterion_schematic}. In both cases, the lowest values of the bright pixel list are compared to the nine pixels in the area set by the red square. The lowest value of the \verb+Bp+ list is set to 6. Using \verb+DETECT_THRESH = 2+, for the central pixel to be masked, none of the values in the red area must be strictly less than \verb+min(Bp) / DETECT_THRESH = 3+. However, for the central pixel to remain unmasked, only one pixel in the red area has to be strictly less than 3, which is true for three pixels on the example on the right.

An example of RMS maps, filtered image, and mask produced for a given source is provided on Figure \ref{fig:2d_mask_example}. The RMS and filtered maps are directly produced by \verb+SExtractor+. The bright pixels determined on the filtered image  are compared to the RMS map to produce the mask according to the method presented above.

This exercise can be used to simulate the detectability of a given source in an image completely independent of the input source. This is useful, for example, in the case of a survey that consists of different and independent FoVs. In that situation, the differences in seeing condition have to be accounted for when measuring the bright pixel profile of the source. This can be achieved through convolution or deconvolution of the original image of the source. An example of how the seeing affects the determination of the bright pixel profiles is shown on Fig. \ref{fig:bright_pixels}.

\begin{figure}
  \centering
  \includegraphics[width=\hsize]{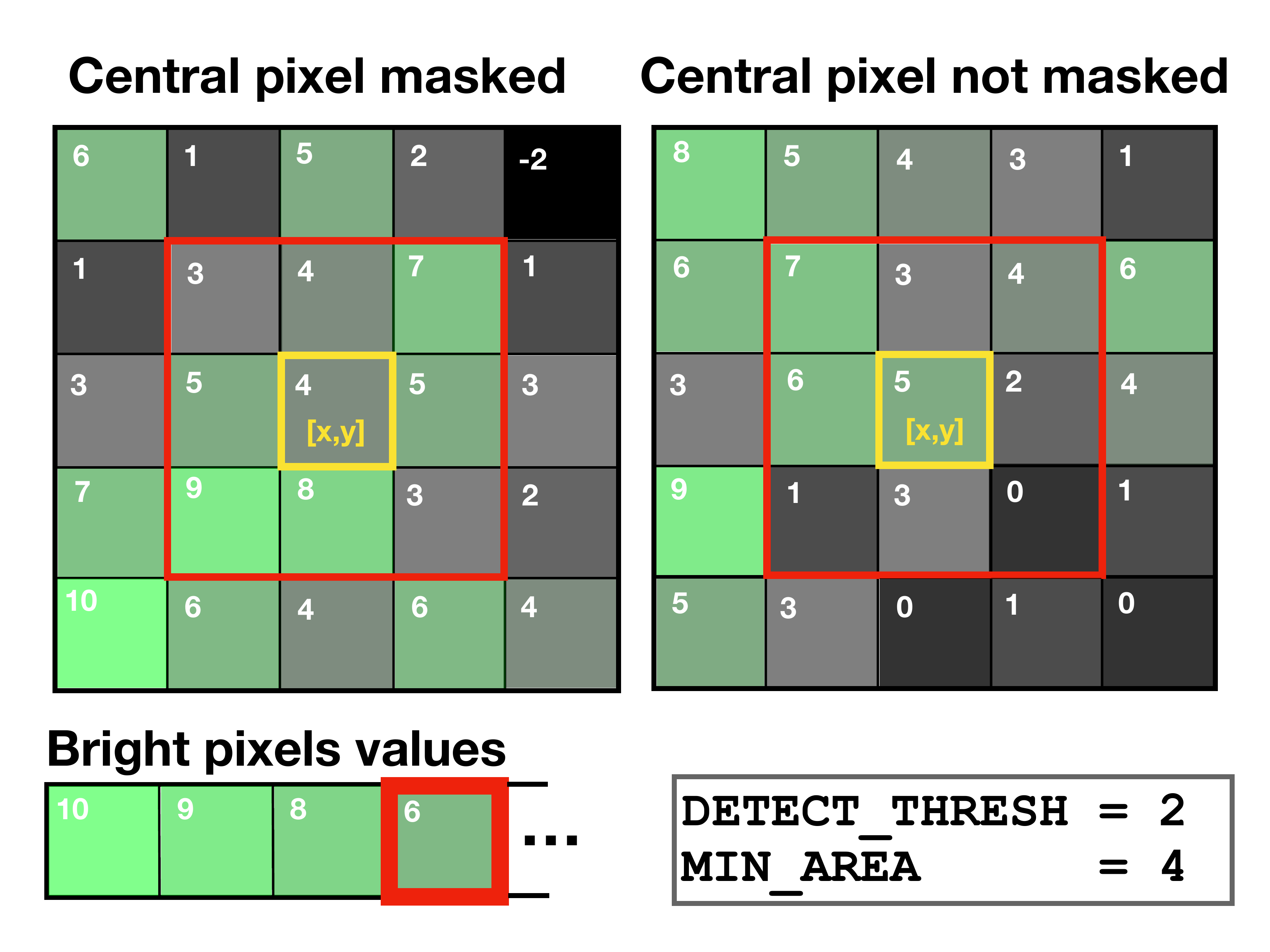}

  \cprotect\caption{Illustration of the criterion used to create the mask. The grid represents part of an RMS map. To determine whether the central pixel [x,y] is masked or not, the bright pixels values shown on the bottom left are used; in this example, only the \verb+MIN_AREA+-th pixel value $= 6$ is used to compare with the local noise. Considering the central pixel [x,y], the comparison to the local noise is only done for the 9 pixels adjacent pixels (i.e. red square). The values used for the detection threshold  and the minimal area in this example are 2 and 4, respectively. On the left, none of the pixels in the red area have values that are strictly less than \verb|min(Bp) / DETECT_THRESH| = 3, which results in the central pixel being masked. On the right panel, three pixels fulfil the condition and the central pixel is not masked.}

  \label{fig:pixel_criterion_schematic}
\end{figure}

\begin{figure*}
 \centering
 \includegraphics[width=\hsize/4]{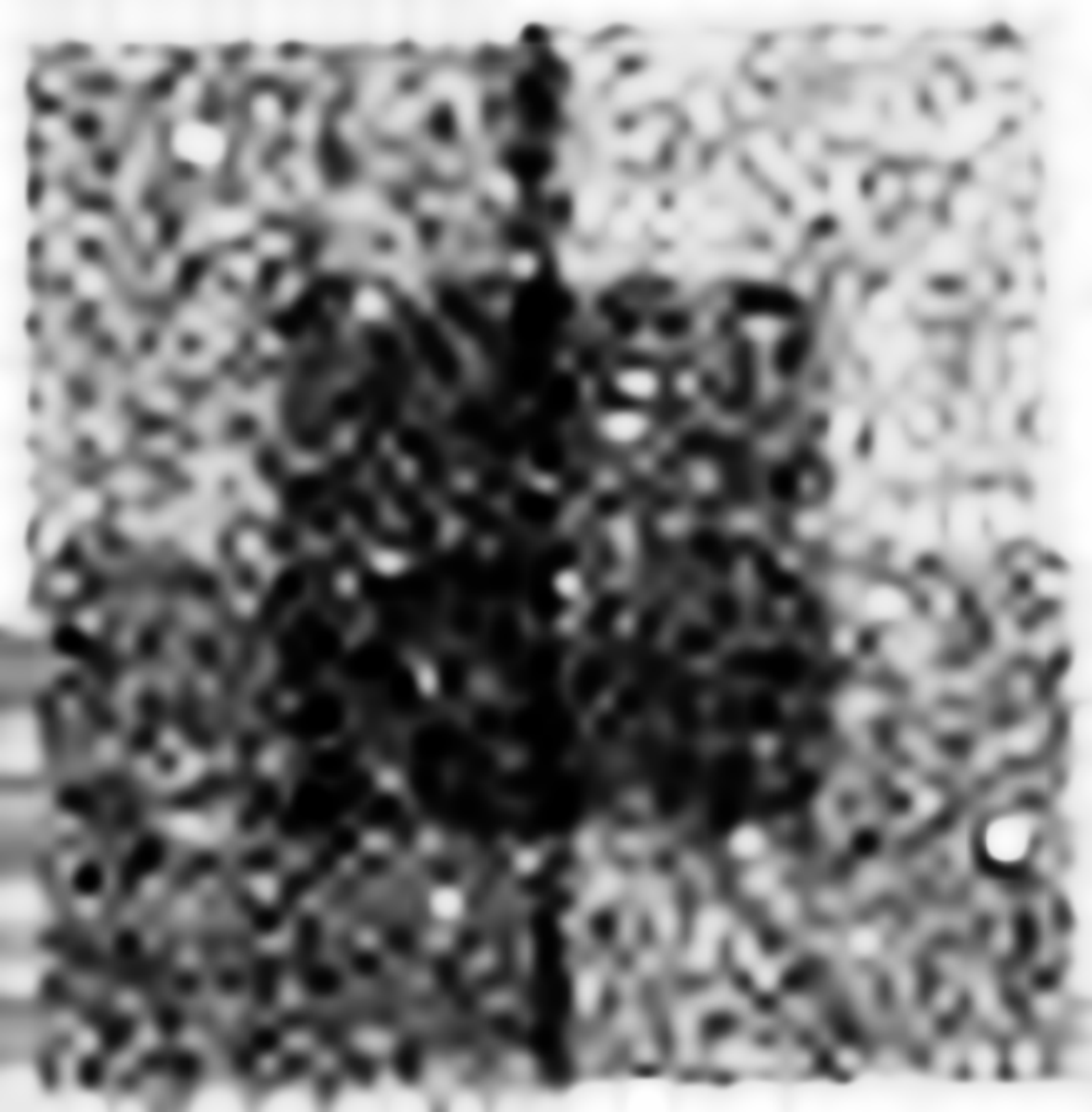}
 \includegraphics[width=\hsize/4]{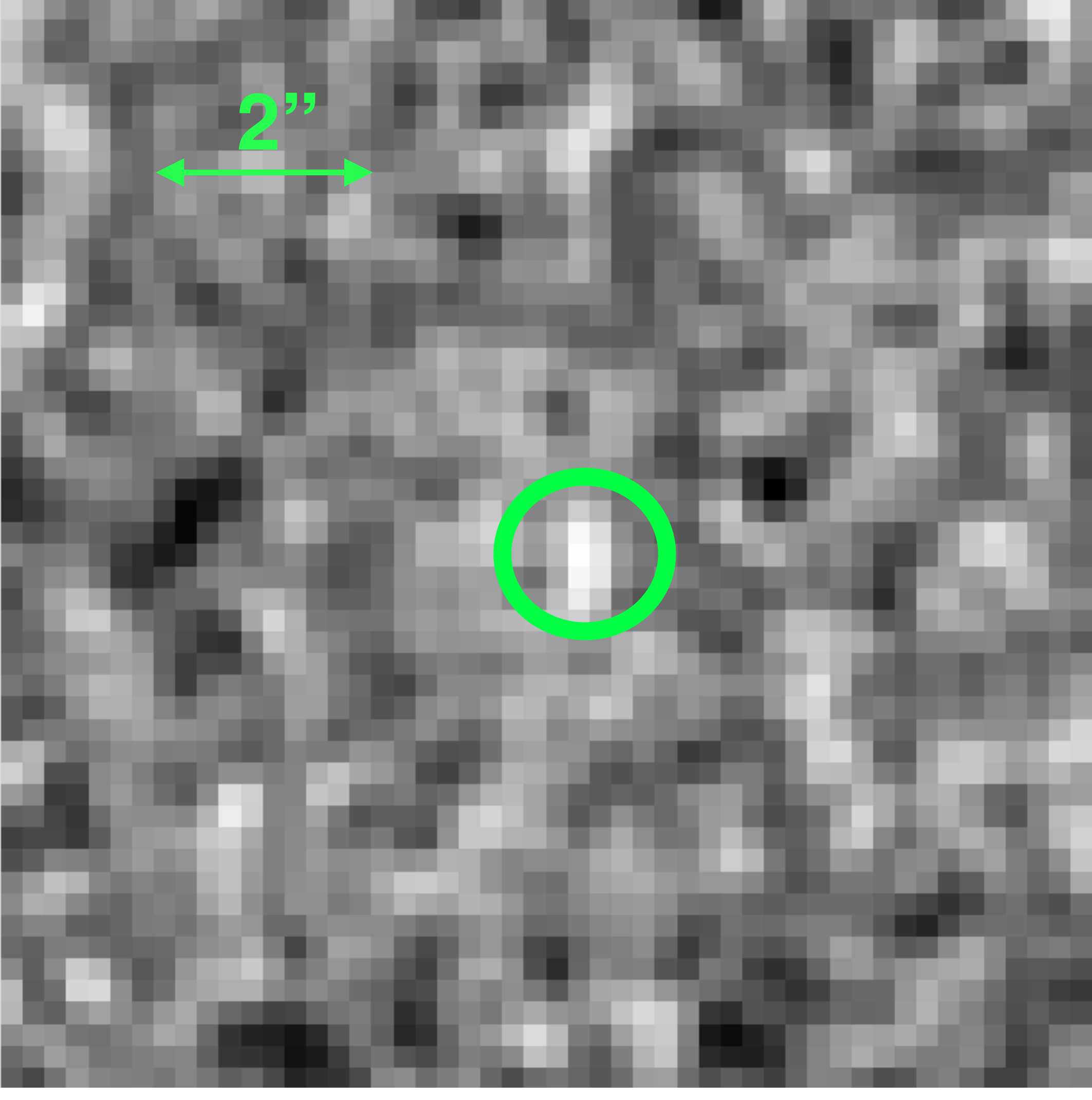}
 \includegraphics[width=\hsize/4]{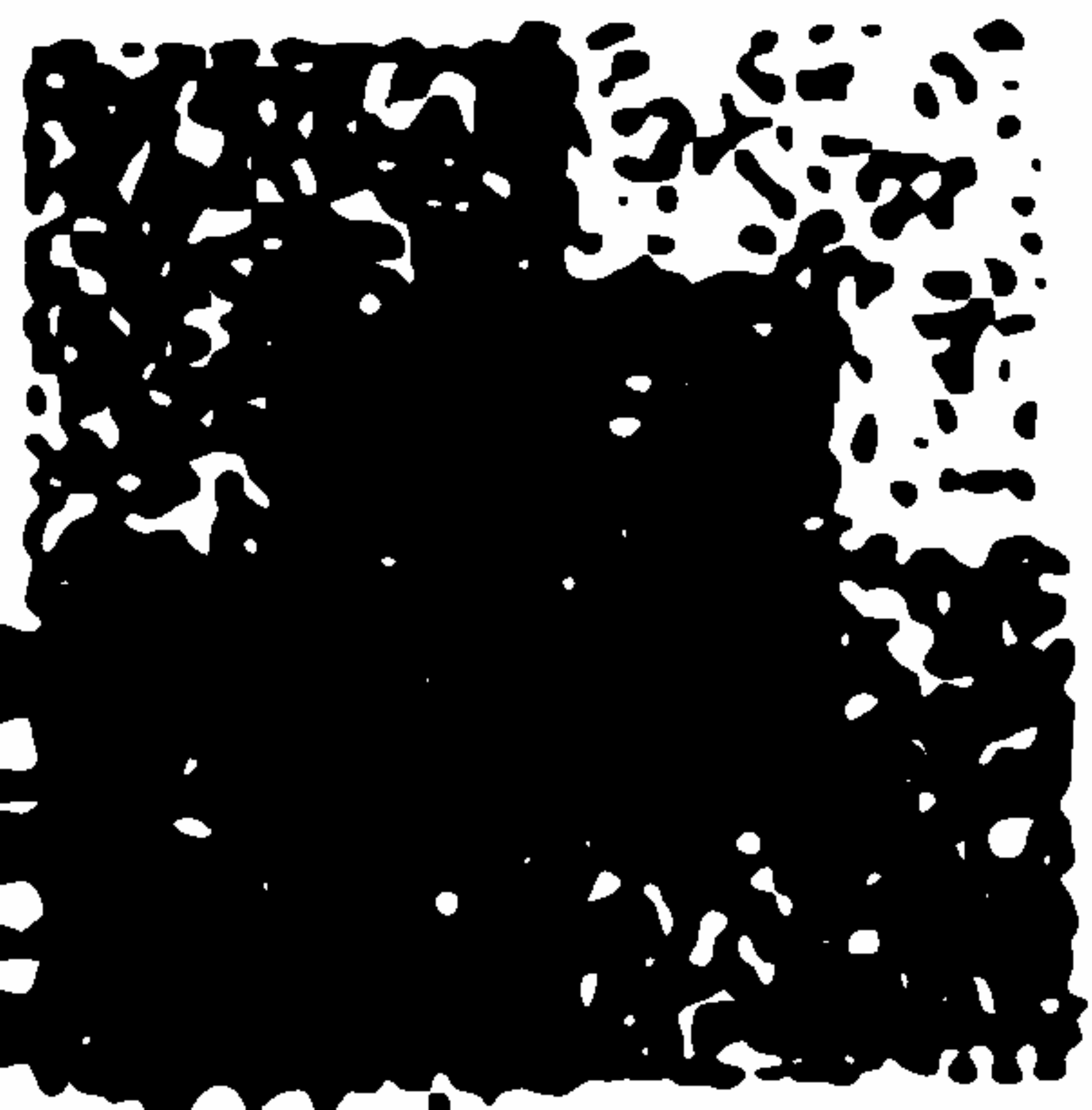}
 \cprotect\caption{Left panel: example of RMS maps produced from one slice of the A2744 cube. The large-scale patterns are due to the different exposure times for the different
   parts of the mosaic. In the deepest part of this field, the noise is reduced because of a longer integration time. Middle panel: filtered image centred on  one of the faint LAE in the A2744 field. The brightest pixels \verb+Bp+ were defined from this image. The size of the field is $\sim 10''$. Right panel: mask produced by this method for the source shown in the middle panel, the masked pixels are shown in white. We can see on this image that the mask patterns closely follow the RMS map.}
 \label{fig:2d_mask_example}
\end{figure*}

\section{Mask examples using median RMS maps}
\label{sec:masks_exemples}

In this section we illustrate the results found when applying the method presented in \ref{annex:create_mask_from_2d_image} to the different cubes, for LAEs detected with different S/N values. A sample of representative masks is presented on Fig. \ref{fig:mask_mosaic}. These masks were used for masking the 3D cubes during the volume computation. They were created with the method described in Sect. \ref{subsubsec:masking_3d_cubes}, including
 a median RMS map for each data cube and a median bright pixel profile to be rescaled in agreement with the actual S/N of the source.
The S/N values used to build the masks increase from left to right. We note that, in this case, this is not a real S/N but a proxy (see Sect. \ref{subsubsec:masking_3d_cubes} for details).

We see that at lower S/N values, the masks are efficient to retrieve the instrumental patterns.  At higher S/N values, these patterns disappear, and only the bright galaxies and the edge of the FoVs remain masked.
For A2744, we see that the masks are very efficient to account for the difference in exposure time in the mosaic. The central quadrant of the mosaic, being  the deepest,  is mostly not masked, whereas the upper right quadrant, being the shallowest, is only unmasked for the highest S/N values.

\begin{figure*}
 \centering
 \includegraphics[width=\hsize]{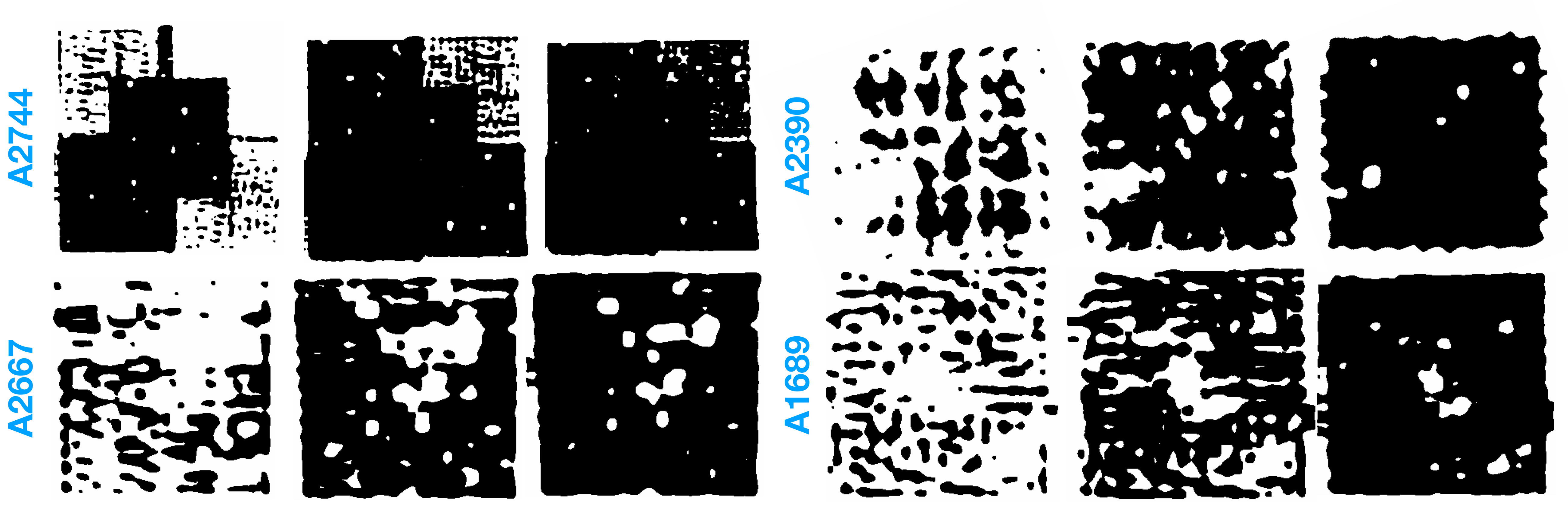}
 \caption{Representative examples of masks obtained in the different fields for different S/N values. The masked pixels are shown in white. For each field, the S/N values used to build the mask increase from left to right.}
 \label{fig:mask_mosaic}
\end{figure*}

\section{Comparison of the different volume computation methods}
\label{sec:volume_comparison}

In this section we compare the results obtained when computing the $V_{\rm max}$ using the method adopted in this study to the classical integration based on a unique mask.
We present in Fig. \ref{fig:volume_comparison} the comparison between the $V_{\rm max}$  values obtained from these two different methods. The first (on the y-axis) is used in this project, based on 3D masks, following the noise variation through the MUSE cubes. The second (on the x-axis) uses a mask generated from a unique \verb+SExtractor+ segmentation map, which is replicated across the spectral dimension. An example of such a mask is provided in Fig. \ref{fig:SE_mask}. It is mostly efficient to mask the brightest sources and haloes on the image. Comparing this mask to the masks presented in Fig. \ref{fig:mask_mosaic}, we see that they are completely different. Whereas the 3D masks adopted in this paper are able to follow the differences in exposure time while encoding the instrumental noise patterns, the simple masks provide a unique pattern for all sources, irrespective of their S/N values. This results in the following effects as seen in Fig. \ref{fig:volume_comparison}: 
First, a unique mask translates into a unique $V_{\rm max}$ value for a large number of sources, as only the lensing effects play a role in the determination of $V_{\rm max}$. This corresponds to the vertical pattern on the right-hand side of Fig. \ref{fig:volume_comparison}. Second, using the adaptive mask method, systematically lower $V_{\rm max}$ values are obtained. And more interestingly, for sources in A1689, A2390, and A2667, we see that the differences are less pronounced (or even not significant for some sources)  than for the sources in the A2744 mosaic.

To explain the first point, it is important to understand that when using a single mask, the only factor that could influence the $V_{\rm max}$ value is the limit magnification $\mu_{\rm lim}$ (see Sect. \ref{subsubsec:volume_integration}). A source with a higher $\mu_{\rm lim}$ value would end up with a smaller $V_{\rm max}$ as the area of the survey with large magnification is smaller. For the bright sources of the sample, it could be that the computed  $\mu_{\rm lim}$ would be under the lower magnification reached on the survey. For those sources, the volume was integrated on the entire survey area.
Using the 3D mask method, $\mu_{\rm lim}$ still plays a role but it is no longer the only factor affecting the final volume value and the local noise level is properly taken into account.

To explain the second point and to  illustrate the systematical difference between the two methods, we can consider a faint source detected in one of the  deepest parts of the A2744 mosaic. When comparing the source to the noise level in the rest of the mosaic, the quadrants  with the lower integration time end up being completely masked. As for the three other cubes, their contribution is zero as they have even less integration time. In that case, only a small portion of the mosaic has a significant contribution to the $V_{\rm max }$ value and it results in a low $V_{\rm max}$. However, all sources detected in A1689, A2390, or A2667 could have been detected anywhere in the A2744 mosaic. Because the A2744 FoV accounts for 80\% of the total volume, only $\mu_{\rm lim}$ affects the final contribution of A2744, and the contribution of the smaller fields is not that significant. This explains the correlation between the two methods for the sources detected in the three shallower fields.

\begin{figure}
 \centering
 \includegraphics[scale=0.5]{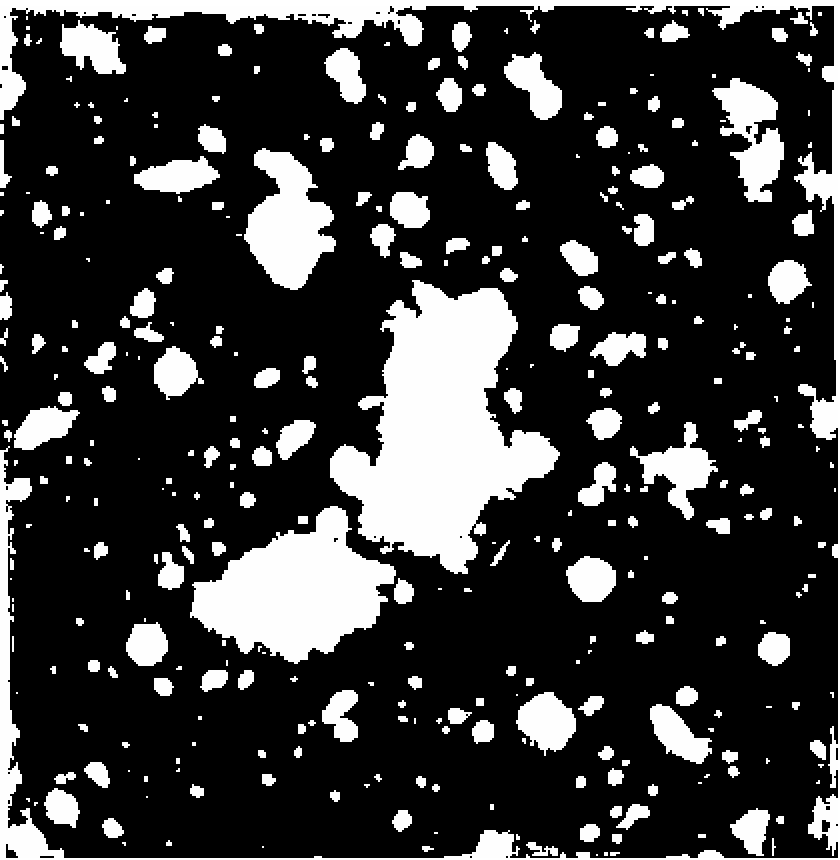}
 \cprotect\caption{Mask of the A2744 FoV, created from a MUSE white light image of the cluster using a \verb+SExtractor+ segmentation map. The masked pixels are shown in white. This type of mask is mostly efficient to mask the brightest sources and haloes.}
 \label{fig:SE_mask}
\end{figure}

\begin{figure}
 \centering
 \includegraphics[width=\hsize]{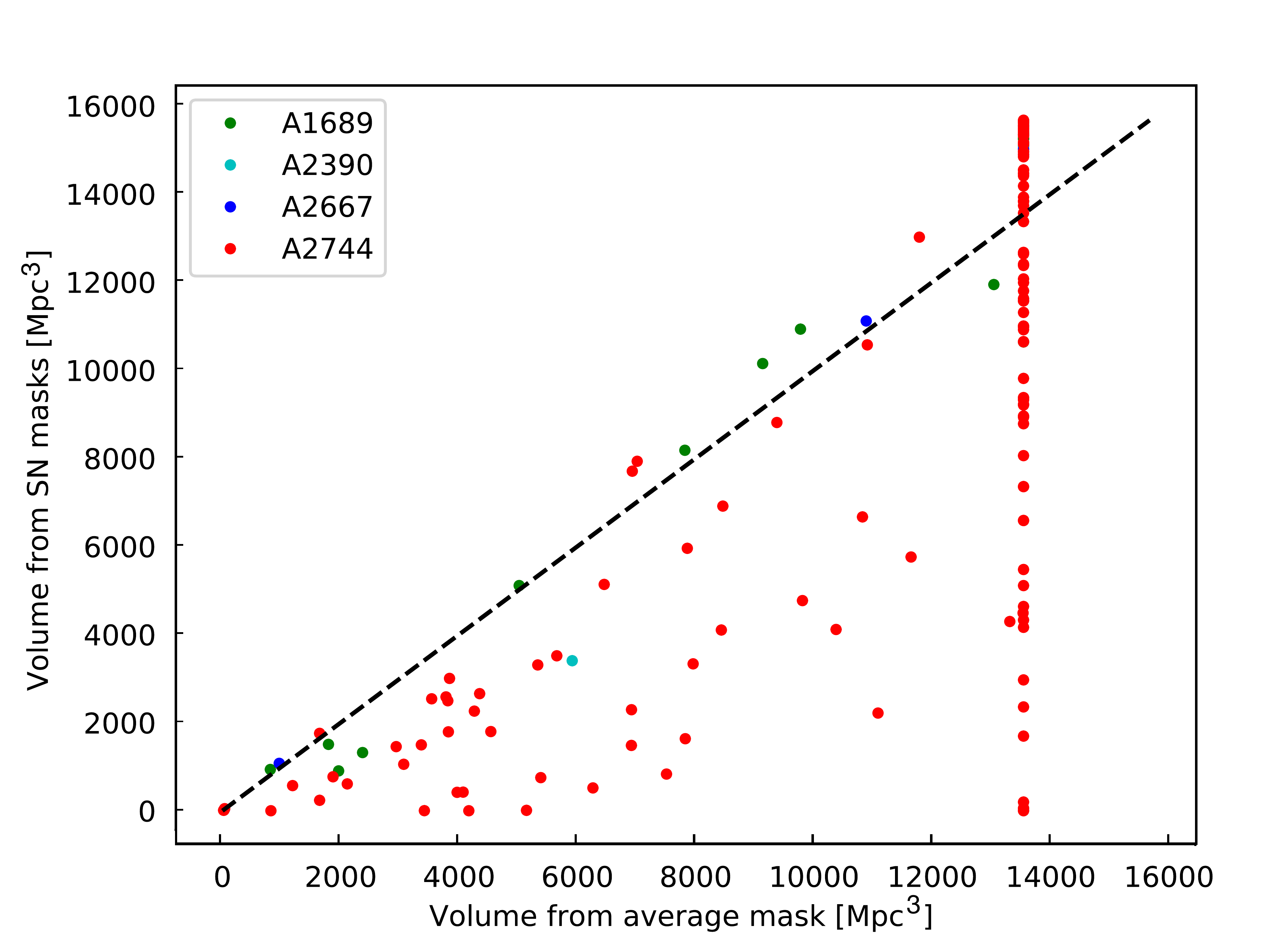}
 \cprotect\caption{Comparison of the results of $V_{\rm max}$ computation using the average mask obtained from a unique \verb+SExtractor+ segmentation map (x-axis) and the 3D masks adopted in this paper, following the evolution of noise through the MUSE cubes (y-axis). See text for details.}
 
 \label{fig:volume_comparison}
\end{figure}

\section{Detailed procedure for volume computation in lensed MUSE cubes}
\label{sec:detailed_volume_schematic}

In this appendix, we provide an overview and a quick description of all the steps needed to compute $V_{\rm max}$. The details are explained in the main text. The goal of this section is to provide a synthetic view to explain the method. The numbers on the notes below refer to the steps listed in Fig. \ref{fig:flow_chart_volume} as follows:

\begin{itemize}
\item[-] (0) The NB cubes consist of all the NB images produced by \verb+Muselet+. All LAEs were detected on those NB images. Details on those NB images are provided in Sect. \ref{subsec:source_detection}  
\item[-] (1.1) Background RMS maps produced separately by \verb+SExtractor+ and assembled into a RMS cube. The RMS cube are cubes of noise that are used to track the spectral evolution of noise levels in cubes.
\item[-] (1.2) Median of the RMS cubes along the spectral axis. One median RMS image is obtained per cube. They are used to mock the 2D \verb+SExtractor+ detection process.
\item[-] (1.3) Set of S/N values designed to encompass all possible values in the LAE sample. The definition used for S/N is provided in Eq. \ref{eq:sn_definition}.
\item[-] (1.4) Using a generalized bright-pixels profile (see Fig. \ref{fig:bright_pixels}) and the median RMS maps, a 2D detection mask is built for each value of the S/N set and for each cube; the method is described in appendix  \ref{annex:create_mask_from_2d_image}.
\item[-] (1.5) Redshift values used to sample the evolution of the source plan projections and magnification maps.
\item[-] (1.6) Source plan projection of the set of 2D masks combined with magnification maps for different redshift.
\item[-] (1.7) For each LAE, the final 3D survey masks are assembled from the set of source plane projections. The procedure browse the S/N curves (see Fig. \ref{fig:sn_evolution}, and picks the pre-computed 2D source plane projection computed from the correct S/N value and the appropriate redshift value. Details on this can be found in Sect. \ref{subsubsec:masking_3d_cubes} and Sect. \ref{subsubsec:volume_integration}).

\item[-] (1.8) Minimal magnification to allow the detection of a given LAE in its parent cube. This first value is computed from the error on the flux detection, which is indicative of the local noise level. See definition in Eq. \ref{eq:mu_lim_definition}.
\item[-] (1.9) A rescaled limit magnification (see definition in Eq. \ref{eq:mu_lim_rescaled}) is computed for each LAE and for the three additional cubes. This is done to account for the differences in both seeing and exposure time. All the details about limiting magnification are explained in \ref{subsubsec:volume_integration}. For each LAE, the four $\mu_{\rm lim}$ values are used to restrict the volume computation to the areas of the source plan projection with a magnification high enough to allow the detection of this LAE.
\item[-] (1.10) Volume of the survey where a given source could have been detected. For one LAE, this volume is computed from the source plane projected 3D masks, on the pixels with a high enough magnification.  
\item[-] (2.1) For each LAE, the NB containing the max of its Lyman-alpha emission is selected. The cleanest detection was obtained on this slice of the NB cube.
\item[-] (2.2) Filtered map produced  with \verb+SExtractor+. See Appendix \ref{annex:create_mask_from_2d_image} for details.

\item[-] (2.3) From the original filtered map produced for each LAE in the parent cube, three additional images are produced to the resolution of the additional cubes the LAE does not belong to using convolution or deconvolution.
\item[-] (2.4) Individual bright-pixel profiles are retrieved for the four different seeing conditions from the filtered images and the three additional images produced in the previous step. The bright-pixel profiles contain the information related to the spatial profile of the LAEs.
\item[-] (2.5) The four generalized bright-pixel profiles are the median of the  individual bright-pixel profiles computed for each seeing condition (see Fig. \ref{fig:bright_pixels}). These generalized profiles are used to limit the number of mask computed and simplify the production of 3D masks.
\item[-] (3.1) The noise level in cubes is an average measure of noise in a given slice of a cube. It is defined in Eq. \ref{eq:sn_definition} and an example is provided in Fig. \ref{fig:evolution_of_RMS_level}.
\item[-] (3.2) Combining the definition of noise levels and the individual bright-pixels profiles, the evolution of S/N for individual sources is computed through the cubes with Eq. \ref{eq:sn_computation} (see Sect. \ref{subsubsec:masking_3d_cubes} and Fig. \ref{fig:sn_evolution}).

\end{itemize}

\endgroup

\end{appendix}

\setcounter{table}{3}
\onecolumn
\renewcommand{\arraystretch}{1.3} 
 \begin{longtable}{l c c c c c c c c}                              
\caption{\label{tab:sample_tab}Table with the main characteristics of the 152 LAEs used to build the LFs.\ The value $F_{Ly_{\alpha}}$ is the detection flux of the LAE, expressed in $10^{-18}$ units, $\mu$ is the flux weighted magnification of the source, and the error bars correspond to the 68\% asymmetric errors computed from $P{\mu}$,$\log{Ly_{\alpha}}$ is the Lyman-alpha luminosity corrected for magnification. No error bars are associated with the luminosity value, as this uncertainty is accounted for during the MC iterations needed to build the LFs. `Comp' is the completeness expressed in percentage. The $V_{\rm max}$ value given in this table are computed for $2.9 < z < 6.9$ } \\ 

\hline \hline                         
\multicolumn{1}{l}{Id} & \multicolumn{1}{c}{$z$} & 
 \multicolumn{1}{c}{$F_{Ly_{\alpha}}$} & 
 \multicolumn{1}{c}{$\mu $} & 
 \multicolumn{1}{c}{$\log(Ly_{\alpha})$} & 
\multicolumn{1}{c}{Comp} & 
\multicolumn{1}{c}{$V_{\rm max}$} & 
\multicolumn{1}{c}{Ra} & \multicolumn{1}{c}{Dec} \\ 
  &  & 
 erg s$^{-1}$ cm$^2$ & 
 \multicolumn{1}{c}{ } & 
 \multicolumn{1}{c}{erg s$^{-1}$} & 
 & Mpc$^3$ & 
\multicolumn{1}{c}{$^\circ $} & \multicolumn{1}{c}{$^\circ $} \\ 
\hline    
\endfirsthead 

 \multicolumn{9}{c}%
{{\bfseries \tablename\ \thetable{} -- continued from previous page}}\\ 
\hline 
 \multicolumn{1}{l}{Id} & 
\multicolumn{1}{c}{z} & 
\multicolumn{1}{c}{$F_{Ly_{\alpha}} $} & 
\multicolumn{1}{c}{$\mu $ } & 
\multicolumn{1}{c}{$\log(Ly_{\alpha})$} & 
\multicolumn{1}{c}{Comp} & 
\multicolumn{1}{c}{$V_{\rm max}$} & 
\multicolumn{1}{c}{Ra} & 
 \multicolumn{1}{c}{Dec} \\ \hline 
\endhead 

\hline \multicolumn{9}{c}{{Continued on next page}} \\ \hline 
\endfoot 

\hline \hline 
 \endlastfoot 

A1689, 619 & $3.0446$ & $102.06\pm 6.27$ & $7.95_{- 0.25}^{+ 0.60}$ & $42.01$ & $73.3\pm1.7$ & 16015.9 & $197.874204$ & $-1.351669$\\  
A1689, 1028 & $3.1109$ & $119.36\pm 3.36$ & $26.83_{- 0.90}^{+ 2.80}$ & $41.58$ & $100.0\pm0.0$ & 15913.4 & $197.881592$ & $-1.344253$\\  
A1689, LN9 & $3.1789$ & $44.72\pm 3.75$ & $7.69_{- 0.52}^{+ 0.55}$ & $41.71$ & $96.4\pm0.7$ & 15946.6 & $197.875790$ & $-1.349321$\\  
A1689, 1404 & $3.1800$ & $11.99\pm 1.84$ & $5.90_{- 0.38}^{+ 0.22}$ & $41.26$ & $12.6\pm1.5$ & 15791.5 & $197.879760$ & $-1.336681$\\  
A1689, 835 & $3.1806$ & $27.48\pm 2.48$ & $11.84_{- 1.23}^{+ 0.66}$ & $41.31$ & $93.2\pm1.0$ & 15835.8 & $197.878000$ & $-1.348089$\\  
A1689, LN10 & $3.4182$ & $16.84\pm 1.36$ & $52.42_{- 10.64}^{+ 44.51}$ & $40.53$ & $99.4\pm0.3$ & 15698.1 & $197.870362$ & $-1.347675$\\  
A1689, LN26 & $4.0541$ & $9.44\pm 1.29$ & $8.51_{- 0.40}^{+ 0.54}$ & $41.25$ & $62.2\pm2.0$ & 15805.0 & $197.870413$ & $-1.352380$\\  
A1689, LN13 & $4.0548$ & $24.66\pm 1.82$ & $8.82_{- 0.66}^{+ 0.69}$ & $41.65$ & $98.1\pm0.6$ & 15943.8 & $197.871113$ & $-1.349303$\\  
A1689, LN14 & $4.1038$ & $19.34\pm 2.37$ & $5.66_{- 0.21}^{+ 0.35}$ & $41.75$ & $98.9\pm0.5$ & 15930.8 & $197.879200$ & $-1.337292$\\  
A1689, LN25 & $4.8426$ & $4.12\pm 0.66$ & $18.74_{- 1.65}^{+ 2.84}$ & $40.73$ & $38.9\pm1.9$ & 15509.8 & $197.869410$ & $-1.348497$\\  
A1689, LN15 & $4.8668$ & $5.75\pm 0.92$ & $4.92_{- 0.32}^{+ 0.38}$ & $41.46$ & $68.9\pm1.8$ & 15851.1 & $197.876460$ & $-1.352164$\\  
A1689, 1379 & $4.8734$ & $91.53\pm 2.22$ & $5.68_{- 0.18}^{+ 0.38}$ & $42.60$ & $99.9\pm0.2$ & 16352.6 & $197.877970$ & $-1.336814$\\  
A1689, LN17 & $5.0117$ & $4.46\pm 0.56$ & $8.28_{- 0.45}^{+ 0.46}$ & $41.15$ & $84.5\pm1.4$ & 15818.2 & $197.870830$ & $-1.352020$\\  
A1689, LN18 & $5.7369$ & $6.16\pm 0.83$ & $18.22_{- 1.22}^{+ 1.44}$ & $41.08$ & $50.1\pm2.0$ & 15711.5 & $197.880900$ & $-1.345920$\\  
A1689, LN19 & $6.1752$ & $6.98\pm 1.00$ & $7.49_{- 0.56}^{+ 0.24}$ & $41.60$ & $97.8\pm0.7$ & 15835.2 & $197.876070$ & $-1.350196$\\  
A2390, L1 & $4.0454$ & $207.18\pm 6.97$ & $19.81_{- 0.53}^{+ 1.22}$ & $42.22$ & $97.6\pm0.9$ & 15832.3 & $328.390790$ & $17.701650$\\  
A2390, 96 & $4.0475$ & $544.64\pm 6.51$ & $11.22_{- 0.33}^{+ 0.55}$ & $42.89$ & $99.2\pm0.8$ & 16246.7 & $328.396350$ & $17.692954$\\  
A2390, 134 & $4.7210$ & $16.75\pm 1.74$ & $24.27_{- 0.32}^{+ 3.28}$ & $41.20$ & $30.7\pm2.5$ & 15010.8 & $328.391020$ & $17.697558$\\  
A2390, 71 & $4.8773$ & $20.70\pm 1.97$ & $7.12_{- 0.24}^{+ 0.25}$ & $41.85$ & $99.4\pm0.3$ & 15810.7 & $328.400050$ & $17.689222$\\  
A2390, 243 & $5.7574$ & $2.69\pm 0.57$ & $21.33_{- 0.74}^{+ 1.26}$ & $40.66$ & $34.4\pm2.5$ & 13282.3 & $328.405510$ & $17.698954$\\  
A2667, 24 & $3.7872$ & $16.54\pm 1.52$ & $9.32_{- 0.34}^{+ 1.16}$ & $41.38$ & $99.2\pm0.4$ & 15732.7 & $357.917309$ & $-26.082718$\\  
A2667, 25 & $3.7872$ & $36.51\pm 2.85$ & $2.96_{- 0.06}^{+ 0.08}$ & $42.22$ & $89.4\pm1.3$ & 15869.4 & $357.906046$ & $-26.078152$\\  
A2667, 30 & $3.9743$ & $59.56\pm 3.40$ & $46.08_{- 6.34}^{+ 24.71}$ & $41.29$ & $94.2\pm0.9$ & 14522.8 & $357.920596$ & $-26.079189$\\  
A2667, 33 & $4.0803$ & $39.13\pm 3.63$ & $12.50_{- 0.49}^{+ 0.88}$ & $41.70$ & $96.1\pm0.8$ & 15696.9 & $357.910908$ & $-26.080737$\\  
A2667, 38 & $4.9467$ & $30.77\pm 3.07$ & $16.22_{- 1.04}^{+ 2.42}$ & $41.68$ & $85.2\pm1.5$ & 15368.8 & $357.919470$ & $-26.082619$\\  
A2667, 41 & $5.1993$ & $18.18\pm 1.30$ & $3.25_{- 0.07}^{+ 0.10}$ & $42.20$ & $99.9\pm0.1$ & 15939.4 & $357.906303$ & $-26.078569$\\  
A2667, 62 & $5.5003$ & $6.52\pm 1.16$ & $43.08_{- 4.85}^{+ 10.58}$ & $40.69$ & $88.1\pm1.4$ & 2002.1 & $357.906020$ & $-26.091870$\\  
A2744, 8683 & $2.9315$ & $25.86\pm 2.33$ & $3.22_{- 0.08}^{+ 0.12}$ & $41.77$ & $96.4\pm0.8$ & 15527.9 & $3.572765$ & $-30.394612$\\  
A2744, 11626 & $2.9422$ & $4.59\pm 0.93$ & $1.75_{- 0.03}^{+ 0.06}$ & $41.29$ & $68.5\pm1.7$ & 13744.9 & $3.606868$ & $-30.385573$\\  
A2744, 5005 & $2.9513$ & $9.71\pm 0.87$ & $18.10_{- 0.82}^{+ 1.63}$ & $40.60$ & $98.9\pm0.5$ & 11423.4 & $3.595135$ & $-30.404478$\\  
A2744, 4010 & $2.9986$ & $4.15\pm 1.34$ & $2.17_{- 0.04}^{+ 0.04}$ & $41.17$ & $21.9\pm1.7$ & 12801.1 & $3.575187$ & $-30.407353$\\  
A2744, 10544 & $3.0211$ & $2.41\pm 0.46$ & $2.95_{- 0.06}^{+ 0.10}$ & $40.81$ & $68.6\pm1.9$ & 13832.2 & $3.592539$ & $-30.387649$\\  
A2744, M10 & $3.0213$ & $2.06\pm 0.53$ & $2.11_{- 0.05}^{+ 0.04}$ & $40.88$ & $21.1\pm2.0$ & 12606.1 & $3.568189$ & $-30.400041$\\  
A2744, M11 & $3.0234$ & $1.34\pm 0.36$ & $3.48_{- 0.12}^{+ 0.07}$ & $40.48$ & $26.9\pm2.1$ & 13373.2 & $3.581978$ & $-30.408336$\\  
A2744, M12 & $3.0337$ & $4.00\pm 0.91$ & $2.34_{- 0.04}^{+ 0.05}$ & $41.13$ & $11.6\pm1.5$ & 12826.9 & $3.573038$ & $-30.401722$\\  
A2744, 3424 & $3.0511$ & $7.76\pm 1.00$ & $9.70_{- 0.55}^{+ 0.41}$ & $40.81$ & $95.4\pm0.9$ & 14816.9 & $3.593917$ & $-30.409719$\\  
A2744, M24 & $3.0532$ & $14.55\pm 1.16$ & $12.90_{- 0.73}^{+ 0.92}$ & $40.96$ & $99.8\pm0.2$ & 15480.4 & $3.590349$ & $-30.410597$\\  
A2744, 11701 & $3.0543$ & $18.54\pm 1.44$ & $4.80_{- 0.12}^{+ 0.12}$ & $41.49$ & $98.4\pm0.5$ & 15555.7 & $3.585514$ & $-30.385878$\\  
A2744, 7858 & $3.1291$ & $82.08\pm 4.02$ & $3.47_{- 0.08}^{+ 0.11}$ & $42.31$ & $100.0\pm0.1$ & 15869.5 & $3.574989$ & $-30.396797$\\  
A2744, 7721 & $3.1295$ & $138.50\pm 5.81$ & $2.78_{- 0.05}^{+ 0.10}$ & $42.63$ & $100.0\pm0.0$ & 15962.4 & $3.571429$ & $-30.396950$\\  
A2744, 11196 & $3.1508$ & $6.72\pm 1.55$ & $3.31_{- 0.09}^{+ 0.12}$ & $41.25$ & $53.1\pm2.2$ & 13573.4 & $3.578329$ & $-30.383213$\\  
A2744, 6876 & $3.1900$ & $1.68\pm 0.32$ & $2.21_{- 0.06}^{+ 0.05}$ & $40.83$ & $64.0\pm2.3$ & 13791.6 & $3.568627$ & $-30.399395$\\  
A2744, M13 & $3.2034$ & $1.98\pm 0.40$ & $4.06_{- 0.12}^{+ 0.08}$ & $40.64$ & $56.8\pm2.2$ & 12840.3 & $3.587266$ & $-30.385496$\\  
A2744, M14 & $3.2034$ & $1.32\pm 0.26$ & $2.32_{- 0.04}^{+ 0.04}$ & $40.71$ & $10.3\pm1.6$ & 10860.6 & $3.603810$ & $-30.400797$\\  
A2744, 2754 & $3.2075$ & $6.29\pm 1.08$ & $8.53_{- 0.48}^{+ 0.47}$ & $40.83$ & $65.3\pm2.2$ & 11925.8 & $3.589229$ & $-30.411825$\\  
A2744, 11806 & $3.2356$ & $3.92\pm 0.68$ & $1.97_{- 0.05}^{+ 0.06}$ & $41.27$ & $47.8\pm2.2$ & 12576.7 & $3.600328$ & $-30.386748$\\  
A2744, 4933 & $3.2466$ & $21.69\pm 1.54$ & $2.46_{- 0.05}^{+ 0.05}$ & $41.92$ & $99.8\pm0.2$ & 15817.9 & $3.604574$ & $-30.404791$\\  
A2744, 3000 & $3.3161$ & $17.87\pm 1.94$ & $1.68_{- 0.03}^{+ 0.02}$ & $42.02$ & $98.8\pm0.5$ & 15504.2 & $3.568377$ & $-30.410915$\\  
A2744, 3759 & $3.3576$ & $2.64\pm 0.36$ & $1.72_{- 0.04}^{+ 0.03}$ & $41.19$ & $84.5\pm1.6$ & 14193.9 & $3.566861$ & $-30.408027$\\  
A2744, 11033 & $3.3788$ & $25.15\pm 1.74$ & $2.64_{- 0.09}^{+ 0.10}$ & $41.99$ & $98.9\pm0.5$ & 15586.4 & $3.593887$ & $-30.383222$\\  
A2744, M7 & $3.4072$ & $31.08\pm 1.14$ & $41.81_{- 3.16}^{+ 53.90}$ & $40.89$ & $100.0\pm0.0$ & 12532.9 & $3.581197$ & $-30.398708$\\  
A2744, M15 & $3.4337$ & $0.62\pm 0.25$ & $1.91_{- 0.06}^{+ 0.07}$ & $40.55$ & $51.1\pm2.1$ & 12429.8 & $3.601463$ & $-30.384161$\\  
A2744, 10382 & $3.4750$ & $8.59\pm 0.49$ & $1.66_{- 0.04}^{+ 0.03}$ & $41.76$ & $100.0\pm0.1$ & 15992.4 & $3.607435$ & $-30.388489$\\  
A2744, 10669 & $3.4757$ & $59.29\pm 2.66$ & $1.90_{- 0.05}^{+ 0.04}$ & $42.54$ & $99.8\pm0.2$ & 15977.6 & $3.601542$ & $-30.387391$\\  
A2744, 9272 & $3.4758$ & $6.50\pm 1.08$ & $1.78_{- 0.02}^{+ 0.04}$ & $41.60$ & $28.1\pm1.8$ & 11796.6 & $3.604649$ & $-30.392232$\\  
A2744, 10725 & $3.4759$ & $6.67\pm 1.08$ & $2.40_{- 0.08}^{+ 0.06}$ & $41.48$ & $66.9\pm2.2$ & 13892.6 & $3.596085$ & $-30.387112$\\  
A2744, 3853 & $3.5415$ & $24.46\pm 1.37$ & $2.97_{- 0.06}^{+ 0.08}$ & $41.98$ & $100.0\pm0.0$ & 15864.7 & $3.604132$ & $-30.407705$\\  
A2744, M16 & $3.5509$ & $3.26\pm 0.61$ & $4.17_{- 0.08}^{+ 0.18}$ & $40.96$ & $55.7\pm2.1$ & 12487.1 & $3.576297$ & $-30.398988$\\  
A2744, 9731 & $3.5510$ & $4.38\pm 0.69$ & $13.85_{- 0.77}^{+ 0.78}$ & $40.56$ & $23.5\pm1.7$ & 4748.5 & $3.588768$ & $-30.390806$\\  
A2744, 5133 & $3.5733$ & $75.75\pm 1.70$ & $9.53_{- 0.78}^{+ 2.40}$ & $41.97$ & $100.0\pm0.1$ & 15822.0 & $3.593486$ & $-30.405044$\\  
A2744, M17 & $3.5756$ & $1.61\pm 0.24$ & $2.41_{- 0.07}^{+ 0.07}$ & $40.90$ & $61.3\pm2.0$ & 13434.5 & $3.595453$ & $-30.386282$\\  
A2744, 10174 & $3.5777$ & $7.84\pm 0.90$ & $5.95_{- 0.15}^{+ 0.13}$ & $41.19$ & $98.2\pm0.6$ & 15075.2 & $3.581085$ & $-30.389094$\\  
A2744, 3423 & $3.5810$ & $23.24\pm 1.82$ & $1.73_{- 0.03}^{+ 0.03}$ & $42.20$ & $86.1\pm1.3$ & 13721.1 & $3.569202$ & $-30.409686$\\  
A2744, 5922 & $3.5931$ & $1.28\pm 0.25$ & $2.13_{- 0.04}^{+ 0.05}$ & $40.85$ & $35.5\pm1.9$ & 11881.1 & $3.570137$ & $-30.401841$\\  
A2744, 9672 & $3.6490$ & $10.42\pm 1.15$ & $1.92_{- 0.04}^{+ 0.04}$ & $41.83$ & $99.4\pm0.4$ & 15536.7 & $3.602504$ & $-30.390868$\\  
A2744, 7737 & $3.6893$ & $25.04\pm 1.68$ & $2.28_{- 0.04}^{+ 0.03}$ & $42.14$ & $100.0\pm0.0$ & 15879.7 & $3.600478$ & $-30.396647$\\  
A2744, 6374 & $3.6913$ & $12.93\pm 0.74$ & $4.10_{- 0.09}^{+ 0.21}$ & $41.60$ & $100.0\pm0.1$ & 15768.8 & $3.597313$ & $-30.400608$\\  
A2744, 2951 & $3.7077$ & $11.74\pm 1.28$ & $1.69_{- 0.03}^{+ 0.02}$ & $41.95$ & $97.6\pm0.6$ & 15061.7 & $3.568234$ & $-30.410972$\\  
A2744, 5625 & $3.7077$ & $5.56\pm 0.60$ & $3.14_{- 0.06}^{+ 0.11}$ & $41.36$ & $97.3\pm0.7$ & 14886.1 & $3.600920$ & $-30.402937$\\  
A2744, M18 & $3.7247$ & $5.17\pm 0.84$ & $1.95_{- 0.03}^{+ 0.03}$ & $41.54$ & $93.0\pm1.2$ & 14470.4 & $3.575449$ & $-30.411075$\\  
A2744, 5624 & $3.7794$ & $64.92\pm 3.14$ & $2.30_{- 0.04}^{+ 0.05}$ & $42.58$ & $100.0\pm0.1$ & 15950.2 & $3.573255$ & $-30.402976$\\  
A2744, 10312 & $3.7866$ & $53.38\pm 2.77$ & $3.96_{- 0.20}^{+ 0.22}$ & $42.26$ & $98.7\pm0.5$ & 14970.5 & $3.570325$ & $-30.388589$\\  
A2744, 2956 & $3.8123$ & $26.26\pm 1.96$ & $2.26_{- 0.05}^{+ 0.03}$ & $42.20$ & $99.5\pm0.3$ & 15517.9 & $3.578298$ & $-30.411327$\\  
A2744, M19 & $3.8790$ & $2.01\pm 0.42$ & $2.01_{- 0.03}^{+ 0.04}$ & $41.16$ & $30.9\pm2.0$ & 9352.5 & $3.575143$ & $-30.409691$\\  
A2744, 8357 & $3.9469$ & $1.81\pm 0.35$ & $1.84_{- 0.03}^{+ 0.03}$ & $41.17$ & $72.1\pm2.2$ & 12209.1 & $3.604823$ & $-30.394963$\\  
A2744, 2104 & $3.9538$ & $3.08\pm 0.30$ & $2.68_{- 0.02}^{+ 0.06}$ & $41.24$ & $85.1\pm1.3$ & 13596.4 & $3.603180$ & $-30.415709$\\  
A2744, 14684 & $3.9619$ & $10.29\pm 1.01$ & $3.21_{- 0.14}^{+ 0.09}$ & $41.68$ & $98.4\pm0.6$ & 15003.1 & $3.577329$ & $-30.381897$\\  
A2744, 3210 & $3.9660$ & $2.16\pm 0.91$ & $1.84_{- 0.02}^{+ 0.04}$ & $41.25$ & $51.1\pm2.0$ & 11563.0 & $3.571654$ & $-30.410013$\\  
A2744, 3986 & $3.9833$ & $3.19\pm 0.58$ & $1.77_{- 0.03}^{+ 0.03}$ & $41.44$ & $22.2\pm1.9$ & 10007.8 & $3.567768$ & $-30.407314$\\  
A2744, 2736 & $4.0207$ & $35.25\pm 1.66$ & $5.99_{- 0.18}^{+ 0.19}$ & $41.96$ & $100.0\pm0.0$ & 15787.9 & $3.600544$ & $-30.412202$\\  
A2744, 2407 & $4.0208$ & $6.50\pm 0.82$ & $2.66_{- 0.04}^{+ 0.10}$ & $41.58$ & $80.9\pm1.7$ & 13299.4 & $3.582264$ & $-30.413744$\\  
A2744, 9303 & $4.0214$ & $10.73\pm 1.16$ & $9.76_{- 0.33}^{+ 0.46}$ & $41.23$ & $36.9\pm2.1$ & 9066.9 & $3.590175$ & $-30.392180$\\  
A2744, 9440 & $4.0214$ & $8.44\pm 1.13$ & $52.96_{- 3.18}^{+ 16.29}$ & $40.40$ & $74.2\pm1.9$ & 486.0 & $3.583412$ & $-30.392082$\\  
A2744, M41 & $4.0214$ & $2.31\pm 0.44$ & $3.41_{- 0.08}^{+ 0.09}$ & $41.02$ & $13.3\pm1.5$ & 10475.7 & $3.576430$ & $-30.400185$\\  
A2744, 6510 & $4.0253$ & $16.92\pm 1.45$ & $2.15_{- 0.05}^{+ 0.05}$ & $42.09$ & $94.6\pm1.0$ & 14159.5 & $3.568214$ & $-30.400358$\\  
A2744, M9 & $4.0280$ & $0.78\pm 0.22$ & $44.55_{- 2.43}^{+ 8.97}$ & $39.44$ & $14.1\pm1.7$ & 124.7 & $3.582152$ & $-30.397957$\\  
A2744, 3672 & $4.0423$ & $22.00\pm 1.66$ & $1.77_{- 0.03}^{+ 0.03}$ & $42.29$ & $100.0\pm0.0$ & 15893.3 & $3.569342$ & $-30.408732$\\  
A2744, 4378 & $4.0450$ & $2.84\pm 0.55$ & $1.82_{- 0.04}^{+ 0.03}$ & $41.39$ & $68.9\pm2.1$ & 12832.3 & $3.567564$ & $-30.406075$\\  
A2744, 1903 & $4.0527$ & $4.59\pm 0.55$ & $3.20_{- 0.04}^{+ 0.05}$ & $41.36$ & $71.1\pm1.7$ & 13441.5 & $3.595858$ & $-30.416496$\\  
A2744, M1 & $4.1924$ & $13.76\pm 0.62$ & $40.04_{- 5.08}^{+ 8.00}$ & $40.77$ & $100.0\pm0.1$ & 9503.7 & $3.591326$ & $-30.398643$\\  
A2744, 10340 & $4.3006$ & $19.82\pm 1.95$ & $8.13_{- 0.28}^{+ 0.39}$ & $41.65$ & $37.8\pm2.0$ & 6647.2 & $3.587131$ & $-30.388782$\\  
A2744, M23 & $4.3088$ & $3.99\pm 0.60$ & $1.98_{- 0.05}^{+ 0.04}$ & $41.57$ & $87.5\pm1.4$ & 13547.6 & $3.601358$ & $-30.388689$\\  
A2744, 5574 & $4.3342$ & $6.55\pm 0.73$ & $2.60_{- 0.05}^{+ 0.06}$ & $41.67$ & $98.6\pm0.5$ & 14004.8 & $3.603312$ & $-30.403131$\\  
A2744, 4926 & $4.3361$ & $139.51\pm 1.82$ & $3.76_{- 0.09}^{+ 0.14}$ & $42.84$ & $99.7\pm0.2$ & 16215.6 & $3.601898$ & $-30.405007$\\  
A2744, 9683 & $4.3602$ & $2.20\pm 0.43$ & $2.06_{- 0.05}^{+ 0.04}$ & $41.30$ & $93.5\pm1.1$ & 13105.2 & $3.600716$ & $-30.390730$\\  
A2744, M25 & $4.3663$ & $2.44\pm 0.34$ & $15.25_{- 0.65}^{+ 0.37}$ & $40.48$ & $80.6\pm1.7$ & 5901.1 & $3.582196$ & $-30.390919$\\  
A2744, 9089 & $4.3748$ & $10.24\pm 0.82$ & $1.95_{- 0.03}^{+ 0.04}$ & $42.00$ & $99.8\pm0.2$ & 15572.1 & $3.602202$ & $-30.392816$\\  
A2744, 3837 & $4.3920$ & $22.47\pm 0.91$ & $2.13_{- 0.06}^{+ 0.02}$ & $42.31$ & $100.0\pm0.0$ & 16051.1 & $3.574057$ & $-30.407694$\\  
A2744, 3275 & $4.4002$ & $10.68\pm 0.94$ & $2.45_{- 0.06}^{+ 0.03}$ & $41.92$ & $99.2\pm0.4$ & 14908.4 & $3.577731$ & $-30.409784$\\  
A2744, 10305 & $4.4013$ & $14.15\pm 1.34$ & $4.43_{- 0.21}^{+ 0.31}$ & $41.79$ & $99.4\pm0.4$ & 14890.4 & $3.571465$ & $-30.388822$\\  
A2744, 4321 & $4.6315$ & $9.11\pm 0.71$ & $1.82_{- 0.04}^{+ 0.03}$ & $42.04$ & $99.7\pm0.3$ & 15121.9 & $3.567559$ & $-30.406253$\\  
A2744, 6505 & $4.6892$ & $6.99\pm 0.55$ & $2.48_{- 0.05}^{+ 0.06}$ & $41.80$ & $99.8\pm0.2$ & 15501.0 & $3.571383$ & $-30.400133$\\  
A2744, 10644 & $4.6974$ & $10.67\pm 0.90$ & $1.80_{- 0.05}^{+ 0.04}$ & $42.12$ & $99.9\pm0.1$ & 15751.0 & $3.604256$ & $-30.387246$\\  
A2744, M26 & $4.7026$ & $3.28\pm 0.38$ & $4.77_{- 0.11}^{+ 0.15}$ & $41.19$ & $87.4\pm1.5$ & 12921.5 & $3.601591$ & $-30.412696$\\  
A2744, 10338 & $4.7125$ & $16.12\pm 1.13$ & $4.46_{- 0.20}^{+ 0.20}$ & $41.91$ & $99.9\pm0.1$ & 15524.2 & $3.574497$ & $-30.388774$\\  
A2744, 2674 & $4.7283$ & $11.13\pm 1.24$ & $1.86_{- 0.03}^{+ 0.03}$ & $42.14$ & $99.1\pm0.4$ & 14054.1 & $3.574354$ & $-30.412531$\\  
A2744, 2874 & $4.7283$ & $5.70\pm 0.60$ & $2.61_{- 0.08}^{+ 0.04}$ & $41.70$ & $99.2\pm0.4$ & 14670.9 & $3.580236$ & $-30.411354$\\  
A2744, M27 & $4.7540$ & $4.06\pm 0.64$ & $5.13_{- 0.09}^{+ 0.18}$ & $41.26$ & $25.2\pm2.0$ & 5747.2 & $3.591995$ & $-30.414036$\\  
A2744, 5488 & $4.7616$ & $4.55\pm 0.85$ & $13.49_{- 0.51}^{+ 0.85}$ & $40.89$ & $15.4\pm1.7$ & 883.2 & $3.585942$ & $-30.403157$\\  
A2744, 2264 & $4.7786$ & $5.11\pm 0.78$ & $4.45_{- 0.07}^{+ 0.11}$ & $41.43$ & $92.1\pm1.1$ & 11664.2 & $3.598817$ & $-30.414598$\\  
A2744, 2077 & $4.7804$ & $13.95\pm 0.73$ & $4.63_{- 0.18}^{+ 0.16}$ & $41.85$ & $100.0\pm0.0$ & 15775.8 & $3.602018$ & $-30.415740$\\  
A2744, 11772 & $4.7984$ & $7.07\pm 0.57$ & $2.40_{- 0.07}^{+ 0.08}$ & $41.84$ & $99.7\pm0.2$ & 15520.6 & $3.595924$ & $-30.386398$\\  
A2744, 10594 & $4.8018$ & $27.00\pm 1.54$ & $5.42_{- 0.12}^{+ 0.16}$ & $42.07$ & $100.0\pm0.1$ & 15738.1 & $3.582351$ & $-30.387678$\\  
A2744, M28 & $4.8660$ & $1.43\pm 0.19$ & $3.51_{- 0.10}^{+ 0.11}$ & $41.00$ & $90.6\pm1.3$ & 13656.0 & $3.583527$ & $-30.381314$\\  
A2744, 3492 & $4.8938$ & $3.50\pm 0.53$ & $2.65_{- 0.13}^{+ 0.07}$ & $41.51$ & $86.1\pm1.5$ & 11999.3 & $3.574447$ & $-30.408904$\\  
A2744, M29 & $4.9020$ & $0.87\pm 0.23$ & $2.03_{- 0.07}^{+ 0.08}$ & $41.03$ & $55.8\pm2.0$ & 10034.8 & $3.599947$ & $-30.383753$\\  
A2744, 10972 & $4.9116$ & $1.88\pm 0.39$ & $3.58_{- 0.10}^{+ 0.12}$ & $41.12$ & $74.6\pm1.8$ & 11445.0 & $3.586659$ & $-30.382469$\\  
A2744, M40 & $4.9139$ & $3.95\pm 0.50$ & $3.77_{- 0.26}^{+ 0.12}$ & $41.42$ & $99.5\pm0.3$ & 14741.6 & $3.605636$ & $-30.385219$\\  
A2744, 11629 & $4.9823$ & $9.05\pm 0.88$ & $2.66_{- 0.07}^{+ 0.10}$ & $41.95$ & $98.9\pm0.5$ & 14720.8 & $3.594384$ & $-30.385804$\\  
A2744, 4946 & $5.0193$ & $4.96\pm 0.69$ & $1.93_{- 0.04}^{+ 0.04}$ & $41.83$ & $52.9\pm2.0$ & 12336.5 & $3.568397$ & $-30.404557$\\  
A2744, 12026 & $5.0537$ & $8.62\pm 1.04$ & $2.52_{- 0.07}^{+ 0.08}$ & $41.96$ & $11.3\pm1.3$ & 9808.2 & $3.595732$ & $-30.386781$\\  
A2744, 12404 & $5.0537$ & $8.55\pm 1.03$ & $2.52_{- 0.07}^{+ 0.08}$ & $41.96$ & $59.1\pm2.0$ & 11332.7 & $3.595425$ & $-30.386816$\\  
A2744, 9377 & $5.1349$ & $12.90\pm 1.33$ & $2.36_{- 0.04}^{+ 0.05}$ & $42.18$ & $94.5\pm0.9$ & 12386.3 & $3.597953$ & $-30.392021$\\  
A2744, 8885 & $5.1879$ & $3.34\pm 0.65$ & $1.75_{- 0.03}^{+ 0.04}$ & $41.73$ & $90.2\pm1.4$ & 14584.2 & $3.606663$ & $-30.393275$\\  
A2744, 4213 & $5.1933$ & $11.08\pm 0.87$ & $1.92_{- 0.03}^{+ 0.04}$ & $42.22$ & $99.4\pm0.3$ & 15508.4 & $3.570431$ & $-30.406540$\\  
A2744, 2821 & $5.2817$ & $3.96\pm 0.61$ & $7.98_{- 0.38}^{+ 0.32}$ & $41.17$ & $52.0\pm1.9$ & 12204.9 & $3.587924$ & $-30.411612$\\  
A2744, 10004 & $5.2896$ & $10.75\pm 1.21$ & $4.00_{- 0.12}^{+ 0.22}$ & $41.90$ & $98.3\pm0.6$ & 14947.1 & $3.572670$ & $-30.389755$\\  
A2744, M30 & $5.4316$ & $4.73\pm 0.64$ & $3.53_{- 0.13}^{+ 0.16}$ & $41.63$ & $71.0\pm2.1$ & 12912.9 & $3.571127$ & $-30.392950$\\  
A2744, M31 & $5.5364$ & $3.44\pm 0.73$ & $3.54_{- 0.08}^{+ 0.12}$ & $41.51$ & $29.2\pm1.9$ & 10209.6 & $3.591857$ & $-30.389259$\\  
A2744, 3306 & $5.5406$ & $2.73\pm 0.70$ & $1.73_{- 0.03}^{+ 0.03}$ & $41.72$ & $39.3\pm2.1$ & 12351.3 & $3.568118$ & $-30.409713$\\  
A2744, M32 & $5.5601$ & $2.58\pm 0.62$ & $3.32_{- 0.08}^{+ 0.09}$ & $41.42$ & $37.1\pm2.0$ & 9668.8 & $3.580342$ & $-30.405810$\\  
A2744, 11194 & $5.6094$ & $9.10\pm 0.91$ & $2.53_{- 0.09}^{+ 0.08}$ & $42.09$ & $99.8\pm0.2$ & 15433.4 & $3.594768$ & $-30.384450$\\  
A2744, 10111 & $5.6218$ & $6.23\pm 0.82$ & $4.99_{- 0.19}^{+ 0.24}$ & $41.63$ & $98.6\pm0.5$ & 14519.9 & $3.575846$ & $-30.389290$\\  
A2744, M3 & $5.6596$ & $8.30\pm 0.62$ & $4.28_{- 0.15}^{+ 0.12}$ & $41.83$ & $99.2\pm0.3$ & 14785.7 & $3.582261$ & $-30.407166$\\  
A2744, M33 & $5.6608$ & $12.41\pm 1.10$ & $149.96_{- 14.99}^{+ 797.40}$ & $40.46$ & $54.8\pm1.8$ & 126.2 & $3.591109$ & $-30.398974$\\  
A2744, 8268 & $5.6618$ & $160.30\pm 2.26$ & $8.50_{- 0.35}^{+ 0.42}$ & $42.82$ & $100.0\pm0.0$ & 15962.9 & $3.590711$ & $-30.395561$\\  
A2744, 5408 & $5.7219$ & $32.37\pm 0.41$ & $28.01_{- 1.88}^{+ 3.18}$ & $41.62$ & $100.0\pm0.0$ & 15772.2 & $3.584398$ & $-30.403397$\\  
A2744, 11559 & $5.7637$ & $4.65\pm 0.68$ & $3.56_{- 0.12}^{+ 0.17}$ & $41.68$ & $93.7\pm1.1$ & 13573.0 & $3.574547$ & $-30.385244$\\  
A2744, 3472 & $5.7648$ & $3.22\pm 0.46$ & $1.80_{- 0.03}^{+ 0.03}$ & $41.81$ & $65.7\pm1.9$ & 12439.3 & $3.569699$ & $-30.409056$\\  
A2744, 11471 & $5.7668$ & $3.80\pm 0.55$ & $2.30_{- 0.07}^{+ 0.08}$ & $41.78$ & $87.3\pm1.3$ & 13356.5 & $3.596271$ & $-30.384448$\\  
A2744, 7747 & $5.7709$ & $4.66\pm 0.67$ & $1.89_{- 0.02}^{+ 0.04}$ & $41.95$ & $97.7\pm0.7$ & 13972.6 & $3.605435$ & $-30.396596$\\  
A2744, 8116 & $5.7751$ & $1.35\pm 0.19$ & $1.82_{- 0.03}^{+ 0.03}$ & $41.43$ & $52.3\pm2.4$ & 10488.7 & $3.606248$ & $-30.395581$\\  
A2744, M34 & $5.8994$ & $2.28\pm 0.37$ & $3.32_{- 0.14}^{+ 0.15}$ & $41.42$ & $92.6\pm1.1$ & 13712.5 & $3.575055$ & $-30.380692$\\  
A2744, M35 & $5.9971$ & $2.06\pm 0.29$ & $2.35_{- 0.06}^{+ 0.06}$ & $41.54$ & $38.9\pm2.3$ & 11330.8 & $3.568441$ & $-30.399065$\\  
A2744, M36 & $6.0938$ & $2.43\pm 0.51$ & $2.13_{- 0.04}^{+ 0.04}$ & $41.67$ & $44.9\pm2.3$ & 14502.1 & $3.578052$ & $-30.413160$\\  
A2744, 2785 & $6.2737$ & $0.57\pm 0.29$ & $1.68_{- 0.03}^{+ 0.03}$ & $41.17$ & $69.9\pm1.8$ & 12638.7 & $3.567632$ & $-30.411871$\\  
A2744, 5353 & $6.3271$ & $6.58\pm 0.63$ & $3.73_{- 0.12}^{+ 0.12}$ & $41.90$ & $94.3\pm1.1$ & 14495.0 & $3.601073$ & $-30.403989$\\  
A2744, 10609 & $6.3755$ & $1.34\pm 0.19$ & $2.28_{- 0.07}^{+ 0.07}$ & $41.43$ & $57.1\pm2.2$ & 12540.9 & $3.598490$ & $-30.387379$\\  
A2744, M37 & $6.5195$ & $1.89\pm 0.43$ & $3.36_{- 0.11}^{+ 0.11}$ & $41.44$ & $20.9\pm1.7$ & 10376.5 & $3.583060$ & $-30.411886$\\  
A2744, M38 & $6.5565$ & $1.48\pm 0.47$ & $3.45_{- 0.11}^{+ 0.09}$ & $41.32$ & $25.3\pm1.8$ & 12082.9 & $3.580148$ & $-30.407903$\\  
A2744, 2115 & $6.5876$ & $12.30\pm 1.27$ & $4.12_{- 0.05}^{+ 0.10}$ & $42.17$ & $58.1\pm2.2$ & 12310.1 & $3.593805$ & $-30.415448$\\  
A2744, M39 & $6.6439$ & $2.39\pm 0.35$ & $3.29_{- 0.08}^{+ 0.14}$ & $41.57$ & $68.5\pm1.8$ & 14415.0 & $3.588970$ & $-30.382048$\\  
\hline 
\end{longtable}       

\end{document}